\documentclass[onecolumn,superscriptaddress,floatfix,preprintnumbers, nofootinbib,hyperref]{revtex4-2} 
\pdfoutput=1
\usepackage[colorlinks=true,breaklinks=true]{hyperref}
\usepackage[normalem]{ulem}
\usepackage[utf8]{inputenc}
\hypersetup{allcolors=[rgb]{0.0 0.0 0.6},linkcolor=[rgb]{0.75 0.05 0.05}}
\usepackage{amsmath,amssymb}
\usepackage{epsfig}  
\usepackage{graphicx}   
\usepackage{slashed}       
\usepackage{tikz}
\usepackage{url}
\usepackage{color}
\usepackage{multirow}
\usepackage{orcidlink}
\usepackage{comment}
\usepackage{amssymb}
\usepackage[shortlabels]{enumitem}

\usepackage{empheq,etoolbox}
\patchcmd{\subequations}
  {\theparentequation\alph{equation}}
  {\theparentequation.\alph{equation}}
  {}{}
  
\DeclareMathOperator{\cm}{cm}

\DeclareMathOperator{\MeV}{MeV}

\DeclareMathOperator{\s}{s}

\DeclareMathOperator{\erg}{erg}

\DeclareMathOperator{\B}{B}

\newcommand{\beq}{\begin{equation}}
\newcommand{\eeq}{\end{equation}}

\allowdisplaybreaks

\bibpunct{[}{]}{,}{n}{}{,}

\begin{document}
\title{Simple fits for the neutrino luminosities from protoneutron star cooling}

\author{Giuseppe Lucente}
\email{giuseppe.lucente@ba.infn.it}
\affiliation{Dipartimento Interateneo di Fisica ``Michelangelo Merlin'', Via Amendola 173, 70126 Bari, Italy.}
\affiliation{Istituto Nazionale di Fisica Nucleare - Sezione di Bari, Via Orabona 4, 70126 Bari, Italy.}
\affiliation{Kirchhoff-Institut f\"ur Physik, Universit\"at Heidelberg, Im Neuenheimer Feld 227, 69120 Heidelberg, Germany}
\affiliation{Institut f\"ur Theoretische Physik, Universit\"at Heidelberg, Philosophenweg 16, 69120 Heidelberg, Germany}

\author{Malte Heinlein}
\email{heinlein@mpa-garching.mpg.de}
\affiliation{Max-Planck-Institut f\"ur Astrophysik, Karl-Schwarzschild-Str.~1, 85748 Garching, Germany}
\affiliation{Technische Universit\"at M\"unchen, TUM School of Natural Sciences,
Physics Department, James-Franck-Str.~1, 85748 Garching, Germany}

\author{Hans-Thomas Janka}
\email{thj@mpa-garching.mpg.de}
\affiliation{Max-Planck-Institut f\"ur Astrophysik, Karl-Schwarzschild-Str.~1, 85748 Garching, Germany}

\author{Alessandro Mirizzi}
\email{alessandro.mirizzi@ba.infn.it }
\affiliation{Dipartimento Interateneo di Fisica ``Michelangelo Merlin'', Via Amendola 173, 70126 Bari, Italy.}
\affiliation{Istituto Nazionale di Fisica Nucleare - Sezione di Bari, Via Orabona 4, 70126 Bari, Italy.}

\begin{abstract}
We propose a simple fit function, $L_{\nu_i}(t) = C\, t^{-\alpha}\, e^{-(t/\tau)^{n}}$,
to parametrize the luminosities of neutrinos and antineutrinos of all flavors during the protoneutron star (PNS) cooling phase at post-bounce times $t \gtrsim 1$\,s. This fit is based on results from a set of neutrino-hydrodynamics simulations of core-collapse supernovae in spherical symmetry. The simulations were performed with an energy-dependent transport for six neutrino species and took into account the effects of convection and muons in the dense and hot PNS interior. We provide values of the fit parameters $C$, $\alpha$, $\tau$, and $n$ for different neutron star masses and equations of state as well as correlations between these fit parameters. Our functional description is useful for analytic supernova modeling, for characterizing the neutrino light curves in large underground neutrino detectors, and as a tool to extract information from measured signals on the mass and equation of state of the PNS and on secondary signal components on top of the PNS's neutrino emission.
\end{abstract}

\maketitle

\section{Introduction}

The neutrino signal from a future Galactic supernova (SN) explosion represents one of the next frontiers of neutrino astrophysics (see, e.g., Refs.~\cite{Scholberg:2012id,Mirizzi:2015eza,
Nakamura:2016kkl,
Roberts:2016rsf,
Horiuchi:2018ofe}). Existing and planned large underground neutrino detectors guarantee that a high-statistics neutrino burst will be collected during such an event (see, e.g., Refs.~\cite{Scholberg:2012id,Mirizzi:2015eza,JUNO:2015zny,Lang:2016zhv,Hyper-Kamiokande:2021frf,Ankowski:2016lab,DUNE:2020zfm,Dighe:2003be}). This detection will be extremely important to probe the SN explosion
mechanism~\cite{Tamborra:2013laa,Muller:2019upo,Nagakura:2020qhb,Shibagaki:2020ksk,Nagakura:2021lma,Nagakura:2021yci,Lin:2022lck,Largani:2023oyk}, neutrino flavor conversions~\cite{Tamborra:2020cul,Volpe:2023met,Dasgupta:2023fdr,Capozzi:2022slf,Sen:2024fxa}, and particle physics~\cite{Raffelt:2011zab,Caputo:2024oqc,Sen:2024fxa} occurring in the deepest stellar regions.

In order to characterize the response of a detector to a SN neutrino burst, one has to rely on the outcome of state-of-the-art numerical SN simulations. Different from the case of solar neutrinos, a standard prediction for SN neutrino fluxes does not exist. Therefore, many current studies employ a parametric approach with a suitable range of variation guided by the results of the simulations.
For this purpose, it has been shown that the time-integrated SN neutrino fluxes are well represented by a spectrum with the functional form \cite{Keil+2003,Tamborra:2012ac}
\begin{equation}
f_\nu(E)\propto E^\beta e^{-(\beta+1)E/\langle E\rangle} \, ,
\end{equation}
where 
$\langle E \rangle$ is the average energy of a given neutrino species.

Concerning the time evolution of the neutrino luminosities, most of the detector characterizations use, for comparison, results obtained from numerical tables provided by simulations (see, e.g., \cite{Totani:1997vj,Fogli:2004ff,Scholberg:2012id,Hyper-Kamiokande:2021frf,JUNO:2023dnp,Kopke:2011xb}). However, in order to extend parametric studies also to the temporal evolution of the neutrino signal, it would be useful to have a simple functional prescription also for the luminosities, based on a few parameters that permit one to cover the range of variation obtained in the numerical SN simulations.
 
In this respect, the seminal work of Loredo and Lamb~\cite{Loredo:2001rx}, concerned with an analysis of SN~1987A data, assumed an exponential cooling 
model.\footnote{A more sophisticated model accounting for both accretion and cooling phase was presented in Ref.~\cite{Pagliaroli:2008ur}.}
This simple recipe was often used (see, e.g., Refs.~\cite{Beacom:1998ya,Gava:2009pj,IceCube:2011cwc})
for schematic estimations. However, as noticed already in Ref.~\cite{Raffelt:1996wa}, protoneutron star (PNS) cooling calculations do not yield exponential neutrino light curves but instead suggest that the neutrino luminosity is better described by a power-law decline~\cite{Woosley:1994ux}. A further step was taken in the recent work of Ref.~\cite{Li:2020ujl}, where on the grounds of an analysis of recent spherically symmetric PNS cooling simulations covering post-bounce times up to $t \sim 70$\,s, a combined ansatz of power-law and exponential behavior of the form $L_{\nu_i}(t) \propto  t^{-1}\, e^{-(t/\tau)^{n}}$ was proposed for the long-time behavior of the neutrino light curve, where $\tau$ is a characteristic timescale of the PNS cooling.\footnote{
Analytical parametrizations of the late-time SN neutrino signal have been also presented in Ref.~\cite{Suwa:2020nee}.}
In that work $\tau \sim 30$ s was adopted to handle the transition to the regime of neutrino transparency, and therefore the functional description of the neutrino light curve follows a simple decline according to $t^{-1}$ in the first 10\,s after core bounce. However, the models considered in Ref.~\cite{Li:2020ujl} did not include the effect of convection in the PNS. This has important consequences for the shape of the neutrino light curve, because the Kelvin-Helmholtz neutrino cooling of the PNS is strongly accelerated if convection persists for many seconds. 

In Ref.~\cite{Roberts:2011yw} it was shown that a kink in the neutrino light curve, when displayed in a doubly logarithmic form, signals the end of convection in the mantle layer of the PNS, while convection still continues in the deeper, high-density PNS core. The duration of PNS mantle convection and therefore the time of the kink depends on the nuclear equation of state (EoS), in particular on the behavior of the nuclear symmetry energy as a function of the density at and above nuclear saturation. The kink is also present in the count rate of a neutrino detection and thus could be easily observed. When it occurs late during the PNS cooling, it resembles a ``knee'' since it is followed by a steep decline when the PNS cools off and gradually becomes transparent to neutrinos. Otherwise the kink transitions into a longer, flatter tail of the light curve. Typically in modern models the knee is witnessed to show up at $t <10$\,s \cite{Fiorillo:2023frv,Pascal+2022}.

Inspired by these previous findings, we perform a detailed analysis of the neutrino luminosities obtained in a set of spherically symmetric PNS cooling models recently described in Ref.~\cite{Fiorillo:2023frv}. These simulations were performed with six-species neutrino transport based on a fully energy-dependent two-moment scheme with a variable Eddington closure obtained from the solution of the Boltzmann transport equation. Moreover, the effects of PNS convection and muons were taken into account. After this analysis we propose an accurate fit for the neutrino luminosities during the PNS cooling evolution at post-bounce times $t \gtrsim 1$\,s, based on the following functional form:
\begin{equation}
L_{\nu_i}(t) = C\, t^{-\alpha}\, e^{-(t/\tau)^{n}}\,,
\label{eq:fit}
\end{equation}
where the parameter $C$ is a normalization constant, $\alpha$ describes the power-law behavior in the early cooling phase, $\tau$ is a characteristic cooling time of the exponential drop at later times, and $n$ determines the steepness of the exponential decline at these late times. Importantly, for models including PNS convection, the value of $\tau$ is much shorter, $\tau< 10$\,s, than assumed in Ref.~\cite{Li:2020ujl}. Therefore, this fit function can be useful as a sensitive probe of the presence and duration of convection in the PNS mantle during the long-time neutrino cooling. Such a probe is complementary to the method of analysis considered in Ref.~\cite{Roberts:2011yw}, where the ratio of neutrino events at early and late times was proposed as a diagnostic measure.

In the present paper, we report our analysis and fitting procedure and provide extended tables with the parameter values for the fit functions of all our models and for interesting correlations between these fit parameters. The outline of the paper is as follows. In Section~\ref{Sec:data} we show the neutrino luminosities from our 1D PNS cooling simulations for different PNS masses and nuclear EoSs. 
In Section~\ref{sec:fit} we describe our analytical fitting of the neutrino and antineutrino luminosities of all flavors based on Eq.~(\ref{eq:fit}) for the models discussed before.
In Section~\ref{sec:depende} we discuss the dependence of the values of the fitting parameters on PNS mass and nuclear EoS and how their determination can help in deducing information on these latter quantities from a neutrino measurement. 
In Section~\ref{sec:correl} we discuss tight correlations of the parameter values on the one hand with the PNS mass (for fixed EoS) and, on the other hand, between each other (for fixed PNS mass). Then, in Section~\ref{sec:muons}, we consider how the previous results change for models that neglect the effects of PNS convection and of muons in the PNS cooling models.
In Section~\ref{sec:count} we show how our analytical recipe can be used to fit the event rate in a large neutrino underground detector.
Finally, in Section~\ref{sec:conclusions} we discuss possible applications of our fitting formula and finish with conclusions. 
In several appendices we provide extended tables with the values obtained by our numerical fitting and needed for practical use.

\section{Neutrino luminosities}
\label{Sec:data}

Our work is based on results from 1D hydrodynamical PNS cooling simulations employing the {\tt PROMETHEUS-VERTEX} neutrino-hydrodynamics code~\cite{Rampp:2002bq}, which solves the fully energy and velocity-dependent neutrino transport for all six species of neutrinos and antineutrinos with a state-of-the-art implementation of the neutrino interactions~\cite{Buras:2005rp,Janka:2012wk,Bollig:2017lki}.
The models we will use are taken from the Garching Core-collapse Supernova Archive~\cite{SNarchive} (all data are available upon request) and they are the same as those considered in the recent study of Ref.~\cite{Fiorillo:2023frv}, to which we refer the interested readers for more information on the model evolution and a discussion of the neutrino emission properties.

\begin{table}[t!]
\centering
 \begin{tabular}{|c|c|c|c|c|c|c|c|}
\hline
Model &$t_{\rm fin}$ [s] & $X_{\nu_e}^{\rm fin}$  &  $X_{\bar{\nu}_e}^{\rm fin}$   & $X_{\nu_\mu}^{\rm fin}$  &  $X_{\bar{\nu}_\mu}^{\rm fin}$ & $X_{\nu_\tau}^{\rm fin}$  &  $X_{\bar{\nu}_\tau}^{\rm fin}$ \\
\hline
1.36-DD2 & 8.69 & 0.023 & 0.015 &0.013 & 0.012 &0.011 &  0.011  \\
1.36-SFHo & 10.50 & 0.034 & 0.027 & 0.026 & 0.023 & 0.018 &  0.018  \\
1.36-SFHx & 10.06 & 0.059 &  0.054 & 0.060 & 0.056 & 0.046 & 0.045\\
1.36-LS220 & 12.36 & 0.078 & 0.065 & 0.075 & 0.067 & 0.064 &  0.062 \\
\hline
1.44-DD2 & 13.72 & 0.004 & 0.002 &0.002 & 0.002 &0.002 &  0.002  \\
1.44-SFHo & 15.00 & 0.006 & 0.004 & 0.003 & 0.003 & 0.003 &  0.003  \\
1.44-SFHx & 11.72 & 0.027 &  0.022& 0.020 & 0.018 & 0.014 & 0.014\\
1.44-LS220 & 14.84 & 0.031 & 0.024 & 0.029 & 0.024 & 0.020 &  0.019 \\
\hline
1.62-DD2 & 10.75 & 0.013 & 0.008 & 0.007  & 0.006 & 0.006 &0.006 \\
1.62-SFHo & 14.26 & 0.010 &  0.007 & 0.006  & 0.005 & 0.005 & 0.004\\
1.62-SFHx & 13.45 &  0.020 & 0.016 & 0.014& 0.013& 0.010 & 0.009\\
1.62-LS220 & 13.58 & 0.123 & 0.111 & 0.115 & 0.107 & 0.115 & 0.112 \\
\hline
1.77-DD2 & 11.26 & 0.015 & 0.009 & 0.009 & 0.008 & 0.007 & 0.007 \\
1.77-SFHo & 13.28 & 0.032 & 0.025 & 0.026  &  0.025 & 0.018 & 0.018\\
1.77-SFHx & 13.91 & 0.030 & 0.026 & 0.025 & 0.024 & 0.017 & 0.017\\
1.77-LS220 & 16.33 & 0.082 & 0.073  & 0.090 & 0.083 & 0.093 & 0.090 \\
\hline
1.93-DD2 & 12.81 & 0.009 & 0.005 & 0.006 & 0.005 & 0.005 & 0.004\\
1.93-SFHo & 15.52 & 0.017 & 0.012 & 0.012 & 0.011 & 0.007 & 0.007\\
1.93-SFHx & 16.38 & 0.016 & 0.013 & 0.011 & 0.010 & 0.007 & 0.007\\
1.93-LS220 & 19.95 & 0.037 & 0.034 & 0.048 & 0.045 & 0.053 & 0.051\\
\hline
1.62-DD2-c & 13.95 & 0.040 & 0.022 & 0.028 & 0.025 & 0.023 & 0.022\\
1.62-SFHo-c & 19.74 & 0.046 & 0.037 & 0.034 & 0.030 & 0.023 & 0.030\\
1.62-SFHx-c & 18.75 & \textbf{0.180} & \textbf{0.172} & $\textbf{0.219}^*$& \textbf{0.212} & \textbf{0.211} & \textbf{0.207}\\
1.61-LS220-c & 20.92 & \textbf{0.172} & 0.146 & $\textbf{0.185}^*$ & \textbf{0.172} & \textbf{0.185} & \textbf{0.181}\\
\hline
1.62-DD2-m & 9.58 & 0.014 & 0.009 & 0.009 & 0.008 & 0.009 & 0.008\\
1.62-SFHo-m & 13.55 & 0.009 & 0.006 & 0.005 & 0.005 & 0.005 & 0.005\\
		\hline
	\end{tabular}
	\caption{Final times $t_{\rm fin}$ and ratios $X_{\nu_i}^{\rm fin}$ [see Eq.~\eqref{eq:xfin}] for all PNS simulations and neutrino species. Values of $X_{\nu_i}^{\rm fin}$ larger than 0.15 are printed in bold and asterisks mark the largest values, $X_{\nu_\mu}^{\rm fin}=0.219$ and $X_{\nu_\mu}^{\rm fin}=0.185$, for models 1.62-SFHx-c and for 1.61-LS220-c, respectively. The detailed numerical neutrino outputs for all the simulations listed here are available at the Garching Core-Collapse Supernova Archive~\cite{SNarchive}.}
	\label{tab:lnufinal}
\end{table}

\begin{figure}[t!]
\centering
\includegraphics[width=0.49\textwidth]{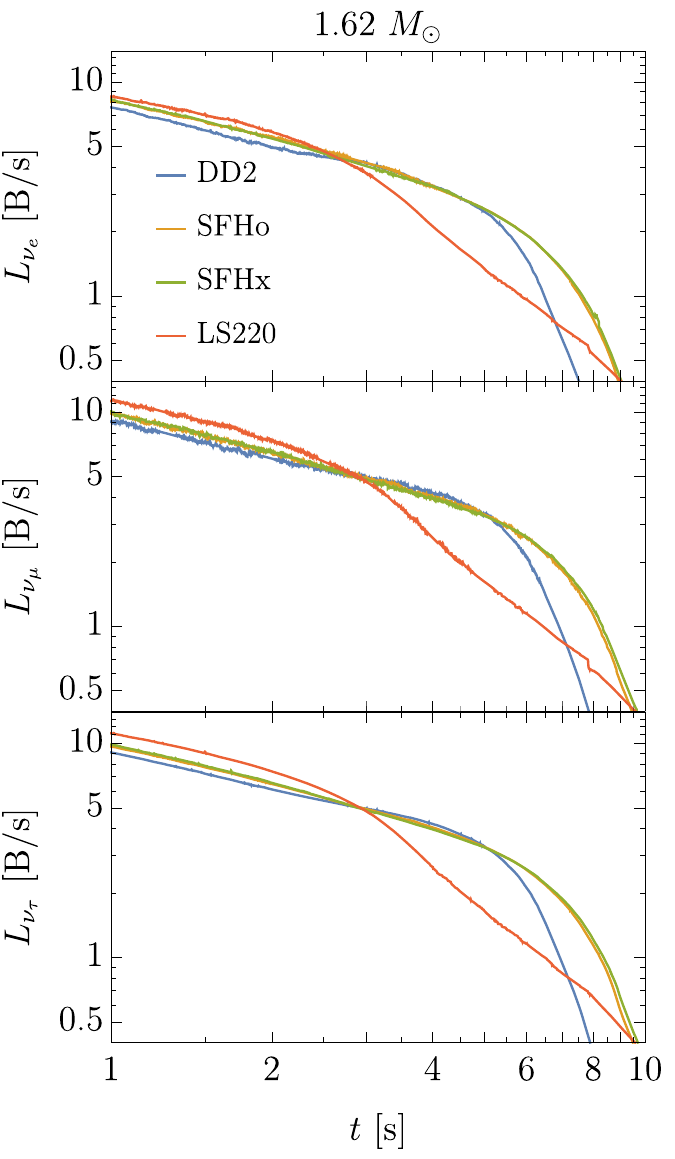}
\includegraphics[width=0.49\textwidth]{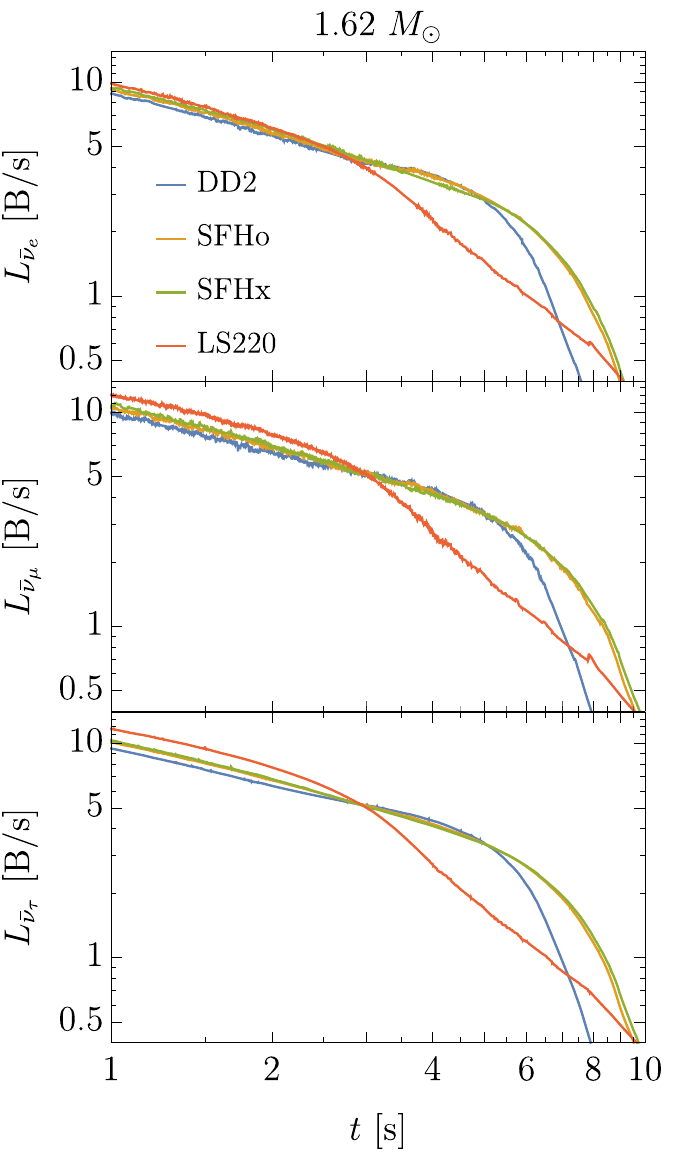}
\caption{Time evolution of the luminosities $L_{\nu_i}$ for neutrinos (left panels) and antineutrinos (right panels) of all flavors for a PNS baryonic mass of $M_{\rm NS}=1.62~M_\odot$ and all considered EoSs: DD2 (blue), SFHo (orange), SFHx (green) and LS220 (red).}
\label{Fig:lnu162}
\end{figure}

A concise and comprehensive summary of the physics inputs in the latest version of the {\tt PROMETHEUS-VERTEX} code is also provided by Ref.~\cite{Fiorillo:2023frv}.  
In particular, the discussed models include a 1D treatment of PNS convection via a mixing-length description of the convective fluxes~\cite{Mirizzi:2015eza} and take into account the presence of muons in the hot PNS, including the corresponding muonic neutrino interactions~\cite{Bollig:2017lki}, although
the differences in the luminosities and spectra of $\mu$ and $\tau$ neutrinos turn out to be relatively small during most of the PNS cooling evolution (unless the PNS is very massive).

We consider simulations that yield baryonic PNS masses\footnote{Since the neutrino emission is tightly correlated with properties of the PNS, we use this quantity as reference for a systematic variation instead of the progenitor mass, which is not monotonically related with the PNS mass.}
of 1.36\,$M_\odot$, 1.44\,$M_\odot$, 1.62\,$M_\odot$, 1.77\,$M_\odot$, and 1.93\,$M_\odot$, in each case computed with four different nuclear EoSs, namely DD2~\cite{Typel+2010,Hempel+2010,Hempel+2012}, SFHo, SFHx~\cite{Steiner+2013,Hempel+2010}, and LS220 with a nuclear incompressibility at saturation density of $K = 220$\,MeV~\cite{Lattimer+1991}. Correspondingly, we follow Ref.~\cite{Fiorillo:2023frv} in our naming convention of the simulations, specifying the PNS mass and the EoS in the model names, e.g. 1.62-DD2. Our standard simulations include PNS convection and muons; those without convection are denoted by a suffix ``-c'' to their names, and those without muons by a suffix ``-m''. In Table~\ref{tab:lnufinal} we list all of the discussed simulations and the final post-bounce times $t_{\rm fin}$ when they were stopped.

In this work we are interested in the post-accretion phase of the PNSs born in core-collapse SNe, for which reason we focus on the evolution of the neutrino signal only at post-bounce times $t \gtrsim 1$\,s. The discussed PNS models result from 1D simulations with initial conditions from several 1D models of stellar progenitors with different pre-collapse masses. These models are artificially exploded at different instants after bounce, which are chosen such that the PNS mass after the end of the post-bounce accretion attains the desired value. The neutrino-driven explosions are triggered by a sufficiently strong reduction of the density and thus of the ram pressure in the infall region upstream of the stalled SN shock (for more details, see Ref.~\cite{Fiorillo:2023frv}). Although it is common knowledge now that SN explosions are a 3D phenomenon and 3D simulations are needed to investigate the explosion mechanism and the post-bounce accretion phase as well as the associated neutrino signal, the late evolution times considered in our study, $t \gtrsim 1$\,s, are sufficiently well represented by our 1D simulations if the PNS does not experience continued accretion that extends beyond 1\,s after bounce. Another prerequisite is that our mixing-length treatment of PNS convection is a good approximation of the 3D hydrodynamics and associated energy and lepton transport that takes place inside the PNS over timescales of many seconds.

\begin{figure}[t!]
\centering
\includegraphics[width=0.49\textwidth]{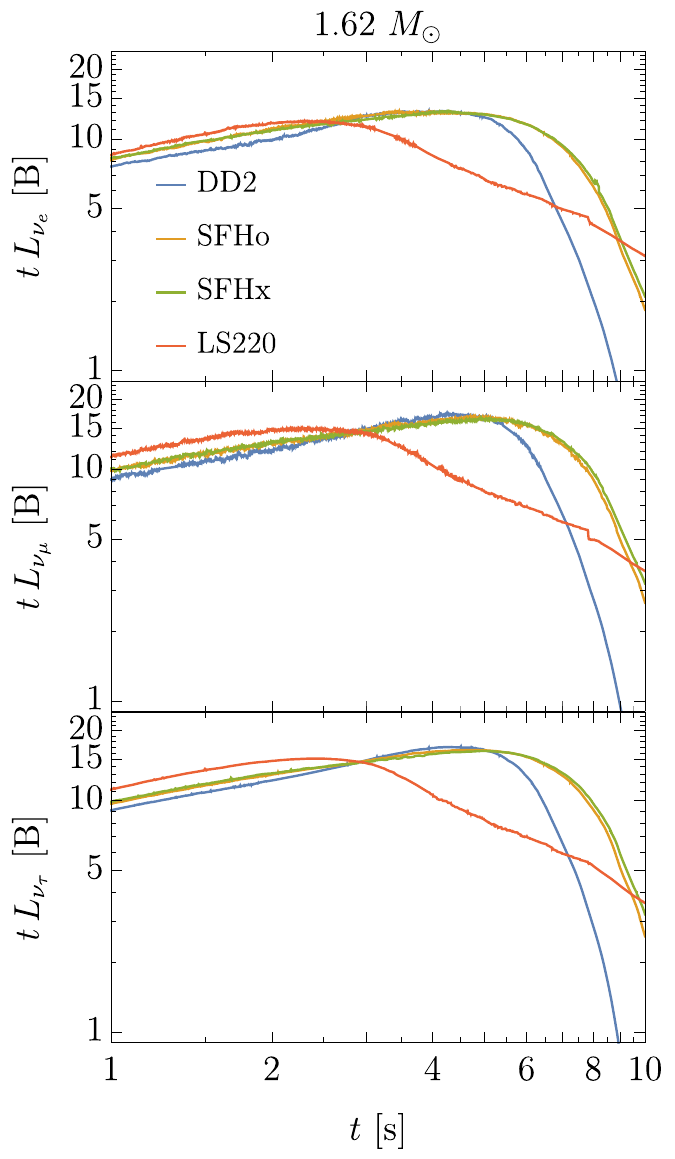}
\includegraphics[width=0.49\textwidth]{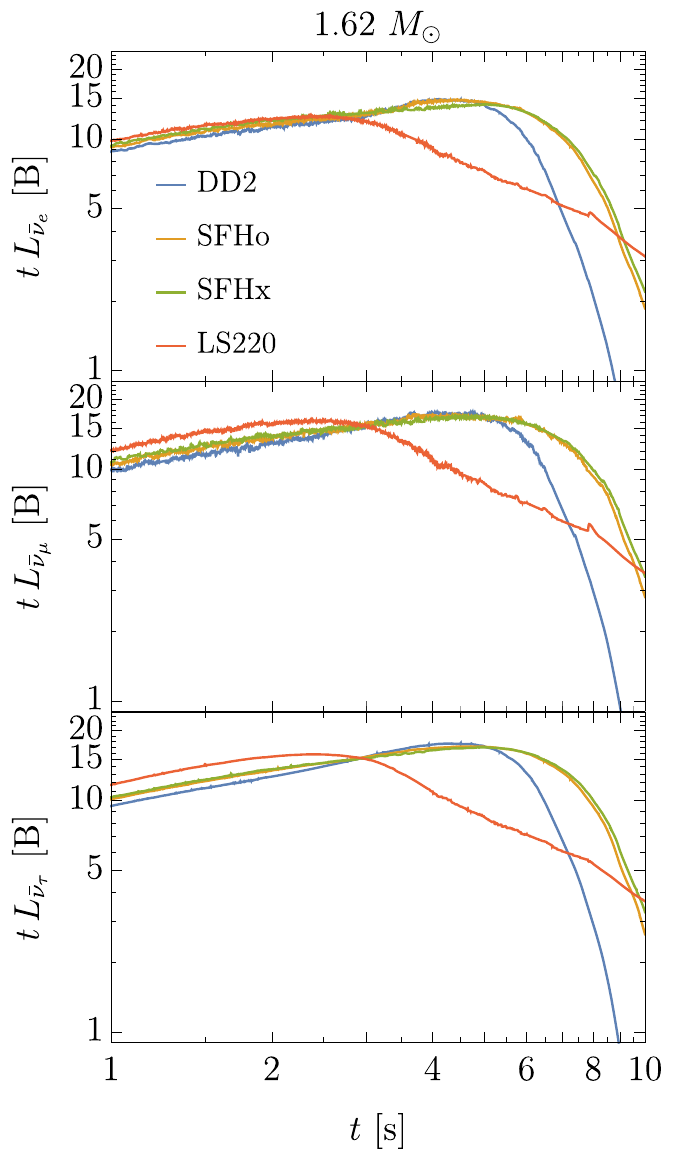}
\caption{Time evolution of the product of time and luminosities, $t\,L_{\nu_i}$, for neutrinos (left panels) and antineutrinos (right panels) of all flavors for $M_{\rm NS}=1.62~M_\odot$ and all considered EoSs: DD2 (blue), SFHo (orange), SFHx (green), and LS220 (red).}
\label{Fig:tlnu162}
\end{figure}

Figure~\ref{Fig:lnu162} displays the time evolution of the luminosities $L_{\nu}(t)$ for all the neutrino (left panels) and antineutrino (right panels) flavors in units of $\mathrm{B/s}$ ($1~\B = 10^{51}~\erg$) in the time interval [1,10]~s for the 1.62~$M_\odot$ models, computed with the set of considered EoSs cases: DD2 in blue, SFHo in orange, SFHx in green, and LS220 in red. From this figure it is evident that the neutrino luminosities exhibit a change of their slope in the time interval considered. Since we expect a power-law behavior of the neutrino luminosities in the early cooling phase and we also want to enhance the visibility of the change in the steepness of the luminosity decline at later times, we take inspiration from Ref.~\cite{Li:2020ujl} and show in Fig.~\ref{Fig:tlnu162} the product of the post-bounce time and the luminosities, $t\,L_{\nu}(t)$, for all of the neutrino (left panels) and antineutrino (right panels) flavors in the same time interval and for the same simulations as in Fig.~\ref{Fig:lnu162}. One witnesses an interesting difference of the results for the employed EoSs. For DD2, SFHo and SFHx EoSs and all of the neutrino species the quantity $t\,L_\nu$ increases as a power-law at $t\gtrsim 1~\s$ to peak around $t\approx 4~\s$, followed by a steep suppression at later times. This behavior produces a prominent knee at about 5--6\,s after bounce. In contrast, for LS220 $t\,L_\nu$ reaches its peak earlier, namely at $t\lesssim 3~\s$, and it features a slower decline later on. Since for each model the luminosities of all heavy-lepton neutrino species show a similar evolution, we will focus exclusively on the $\bar{\nu}_\mu$ signal as representative of all non-electron neutrino species, unless otherwise noted.

\begin{figure}[t!]
\centering
\includegraphics[width=0.45\textwidth]{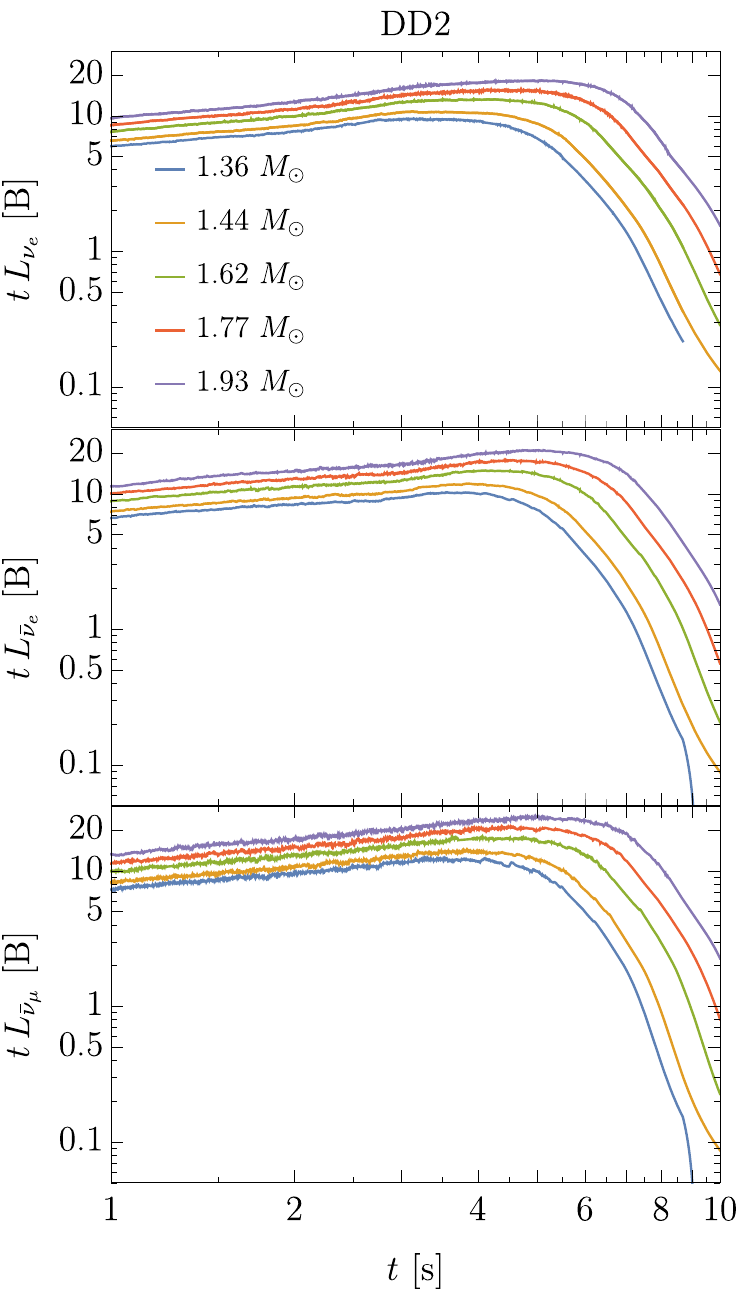}
\includegraphics[width=0.45\textwidth]{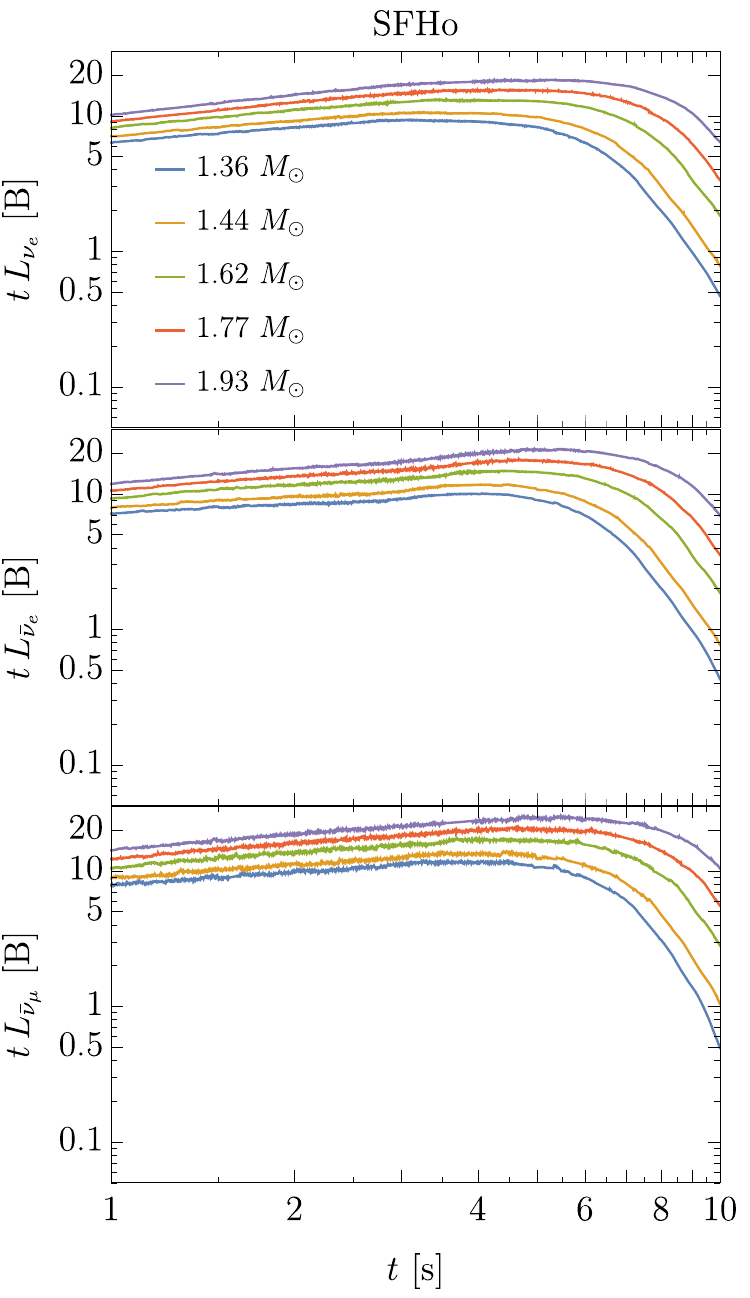}
\caption{Time evolution of the product of time and luminosity, $t\,L_{\nu_i}$, for $\nu_e$ (upper panels), $\bar{\nu}_e$ (middle panels) and $\bar{\nu}_\mu$ (lower panels) for DD2 (left) and SFHo (right) and all investigated PNS masses: $M_{\rm NS}=1.36~M_\odot$ (blue), $M_{\rm NS}=1.44~M_\odot$ (orange), $M_{\rm NS}=1.62~M_\odot$ (green), $M_{\rm NS}=1.77~M_\odot$ (red) and $M_{\rm NS}=1.93~M_\odot$ (purple).}
\label{Fig:tlnudd2}
\end{figure}

\begin{figure}[t!]
\centering
\includegraphics[width=0.45\textwidth]{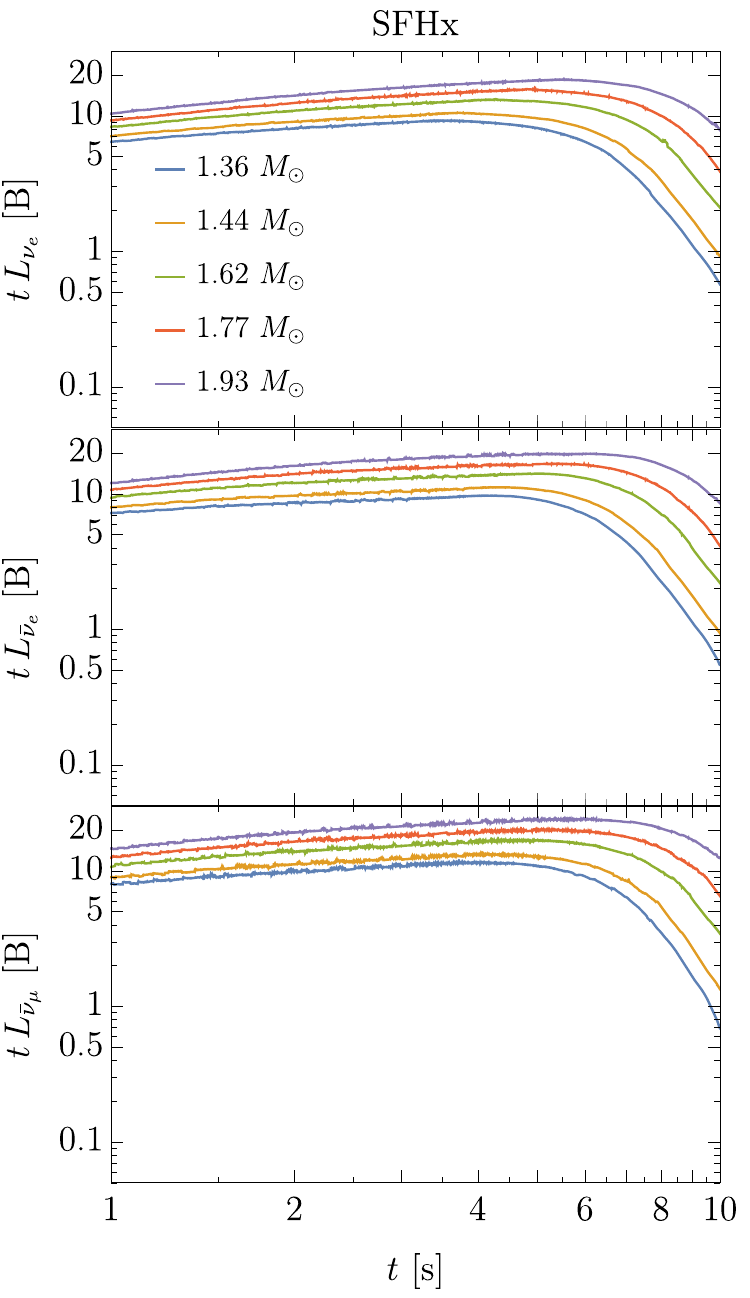}
\includegraphics[width=0.45\textwidth]{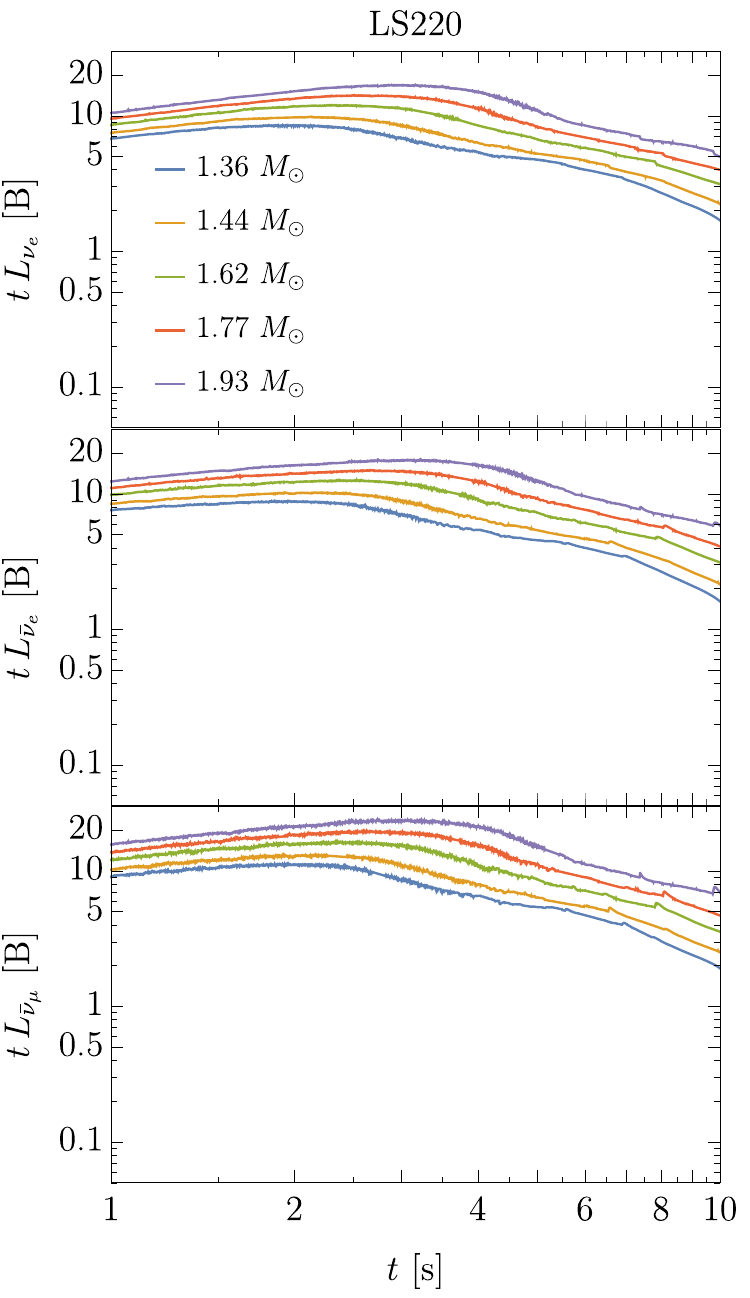}
\caption{Time evolution of the product of time and luminosity, $t\,L_{\nu_i}$, for $\nu_e$ (upper panels), $\bar{\nu}_e$ (middle panels) and $\bar{\nu}_\mu$ (lower panels) for SFHx (left) and LS220 (right) and all investigated PNS masses: $M_{\rm NS}=1.36~M_\odot$ (blue), $M_{\rm NS}=1.44~M_\odot$ (orange), $M_{\rm NS}=1.62~M_\odot$ (green), $M_{\rm NS}=1.77~M_\odot$ (red) and $M_{\rm NS}=1.93~M_\odot$ (purple).}
\label{Fig:tlnusfhx}
\end{figure}

In order to compare the mass dependence for the different EoS cases, we show the time evolution of $t\,L_\nu (t)$ for the different PNS masses keeping the EoS fixed. Specifically, in Fig.~\ref{Fig:tlnudd2} we consider the results for DD2 (left) and SFHo (right), and in Fig.~\ref{Fig:tlnusfhx} we present those for SFHx (left) and LS220 (right). In each panel the $1.36~M_\odot$ (blue), $1.44~M_\odot$ (orange), $1.62~M_\odot$ (green), $1.77~M_\odot$ (red) and $1.93~M_\odot$ (purple) models are plotted for $\nu_e$ (top), $\bar{\nu}_e$ (middle) and $\bar{\nu}_\mu$ (bottom). It is obvious that, at fixed time $t$, the luminosity becomes larger as the PNS mass increases. This dependence is similar for all EoSs and all neutrino species. Additionally, the maximum of $t\,L_\nu$ is shifted to later times and the subsequent decline starts correspondingly later for higher PNS masses. These findings, which are consistent with multi-D results in~\cite{Nagakura:2021yci}, can be explained by the fact that more gravitational binding energy is released in neutrinos when the PNS has a bigger mass.

For each case shown in Figs.~\ref{Fig:tlnu162}, \ref{Fig:tlnudd2}, and \ref{Fig:tlnusfhx}, the quantity  $t\,L_\nu$ has a peak at $t_{\nu_i,\,\rm max}\lesssim 6$\,s, when we define 
${L_{\nu_i,\,\rm max} \equiv L_{\nu_i}(t_{\nu_i\,\rm max})}$. The corresponding values of $t_{\nu_i,\,\rm max}$ are listed in Table~\ref{tab:lnumax} for neutrinos and in Table~\ref{tab:lnubarmax} for antineutrinos in Appendix~\ref{App:tmax}. In Table~\ref{tab:lnufinal} we provide the values of the quantity 
\begin{equation}
X_{\nu_i}^{\rm fin} \equiv \frac{t_{\rm fin}L_{\nu_i,\,\rm fin}}{t_{\nu_i,\rm max}L_{\nu_i,\,\rm max}}\,,
\label{eq:xfin}
\end{equation}
which is a measure for how long our simulations followed the late-time luminosity decline until they were stopped at $t_{\rm fin}$. For our benchmark PNS models (all models including muons and convection) $X_{\nu_i}^{\rm fin} \lesssim 0.15$ holds for all neutrinos and antineutrinos (the largest value is $X_{\nu_e}^{\rm fin}=0.123$ for 1.62-LS220). In order to adopt a well-defined, common final time for all of the neutrino species and simulations when fitting the luminosities in the following, we cut the data at an instant $t_{\nu_i,\rm c}$  when  $X_{\nu_i}^{\rm c}\equiv t_{\nu_i,\rm c}L_{\nu_i,\rm c}/(t_{\nu_i,\rm max}L_{\nu_i,\rm max})=0.15$. In addition to the values of $t_{\nu_i,\rm max}$ we also provide those of $t_{\nu_i,{\rm c}}$ for all neutrino species and simulations in Tables~\ref{tab:lnumax} and~\ref{tab:lnubarmax} in Appendix~\ref{App:tmax}. It can be seen that for a fixed EoS and neutrino species, the values of both of these quantities increase with the PNS mass. In contrast, given the PNS mass and EoS, $t_{\nu_i,\,\rm max}$ and $t_{\nu_i,{\rm c}}$ do not vary much among the different kinds of neutrinos and antineutrinos (e.g., for 1.62-DD2 $t_{\nu_i,{\rm max}}\approx 4\,\s$ and $t_{\nu_i,{\rm c}}\approx 8\,\s$ for all species). However, at fixed PNS mass, $t_{\nu_i,\,\rm max}$ and $t_{\nu_i,{\rm c}}$ feature a strong dependence on the considered EoS. In particular, $t_{\nu_i,{\rm c}}\lesssim 10\,\s$ for all the benchmark models with DD2, whereas SFHo and SFHx lead to slightly larger values of $t_{\nu_i,{\rm c}}$, with $t_{\nu_i,{\rm c}}\lesssim 12.5\,\s$ for all models, and in the cases with LS220 we find the highest values of $t_{\nu_i,{\rm c}}$, with $t_{\nu_i,{\rm c}}\lesssim 15\,\s$ for the largest PNS mass.

\begin{figure}[t!]
\centering
\includegraphics[width=.49\textwidth]{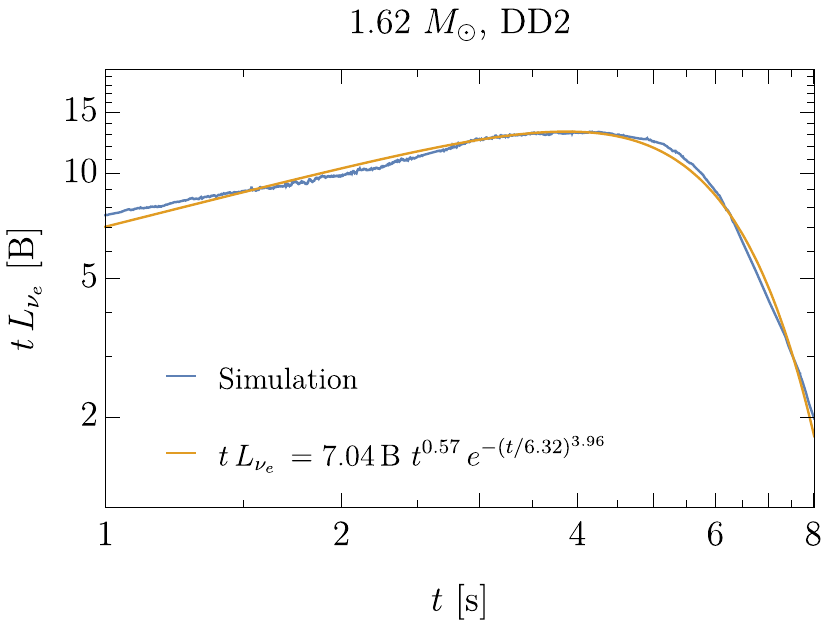}
\includegraphics[width=.49\textwidth]{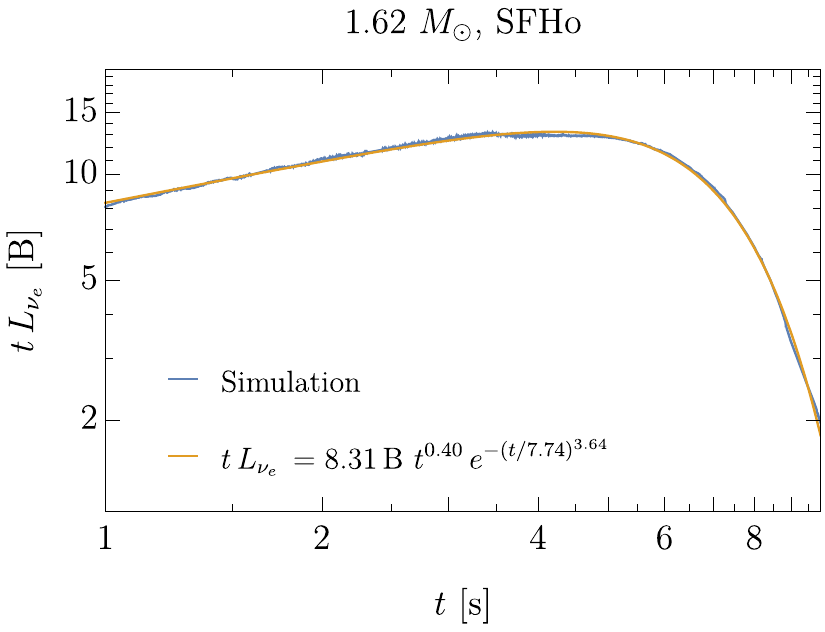}
\includegraphics[width=.49\textwidth]{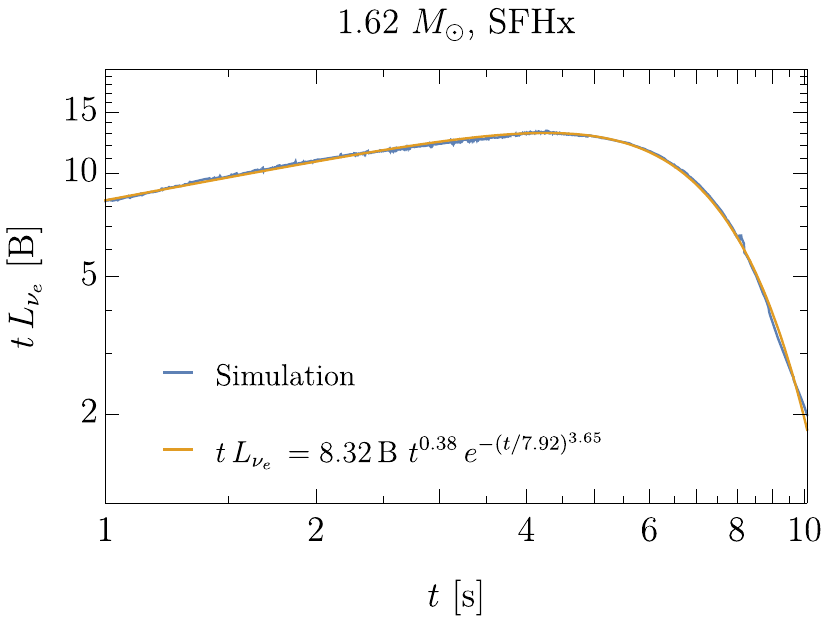}
\includegraphics[width=.49\textwidth]{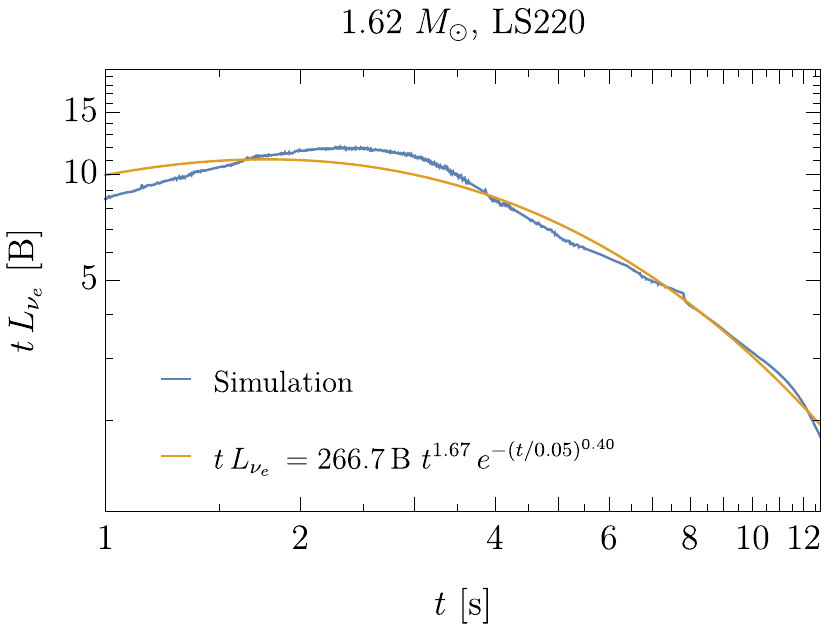}
\caption{Time evolution between 1\,s and $t_{\nu_e,{\rm c}}$ of the product of time and $\nu_e$ luminosity, $t\,L_{\nu_e}$, for simulation data (blue) and their fits (orange) for $M_{\rm NS}=1.62~M_\odot$ and different EoSs: DD2 (upper left), SFHo (upper right), SFHx (lower left) and LS220 (lower right).}
\label{Fig:nuefit162}
\end{figure}
\begin{figure}[t!]
\centering
\includegraphics[width=.49\textwidth]{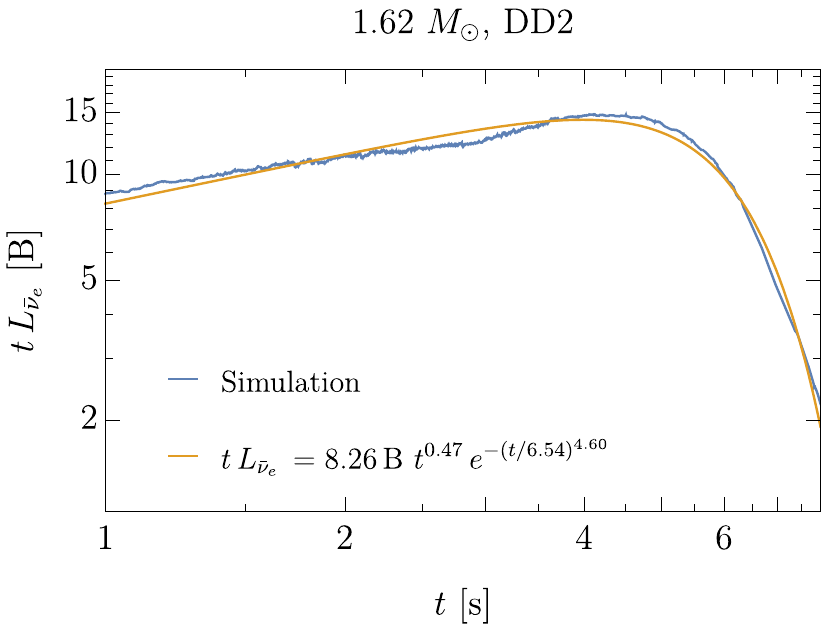}
\includegraphics[width=.49\textwidth]{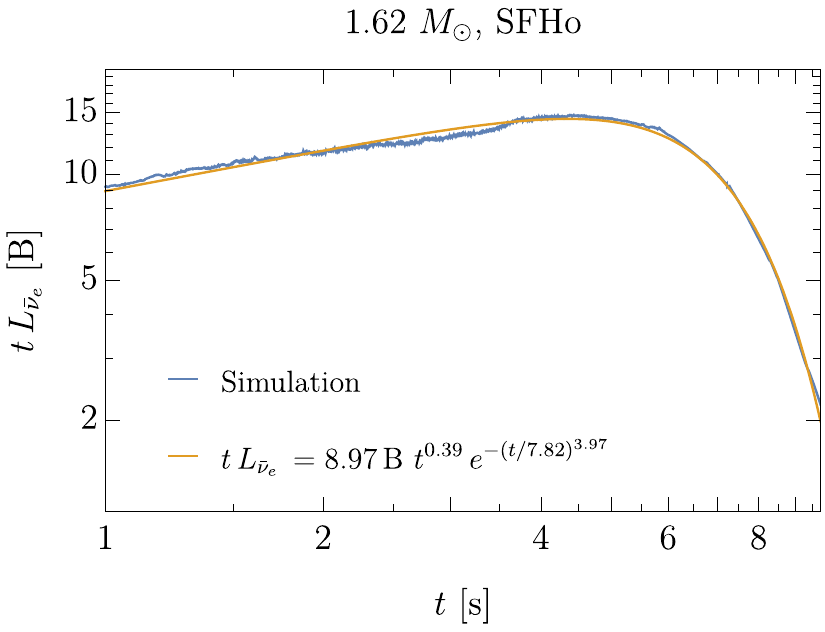}
\includegraphics[width=.49\textwidth]{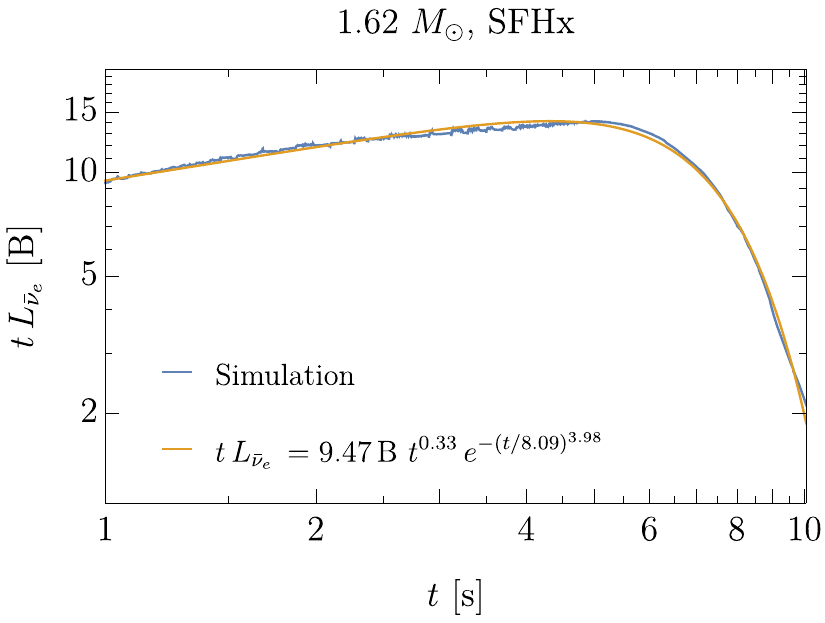}
\includegraphics[width=.49\textwidth]{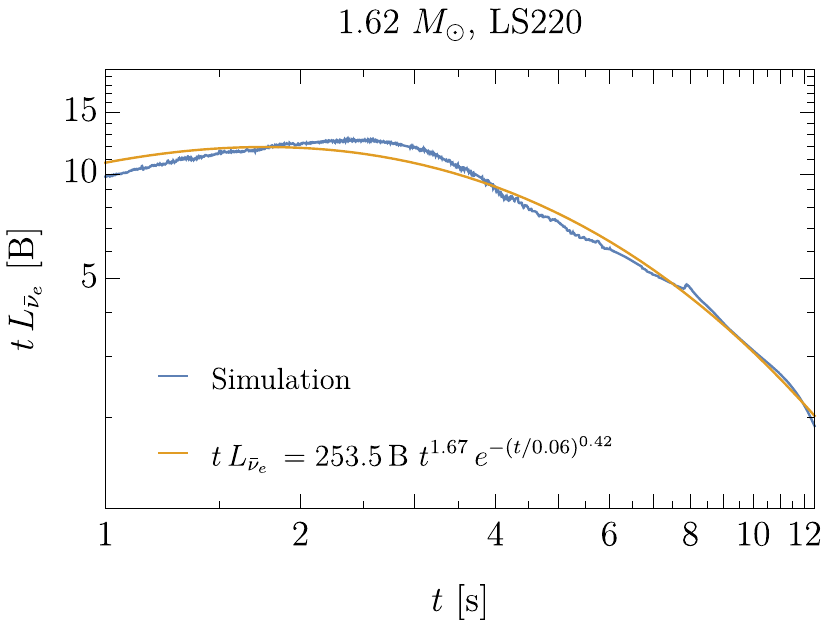}
\caption{Time evolution between 1\,s and $t_{\bar\nu_e,{\rm c}}$ of the product of time and $\bar\nu_e$ luminosity, $t\,L_{\bar\nu_e}$, for simulation data (blue) and their fits (orange) for $M_{\rm NS}=1.62~M_\odot$ and different EoSs: DD2 (upper left), SFHo (upper right), SFHx (lower left) and LS220 (lower right).}
\label{Fig:nuebfit162}
\end{figure}
\begin{figure}[t!]
\centering
\includegraphics[width=.49\textwidth]{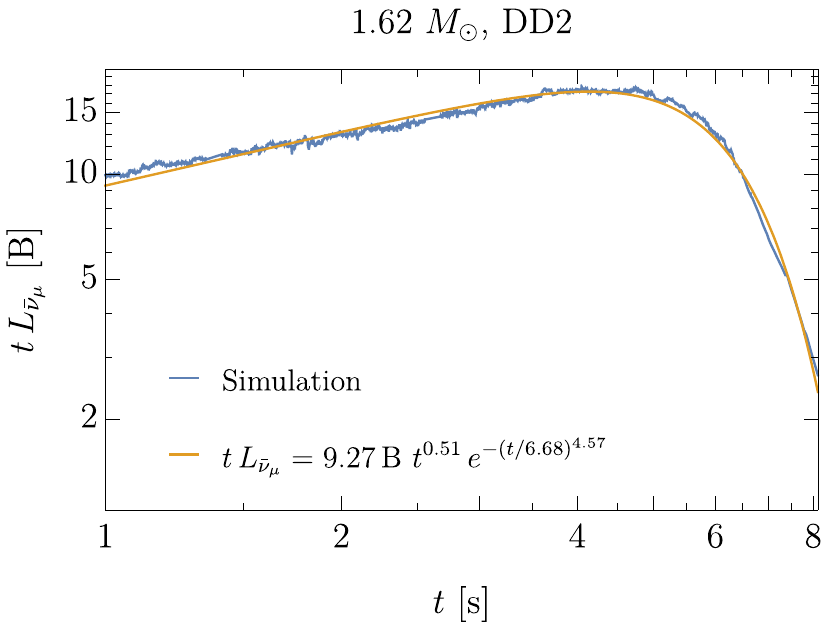}
\includegraphics[width=.49\textwidth]{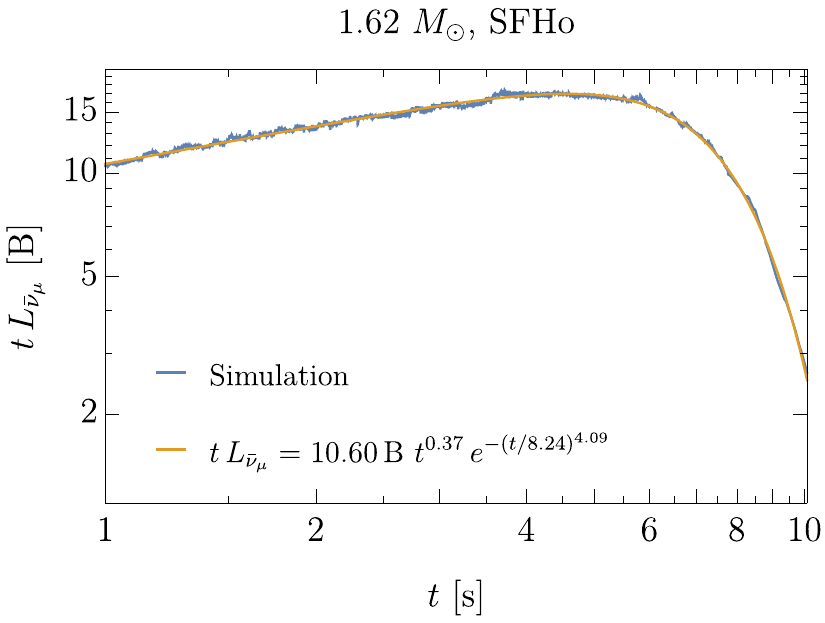}
\includegraphics[width=.49\textwidth]{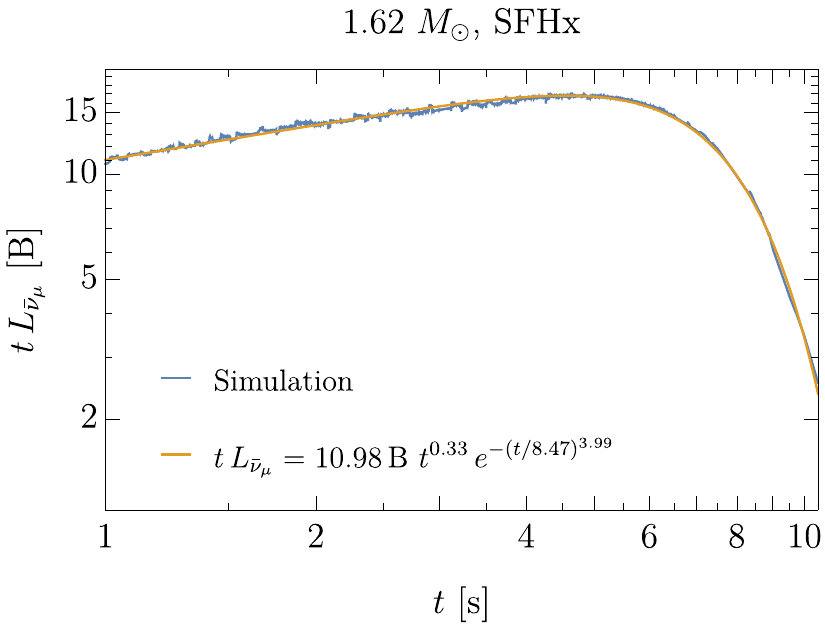}
\includegraphics[width=.49\textwidth]{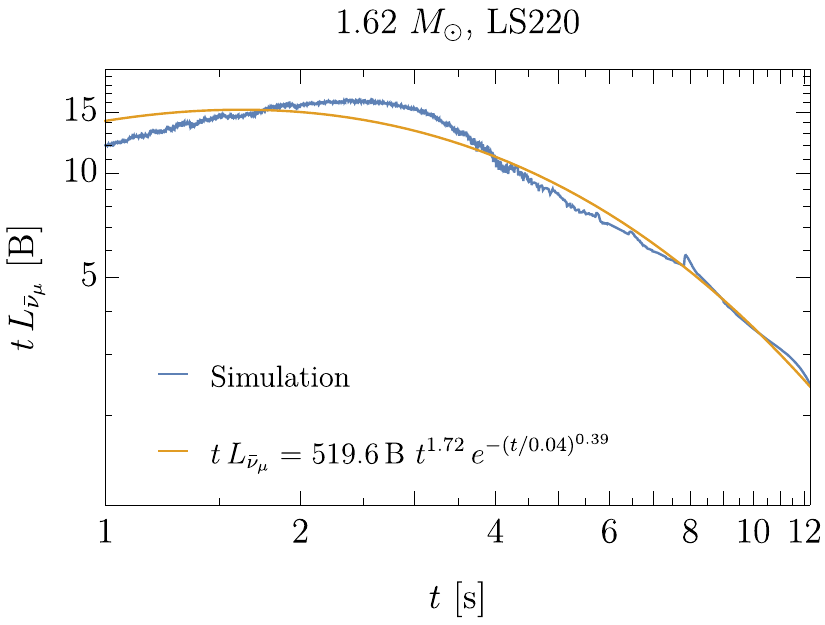}
\caption{Time evolution between 1\,s and $t_{\bar\nu_\mu,{\rm c}}$ of the product of time and $\bar\nu_\mu$ luminosity, $t\,L_{\bar\nu_\mu}$, for simulation data (blue) and their fits (orange) for $M_{\rm NS}=1.62~M_\odot$ and different EoSs: DD2 (upper left), SFHo (upper right), SFHx (lower left) and LS220 (lower right).}
\label{Fig:numbfit162}
\end{figure}

\section{Fit functions}
\label{sec:fit}

After an analysis of the models described above, we propose Eq.~(\ref{eq:fit}) as fit function for the neutrino luminosities in the considered time window:
\begin{equation}
L_{\nu_i}(t) = C\, t^{-\alpha}\, e^{-(t/\tau)^{n}}\,,
\nonumber
\end{equation}
where $C$, $\alpha$, $\tau$ and $n$ are free parameters and $t$ and $\tau$ are measured in seconds.\footnote{We define the general parameters as $C$, $\alpha$, $\tau$ and $n$ and we use $C_{\nu_i}$, $\alpha_{\nu_i}$, $\tau_{\nu_i}$ and $n_{\nu_i}$ when referring specifically to a neutrino species $\nu_i$.}
In Eq.~\eqref{eq:fit} the parameter $C$ is a normalization constant, $\alpha$ describes the power-law behavior at early cooling time deviating from the simple $t^{-1}$ behavior, $\tau$ is a characteristic cooling time for the exponential drop after the peak in $t\,L_\nu$, with $n$ representing the strength of the suppression at late times. We mention here that at the moment there is no straightforward argument explaining the power-law behavior at early cooling time, given that PNS convection plays an important role and it cannot be treated in simple ways. As an example, we show for the $1.62~M_\odot$ model and all EoSs the time evolution of $t\,L_{\nu_i}$ from simulations (blue lines) and its best fit (orange) between $t=1$\,s and $t_{\nu_i,{\rm c}}$, for $\nu_e$ in Fig.~\ref{Fig:nuefit162}, for $\bar{\nu}_e$ in Fig.~\ref{Fig:nuebfit162} and for $\bar{\nu}_\mu$ in Fig.~\ref{Fig:numbfit162}. For DD2, SFHo and SFHx the fit well reproduces the results from the simulations in the considered time interval, with best-fit parameters  in a similar range.
For each EoS, the highest values of the fit parameters $C$ and $\tau$ are obtained for $\bar{\nu}_{\mu}$, while $\nu_e$ show the lowest ones. A similar behavior is found for the other PNS masses. 

The fit for models with EoS LS220 shows worse agreement with the data from simulations (see the lower-right panels in Figs.~\ref{Fig:nuefit162}, \ref{Fig:nuebfit162}, and \ref{Fig:numbfit162}). In particular, for all neutrino species the fit overestimates the luminosity at $t\approx 1\,$s and exhibits a lower value of the peak, whereas it well reproduces the luminosity at later times. In this case, the values of the best-fit parameters are in a completely different range compared to those of the models with the other EoSs. In particular, for each neutrino species $C\approx \mathcal{O}(100)~\B/\s$ and $\alpha <0$, whereas $\tau$ and $n$ adopt much smaller values than for the other considered EoSs, with $\tau\lesssim 0.1\,$s and $n\lesssim0.5$, due to the more shallow decline of the luminosity at late times. The differences in the time dependence of the neutrino luminosities between models with different EoSs can be understood by the different behavior of the nuclear symmetry energy as a function of baryon density for the considered EoS. In the LS220 EoS, the symmetry energy exhibits a steeper increase with baryon density than in all other cases. As discussed in Ref.~\cite{Roberts:2011yw}, a larger positive derivative of the symmetry energy with baryon density can lead to suppressed convection in the PNS mantle at high densities and low electron fraction. This effect happens in the simulations with the LS220 EoS after about 3\,s and gradually quenches PNS mantle convection, thus delaying subsequent PNS neutrino cooling due to reduced neutrino luminosities. In contrast, PNS convection continues to be active in a spatially more extended region including the PNS mantle in simulations with the other three EoSs, for which reason convectively enhanced neutrino luminosities are maintained for a longer period of time until a late, steep decline follows when the PNS has deleptonized and cools off. These differences explain the early kink in the neutrino luminosities for models with the LS220 EoS, whereas a prominent knee-like shape of $L_\nu(t)$ can be witnessed for the DD2, SFHo, and SFHx models (see Fig.~\ref{Fig:lnu162}). These differences motivate us to define DD2, SFHo, and SFHx as members of an EoS-class that we call ``Class~A'', whereas LS220 is a representative of a ``Class~B''. The parameters of the symmetry energies for all the EoS cases used in our study are listed in Table~\ref{tab:symm_en} in Appendix~\ref{App:symm_en}.

In Appendix~\ref{App:bestfit} we report the values of the best-fit parameters and their $1\,\sigma$ uncertainties,\footnote{The best-fit values and errors are obtained with NonlinearModelFit function in \emph{Mathematica}.} for neutrinos and antineutrinos of all flavors and all considered models with different PNS masses and EoSs. For each species, the 1\,$\sigma$ errors on the best-fit parameters are within $\sim\mathcal{O}(0.1-1)\%$ for Class A EoSs, whereas for LS220 the maximum errors are $\sim\mathcal{O}(10)\%$ or even larger, and tend to increase at higher PNS masses, reflecting the lower quality of the fit.

\begin{figure}[t!]
\centering
\includegraphics[width=.49\textwidth]{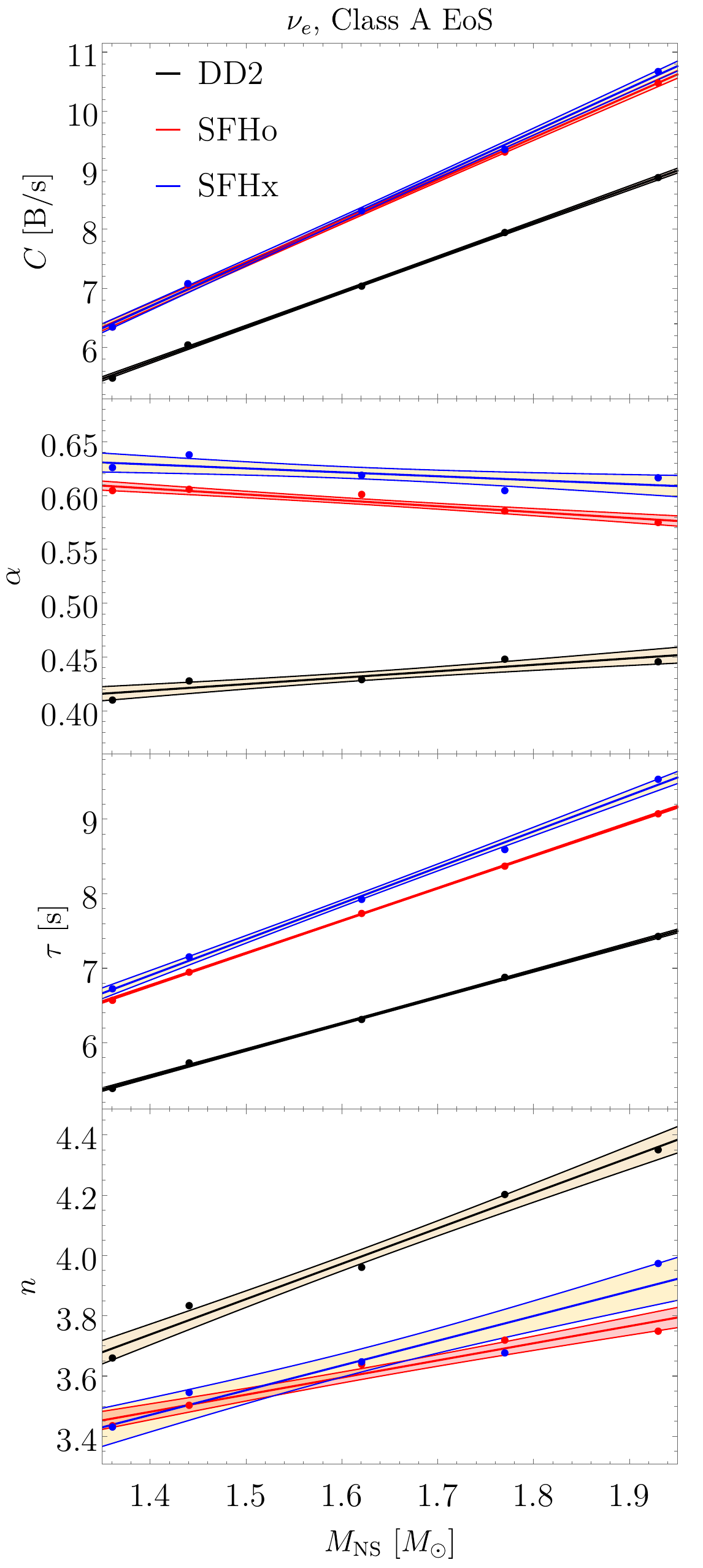}
\includegraphics[width=.49\textwidth]{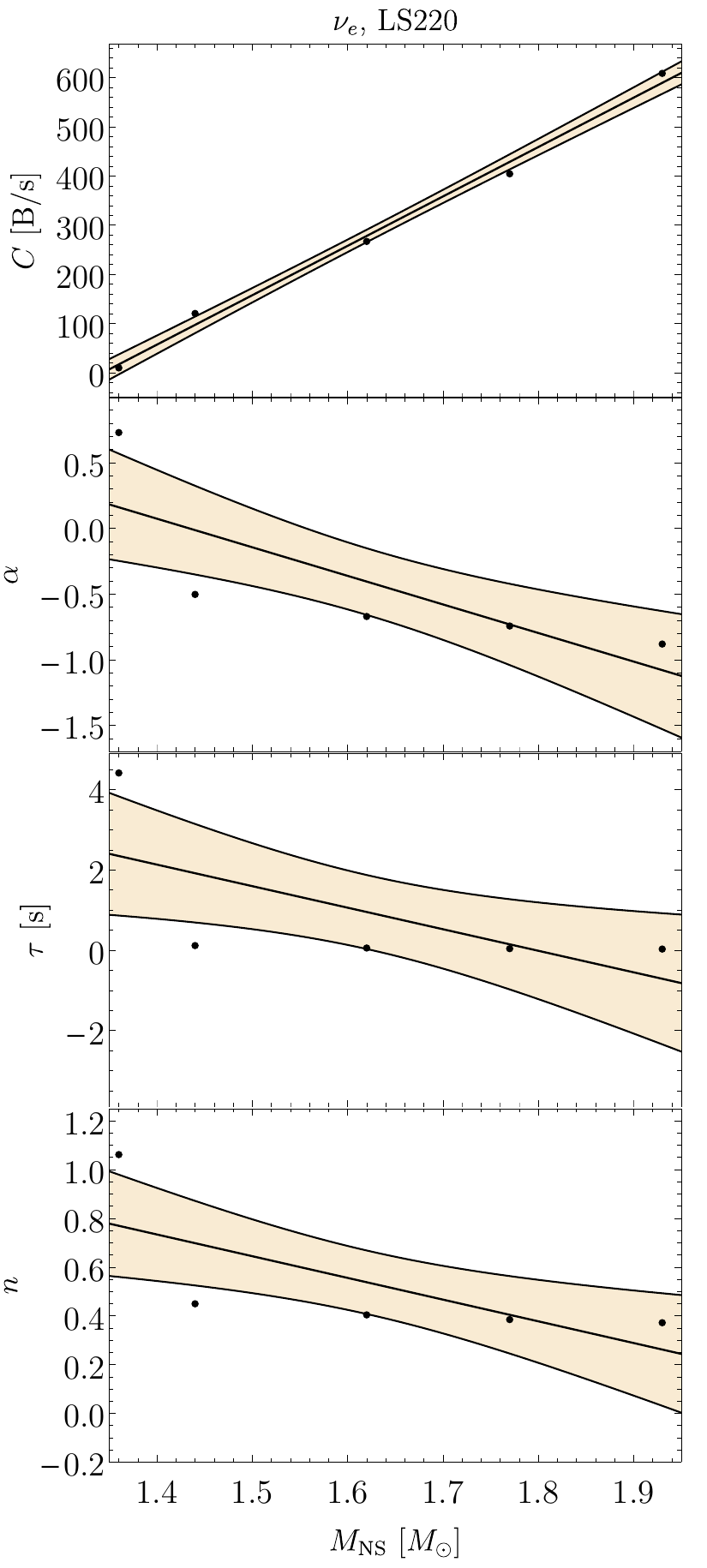}
\caption{Best-fit parameters $C$, $\alpha$, $\tau$, and $n$ as functions of the PNS mass for $\nu_e$ for Class~A EoSs (left panels) and for LS220 (right panels), obtained with data between 1\,s and $t_{\nu_e,{\rm c}}$. The shaded areas represent the $1\,\sigma$ confidence bands.}
\label{Fig:nuefitA}
\end{figure}
\begin{figure}[t!]
\centering
\includegraphics[width=.49\textwidth]{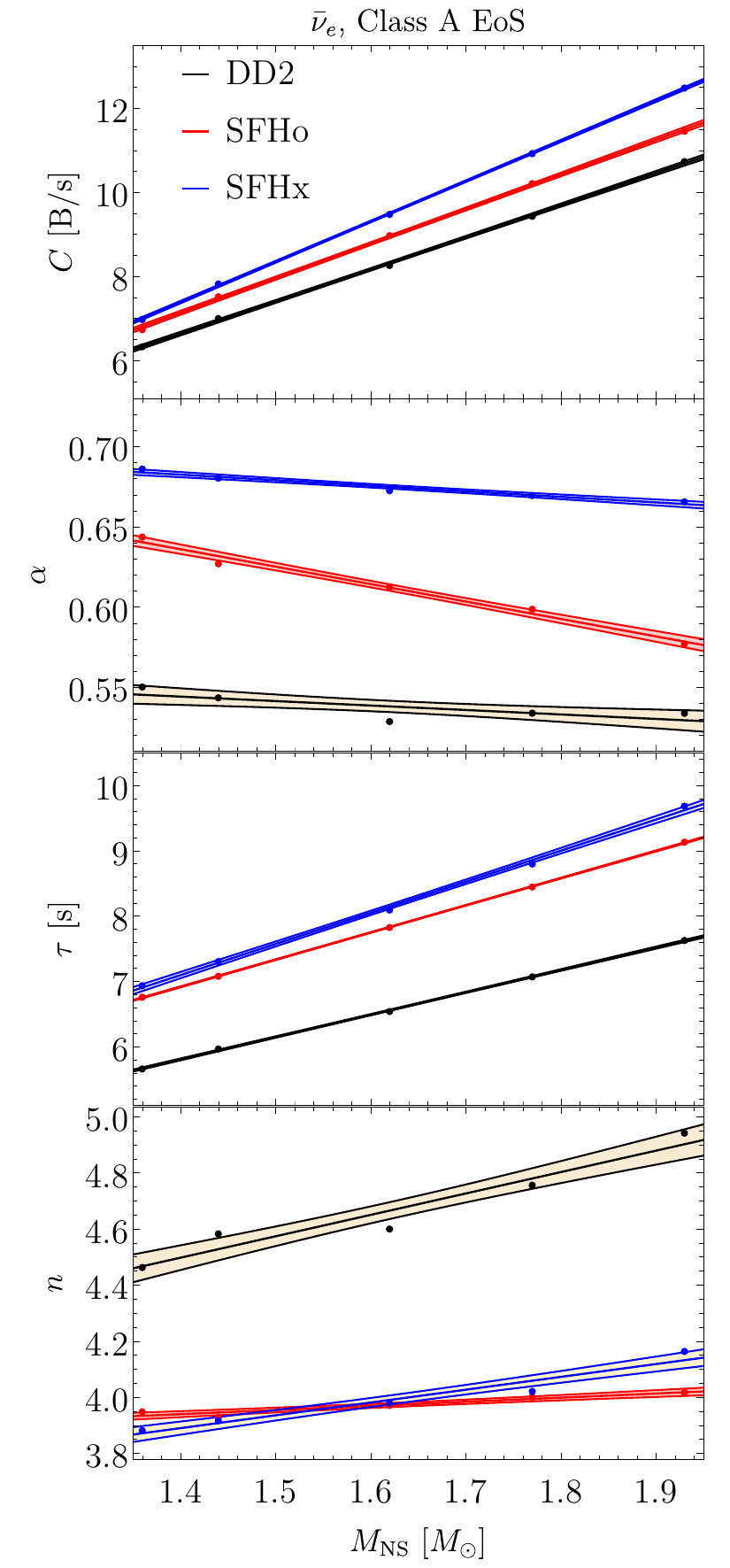}
\includegraphics[width=.49\textwidth]{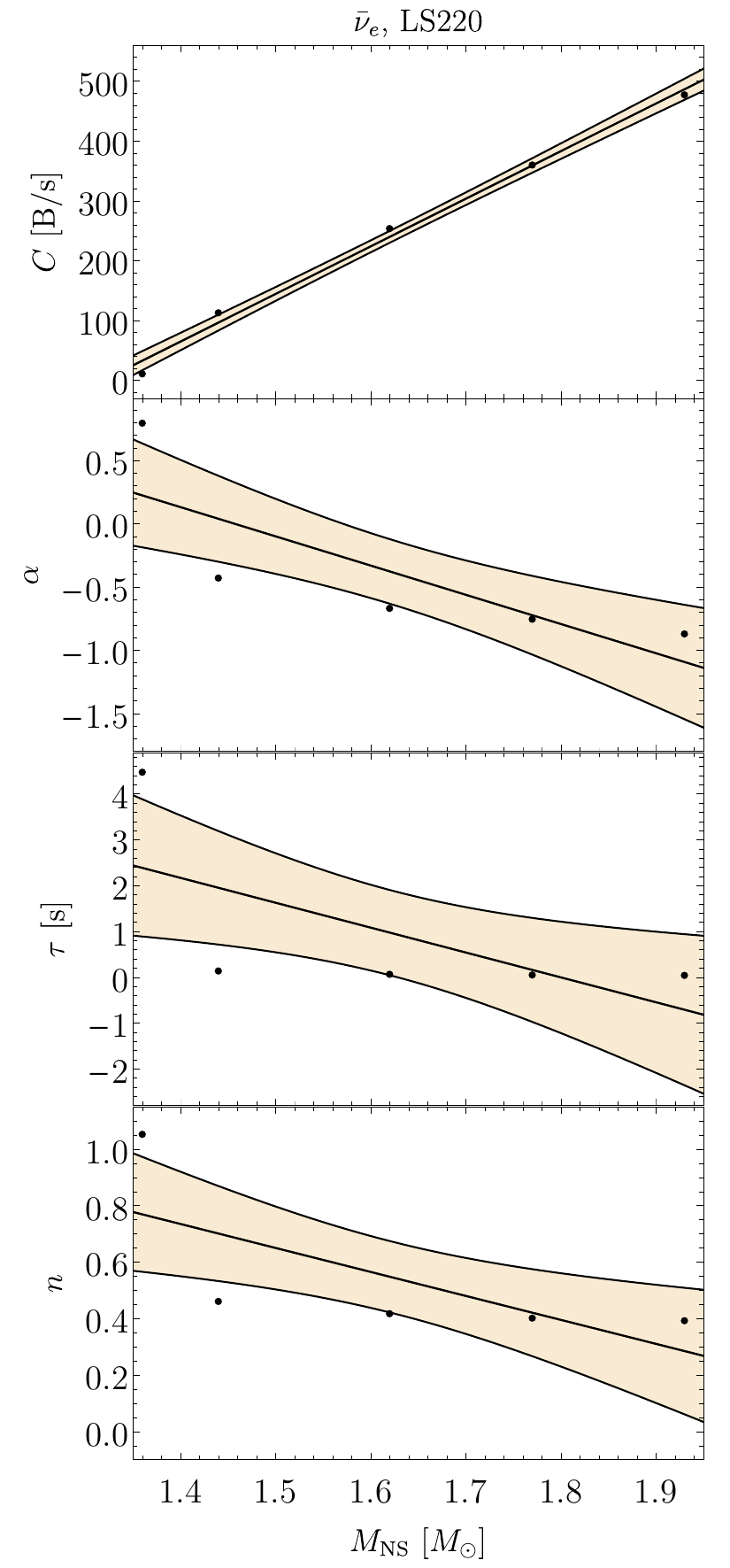}
\caption{Best-fit parameters $C$, $\alpha$, $\tau$, and $n$ as functions of the PNS mass for $\bar{\nu}_e$ for Class~A EoSs (left panels) and for LS220 (right panels), obtained with data between 1\,s and $t_{\bar{\nu}_e,{\rm c}}$. The shaded areas represent the $1\,\sigma$ confidence bands.}
\label{Fig:nuebfitA}
\end{figure}
\begin{figure}[t!]
\centering
\includegraphics[width=.49\textwidth]{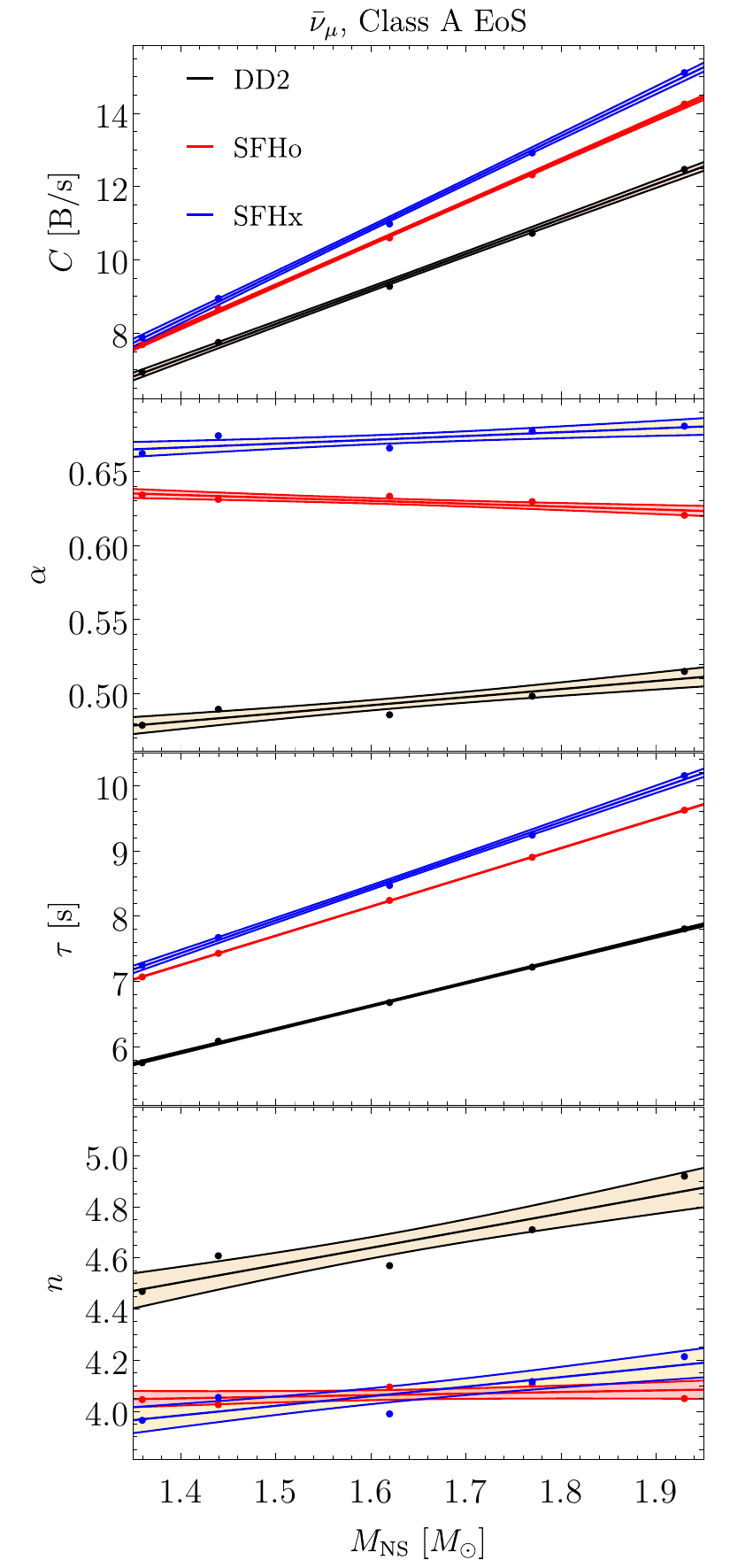}
\includegraphics[width=.49\textwidth]{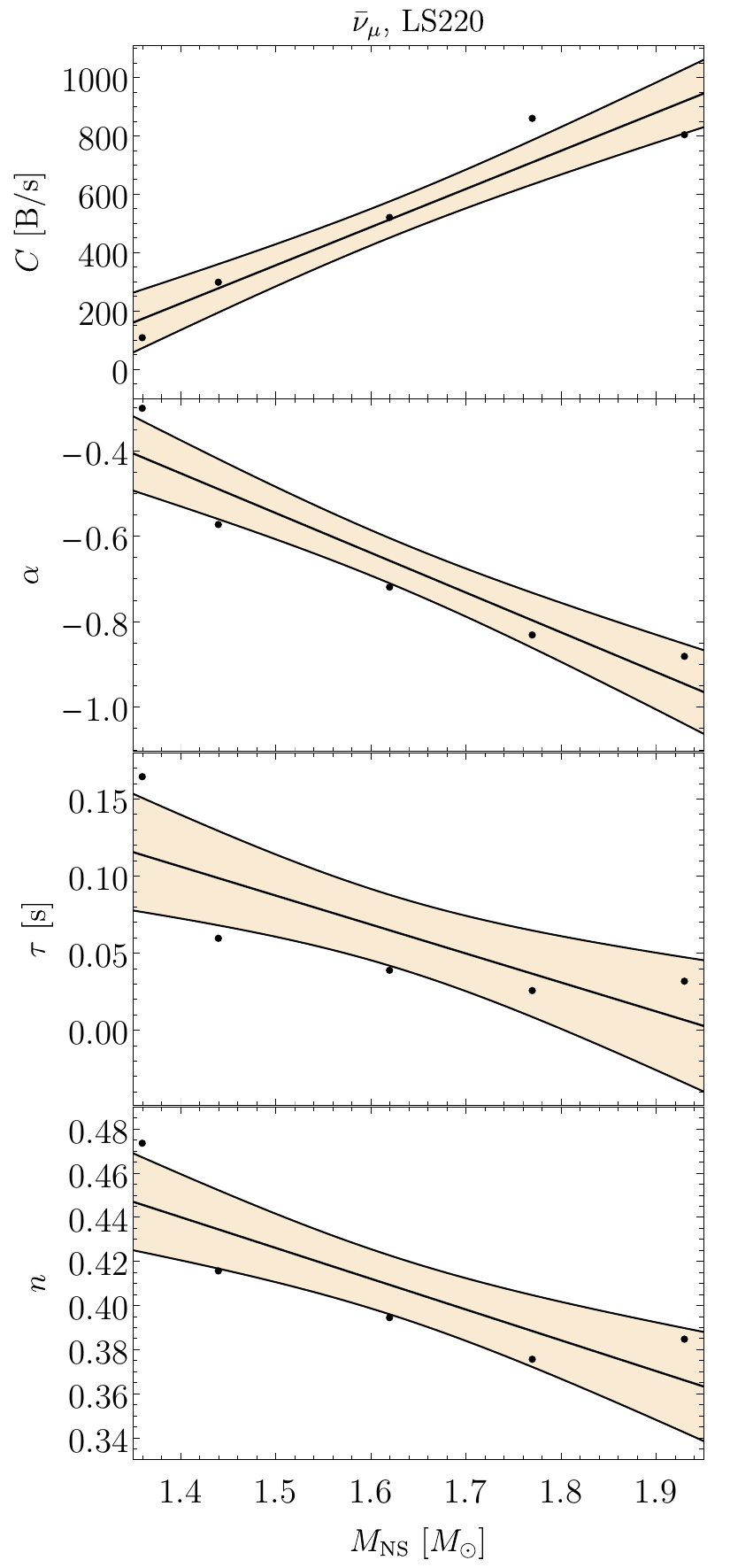}
\caption{Best-fit parameters $C$, $\alpha$, $\tau$, and $n$ as functions of the PNS mass for $\bar{\nu}_\mu$ for Class~A EoSs (left panels) and for LS220 (right panels), obtained with data between 1\,s and $t_{\bar{\nu}_\mu,{\rm c}}$. The shaded areas represent the $1\,\sigma$ confidence bands.}
\label{Fig:numbfitA}
\end{figure}

\section{Dependence on PNS mass and equation of state}
\label{sec:depende}

The best-fit parameters are plotted as functions of the PNS mass in Fig.~\ref{Fig:nuefitA} for $\nu_e$, in Fig.~\ref{Fig:nuebfitA} for $\bar{\nu}_e$ and in Fig.~\ref{Fig:numbfitA} for $\bar{\nu}_\mu$, separately for Class~A EoSs in the left panels and the LS220 case of Class~B in the right panels. 

\subsection{Class A EoS}

For Class~A EoSs and all species of neutrinos and antineutrinos, the behavior of all of the fitting parameters can be well reproduced with linear dependences on the PNS mass, as shown by the black (DD2), red (SFHo), and blue (SFHx) lines in the left panels of Figs.~\ref{Fig:nuefitA}, \ref{Fig:nuebfitA}, and \ref{Fig:numbfitA}. The shaded area around the linear interpolation represents the $1\,\sigma$ confidence band for each linear fit, obtained from the standard mean square uncertainties associated with the linear regression. In Table~\ref{tab:massdep} of Appendix~\ref{App:massdep} we provide the best-fit parameter values that describe the linear dependencies of the fitting parameters $C$, $\alpha$, $\tau$, and $n$ on the PNS mass for all neutrino and antineutrino species and all of the considered EoS.  

\begin{enumerate}[label=(\roman*)]
    \item The ``normalization  parameter  $C$'' (top panels) adopts values of $5\,\B/\s \lesssim C \lesssim 15\,\B/\s$ for all neutrino species and it increases with the PNS mass. In particular, DD2 leads to the lowest values and SFHo and SFHx feature a similar dependence, with SFHx providing the largest values of $C$.
     \item  The ``power-law index'' $\alpha$ (second panels from top) shows a mild dependence on the PNS mass, featuring values of $0.50 \lesssim \alpha \lesssim 0.70$ for all neutrino species, with an increase in the case of DD2 and a slight decrease for SFHo and SFHx. Similar to the normalization parameter, for each kind of neutrino and PNS mass, the largest value of $\alpha$ is obtained with SFHx and the smallest one with DD2.
      \item The ``late-time suppression parameters'' $\tau$ and $n$ (third and fourth panels from top) tend to increase with the PNS mass for all neutrino species, with values of $5\,\s \lesssim \tau \lesssim 10\,\s$ and $3\lesssim n \lesssim 5$. In particular, for all neutrino kinds and PNS masses, DD2 leads to the lowest values of $\tau$ and SFHx to the largest ones. On the other hand, DD2 yields the largest $n$ values  for $\nu$ and $\bar\nu$ of all flavors and all PNS masses, whereas the $1\,\sigma$ confidence bands of $n$ for SFHo and SFHx models overlap in the mass range of $1.4\,M_\odot \lesssim M_{\rm NS}\lesssim 1.7\,M_\odot$, with SFHx showing larger values of $n$ than SFHo at higher PNS masses.
     \end{enumerate}
 
A given value of $C$, $\tau$, and $n$ can be obtained with different combinations of EoS and neutron star (NS) mass. For instance, for $\bar{\nu}_e$, $\tau=7\,\s$ corresponds to $M_{\rm NS}\approx 1.38\,M_\odot$ and SFHx, $M_{\rm NS}\approx 1.45\,M_\odot$ and SFHo, as well as $M_{\rm NS}\approx 1.77\,M_\odot$ and DD2. 
However, the best-fit values of $\alpha$ lie in different ranges for different EoSs. For instance, $0.6 \lesssim \alpha_{\bar{\nu}_e} \lesssim 0.65$ would point to the SFHo EoS, irrespective of the PNS mass. Therefore, in the case of a future observation of a SN neutrino signal, the measurement of $\alpha$ would provide information on the EoS and the PNS mass and, combined with the measurements of the other fit parameters, would allow us to characterize the PNS mass and the EoS. More explicitly, neglecting for simplicity flavor mixing, a combined measurement of $\tau=7\,\s$, $\alpha\approx 0.63$, and $n\approx 3.9$ for the $\bar{\nu}_e$ luminosity would suggest a SN explosion leading to a PNS of mass $1.44~M_\odot$, whose interior properties are described by the SFHo EoS. As further discussed in Sec.~\ref{sec:count}, due to the similarities of the spectra and luminosities of electron antineutrinos and heavy-lepton antineutrinos in our models at times $t\gtrsim 1$\,s, flavor conversions are not a major effect, although they would make the reconstruction of the fit parameters less straightforward, but without spoiling our result. The accurate reconstruction of the parameters is left for future work.

\subsection{LS220 EoS}

A different discussion is required by the LS220 models. In this case, as shown in the right panels of Figs.~\ref{Fig:nuefitA}, \ref{Fig:nuebfitA}, and~\ref{Fig:numbfitA} and in Table~\ref{tab:massdep} of Appendix~\ref{App:massdep}, the lower quality of the fit leads to a slightly different dependence of the fit parameters on the PNS mass.

\begin{enumerate}[(i)]
  \item The ``normalization  parameter''  $C$ (top panels) is  $C\sim O(10-1000)\,{\rm B}/\s$ for all of the neutrino species and it increases linearly with the PNS mass.
     \item  The ``power-law index'' $\alpha$ (second panels from top) 
     tends to decrease as the PNS mass increases. It is negative for all models and neutrino species except for $\nu_e$ and $\bar{\nu}_e$ in the simulation with $M_{\rm NS}=  1.36\,M_\odot$, where $\alpha > 0$. 
      \item The ``late-time suppression parameters'' $\tau$ and $n$ (third and fourth panels from top) tend to decrease with higher PNS mass. For the 1.36-LS220 model we get
       $\tau \sim 4$\,s and $n>1$ for $\nu_e$ and $\bar\nu_e$ (see Table~\ref{tab:lnue} in Appendix~\ref{App:bestfit}), whereas $\tau\lesssim 0.25$\,s and $n\lesssim 0.5$ in all the other cases (Tables~\ref{tab:lnum} and \ref{tab:lnut}).
     \end{enumerate}
Even though the linear interpolations of the mass dependence have a lower quality for the LS220 EoS, the best-fit parameters for this EoS possess values in completely different ranges than for the Class~A EoSs. Thus, the observation of such values would clearly point to a Class~B EoS for the PNS in the discovered SN.

\begin{figure}[t!]
\centering
\includegraphics[width=0.49\textwidth]{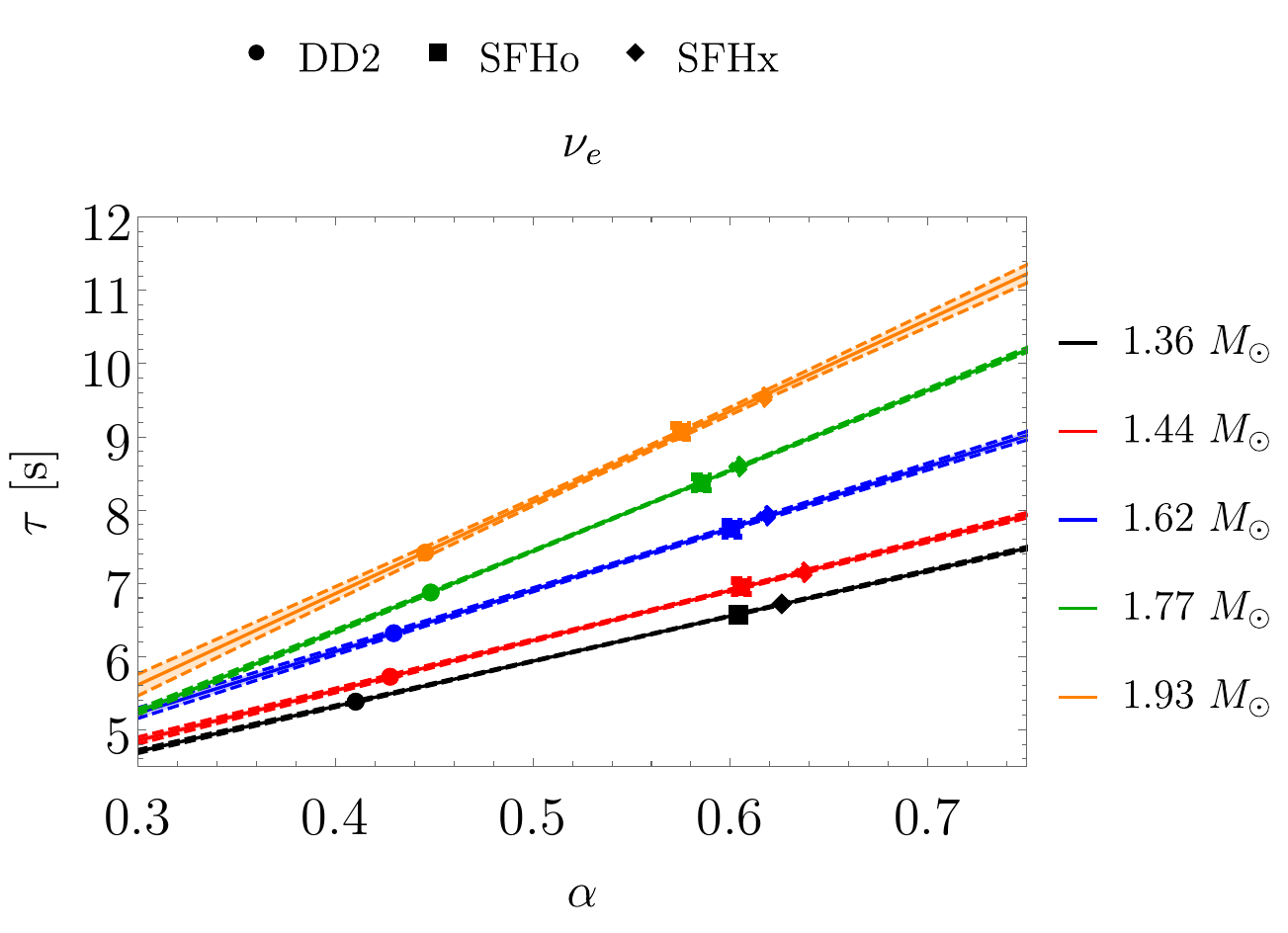}
\includegraphics[width=0.49\textwidth]{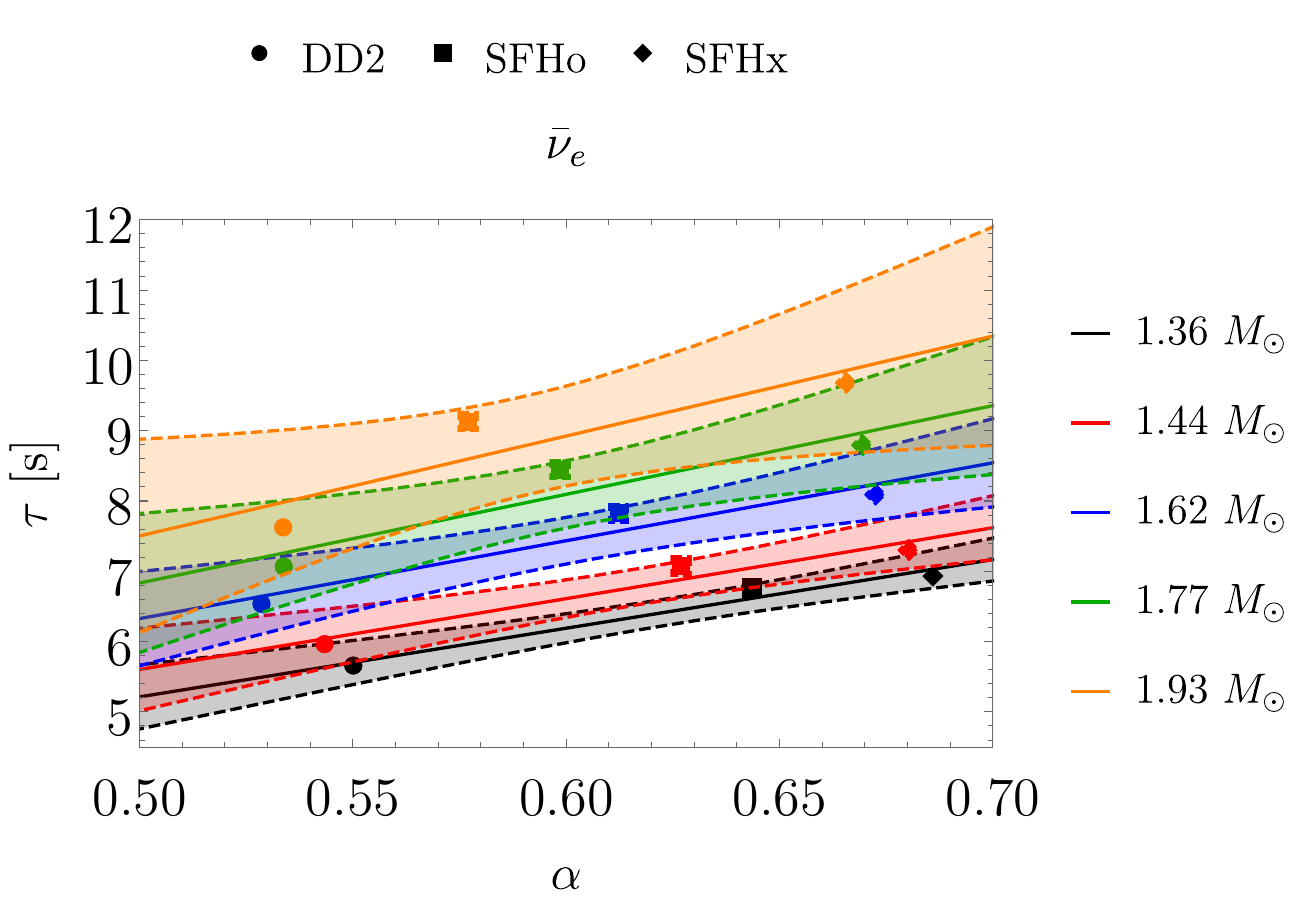}
\includegraphics[width=0.49\textwidth]{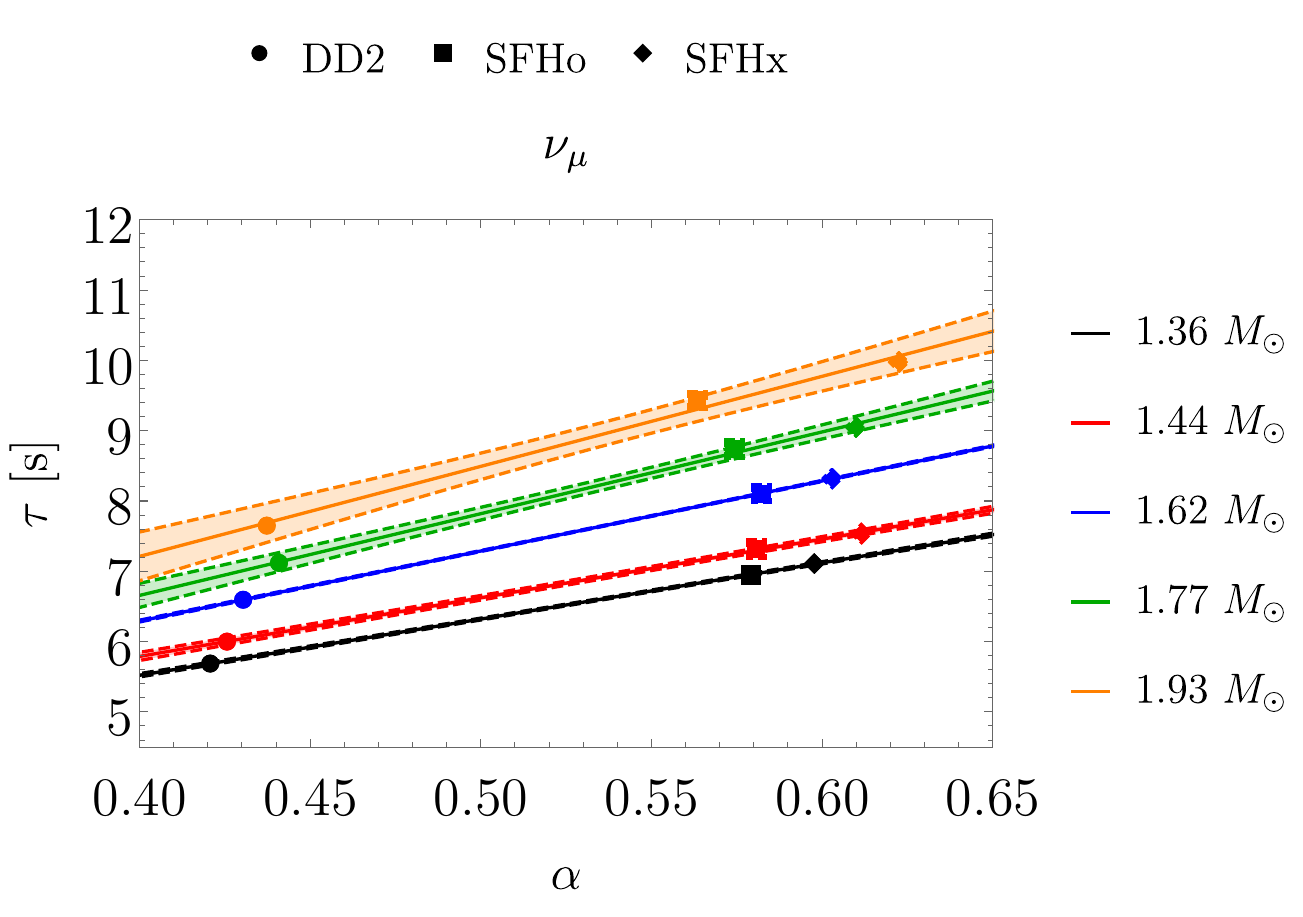}
\includegraphics[width=0.49\textwidth]{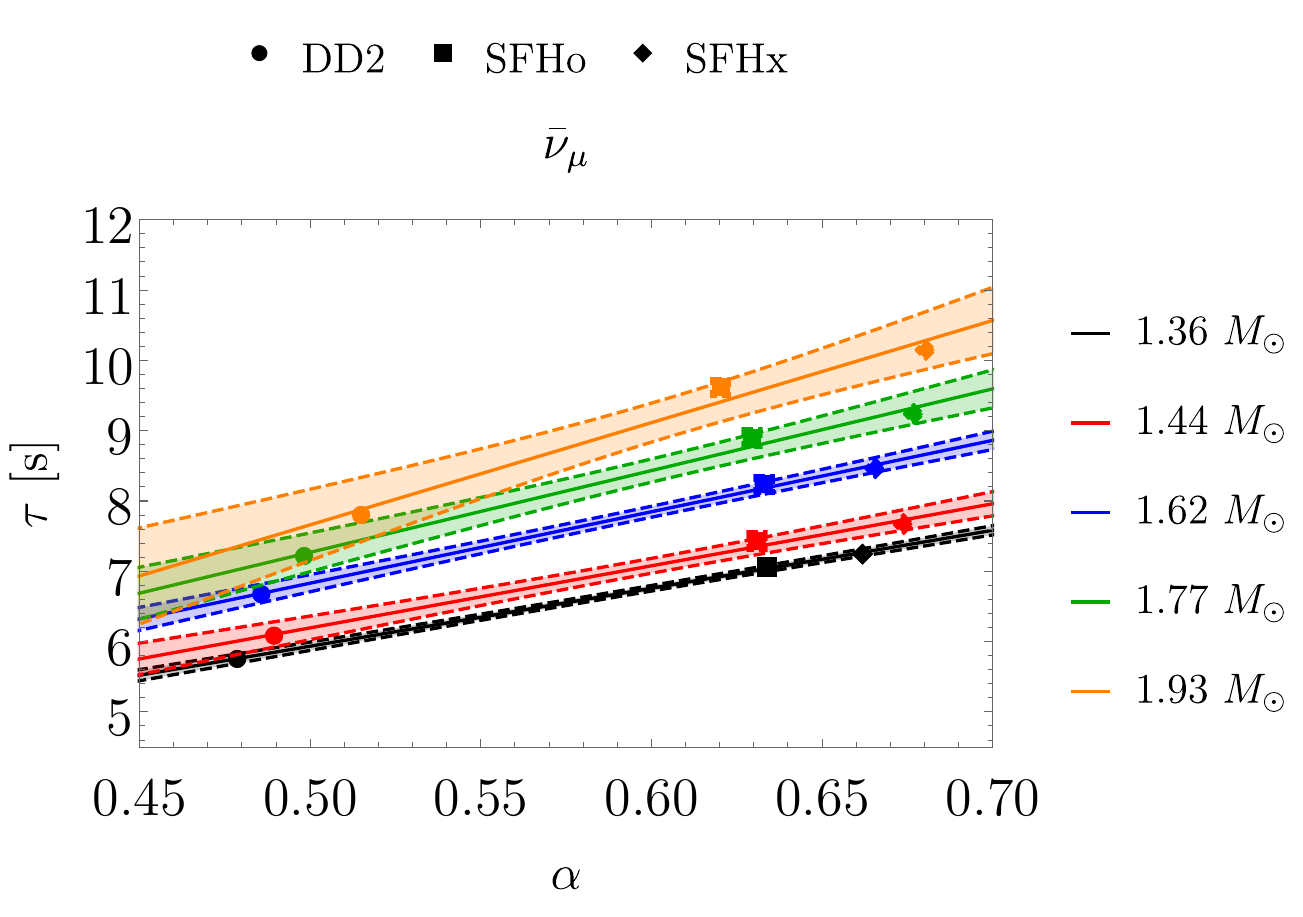}
\includegraphics[width=0.49\textwidth]{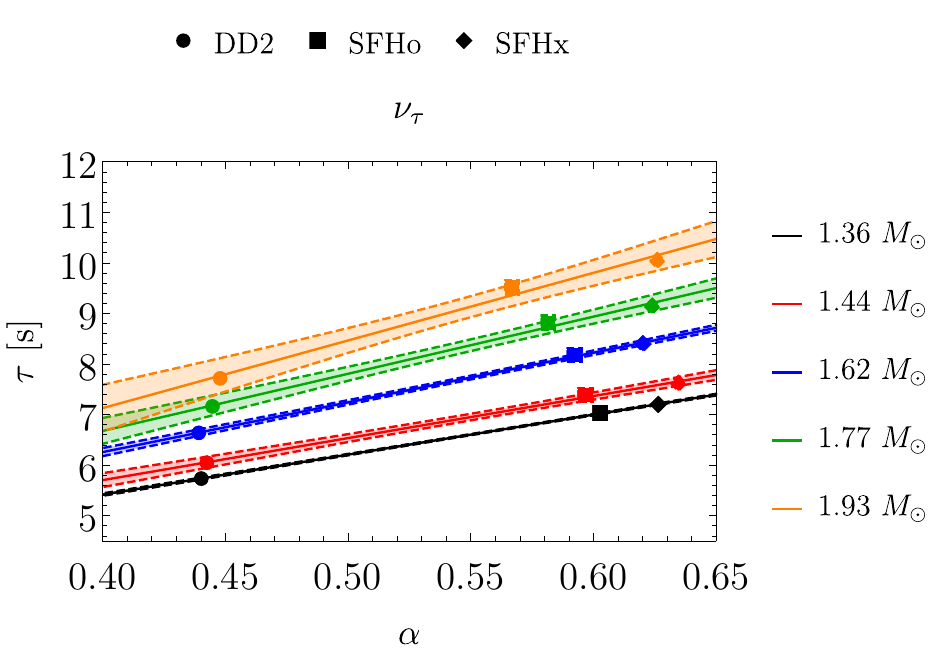}
\includegraphics[width=0.49\textwidth]{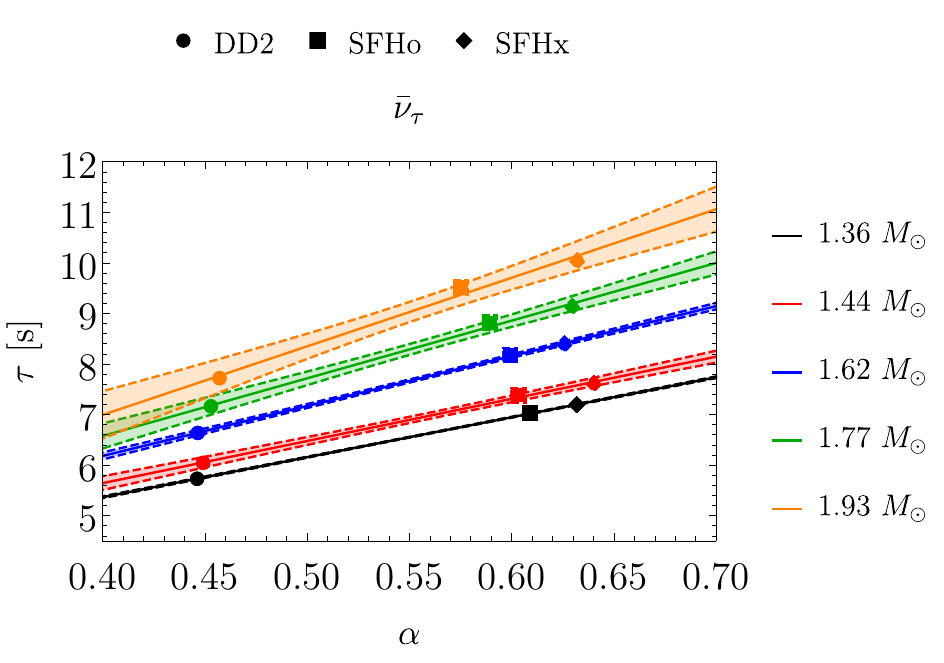}
\caption{Correlations of the fitting parameters $\tau$ (in units of seconds) and $\alpha$ for $\nu_e$ (top left panel), $\bar{\nu}_e$ (top right panel), $\nu_\mu$ (central left panel), $\bar{\nu}_\mu$ (central right panel), $\nu_\tau$ (bottom left panel), $\bar{\nu}_\tau$ (bottom right panel), for all Class~A EoSs and all considered PNS masses, obtained from simulation output data between 1\,s and $t_{\nu_i,{\rm c}}$ for each neutrino species $\nu_i$.}
\label{Fig:scatternu_01}
\end{figure}

\section{Correlation between $\tau$ and $\alpha$}
\label{sec:correl}

Since the fit parameters of all Class~A EoSs have values in similar ranges, it is useful to search for possible relations between them. We disregard the case of LS220 here, because its parameter values lie in completely different regimes. For fixed PNS mass and given EoS, we find that the power-law index $\alpha$ and the suppression time $\tau$ for the Class A EoSs exhibit correlations that can be fairly well described by linear functions,
\begin{equation}
\tau (\s) = A + B\,\alpha\,,
\label{eq:tau-alpha}
\end{equation}
for all neutrino species except $\bar\nu_e$. This is shown in Fig.~\ref{Fig:scatternu_01} for neutrinos (left) and antineutrinos (right) of all flavors and all Class~A EoSs. In Appendix~\ref{App:scatternu}, in Table~\ref{tab:scatternu} we report the best-fit values with errors obtained from linear regression for $A$ and $B$ for all neutrino species. The pairs of values $(\tau,\alpha)$ are represented by filled circles for the DD2 EoS, filled squares for SFHo, and filled diamonds for SFHx in Fig.~\ref{Fig:scatternu_01}. Different colors correspond to the different PNS masses: black for $M_{\rm NS}=1.36~M_{\odot}$, red for $M_{\rm NS}=1.44~M_{\odot}$, blue for $M_{\rm NS}=1.62~M_{\odot}$, green for $M_{\rm NS}=1.77~M_{\odot}$, and orange for $M_{\rm NS}=1.93~M_{\odot}$. 

As a general trend for fixed NS mass, SFHx leads to the largest values of $\tau$ and $\alpha$ for $\nu$ and $\bar\nu$ of all flavors, whereas DD2 yields the smallest values of $\tau$ and $\alpha$.
The linear increase of $\tau$ as a function of $\alpha$ could be understood in simple terms by the fact that a larger value of $\alpha$ causes a faster decline of the luminosity in the first seconds, and therefore, if the initial luminosity (i.e., specifically at 1\,s in our context) were the same, 
less energy is carried away by neutrinos during this early power-law phase of the luminosity. Thus, if the total energy released in neutrinos were fixed, one would expect that a higher value of $\alpha$ leads to a stretching of the subsequent exponential luminosity decrease with a final drop only at later times, implying a larger $\tau$. Although this explanation sounds plausible, it is an oversimplification of the real situation. First, at $t = 1$\,s the luminosities are not the same for a given PNS mass, but smallest for DD2 and largest for SFHx (see Fig.~\ref{Fig:lnu162}). Second, also the total energy (individually for all neutrino species as well as summed up) is different, namely smallest for DD2 and largest for SFHx (see Tables~I and VII in \cite{Fiorillo:2023frv}), which correlates with the final gravitational binding energy of the cold NS (but not strictly with the final NS radius, which is largest for models with the DD2 EoS and smallest with SFHo when $M_\mathrm{NS}\gtrsim 1.2\,M_\odot$). The true reasons for the tight correlation between $\alpha$ and $\tau$ values are therefore more subtle than suggested by the simple argument given above.  
 
Moreover, the quality of the linear relation, Eq.~(\ref{eq:tau-alpha}), differs between different neutrino species. Figure~\ref{Fig:scatternu_01} and the values of the coefficients $A$ and $B$ in Table~\ref{tab:scatternu} in Appendix~\ref{App:scatternu} reveal that the linear relation works better for neutrinos than for antineutrinos.
The good quality of the linear fit especially for $\nu_e$ could be useful to disentangle the NS mass and EoS with a measured pair of values $(\tau,\alpha)$. In this context, it should be mentioned that the linear fits for ${\bar\nu_e}$ are considerably worse than for all other neutrino species (upper right panel in Fig.~\ref{Fig:scatternu_01}); the $1\,\sigma$ confidence bands display substantial overlap in the whole range of $\alpha$ values. In contrast, the linear fit function works better for the antineutrinos of the non-electron flavors, with $\bar{\nu}_\tau$ (bottom right panel) showing slightly narrower $1\,\sigma$ confidence bands than $\bar{\nu}_\mu$ (middle right panel). The reason for the particularly poor quality of the linear fit for $\bar\nu_e$ is the relatively wide separation of the $\alpha$ values for the SFHo and SFHx EoSs, in contrast to $\nu_e$, where the corresponding values are very close to each other and the linear fit is very good. The lower quality of linear $\tau$-$\alpha$ relations for the heavy-lepton neutrinos also confirms this general tendency that such a fit function is less suitable to describe the correlation of both parameters when the distance between $\alpha$ for SFHo and SFHx grows. This leaves the possibility that the linear function works well for $\nu_e$ just because of favorable properties of the SFHo and SFHx EoSs. More investigation with larger sets of different nuclear EoS cases is therefore required before one can rely on the validity of linear $\tau$-$\alpha$ relations.

\begin{figure}[t!]
\centering
\includegraphics[width=0.49\textwidth]{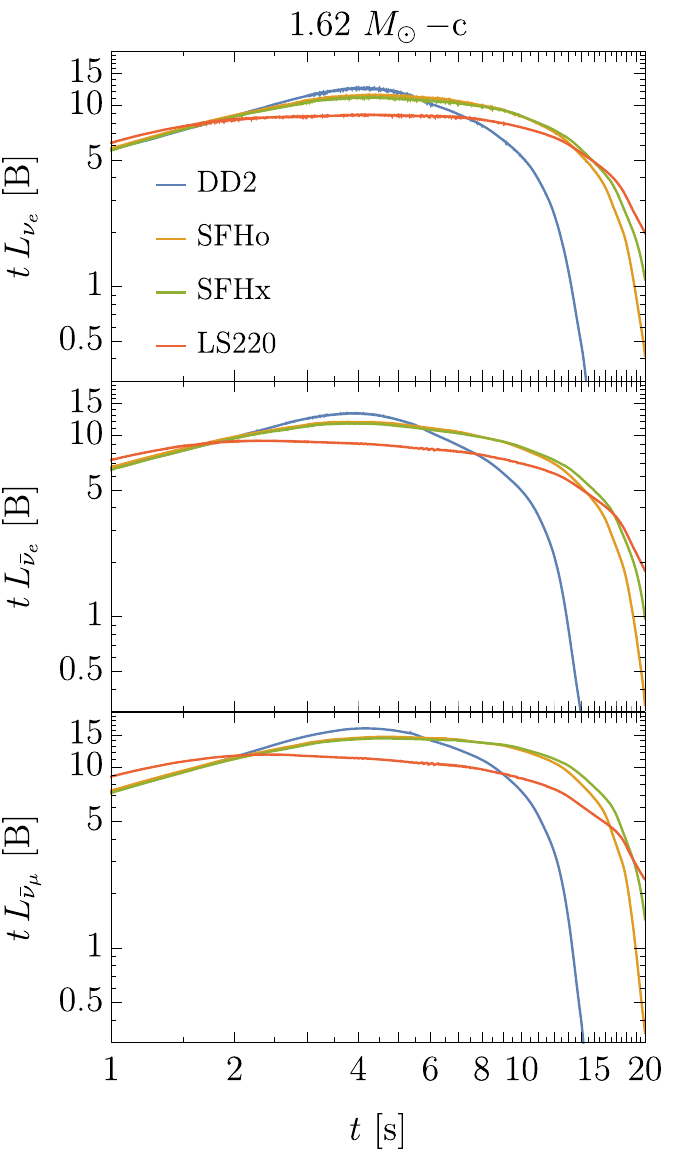}
\includegraphics[width=0.49\textwidth]{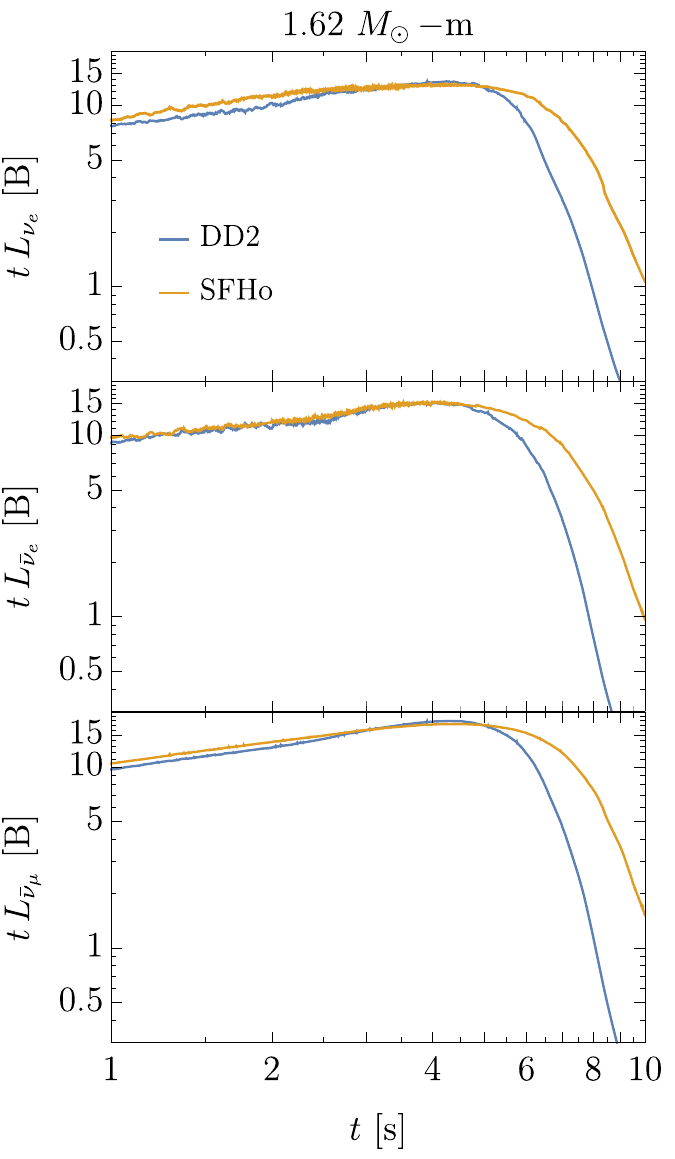}
\caption{Time evolution of the product of time and luminosity, $t\,L_{\nu_i}$, for $\nu_e$ (top panels), $\bar{\nu}_e$ (middle) and $\bar{\nu}_\mu$ (bottom) for a simulations without convection (left panels) and without muons (right panels), with $M_{\rm NS}=1.62\,M_\odot$ and different EoS: DD2 (blue), SFHo (orange), SFHx (green), and LS220 (red). Simulations using SFHx and LS220 without muons are not available.}
\label{Fig:tlnu162cm}
\end{figure}
\begin{figure}[t!]
\centering
\includegraphics[width=0.49\textwidth]{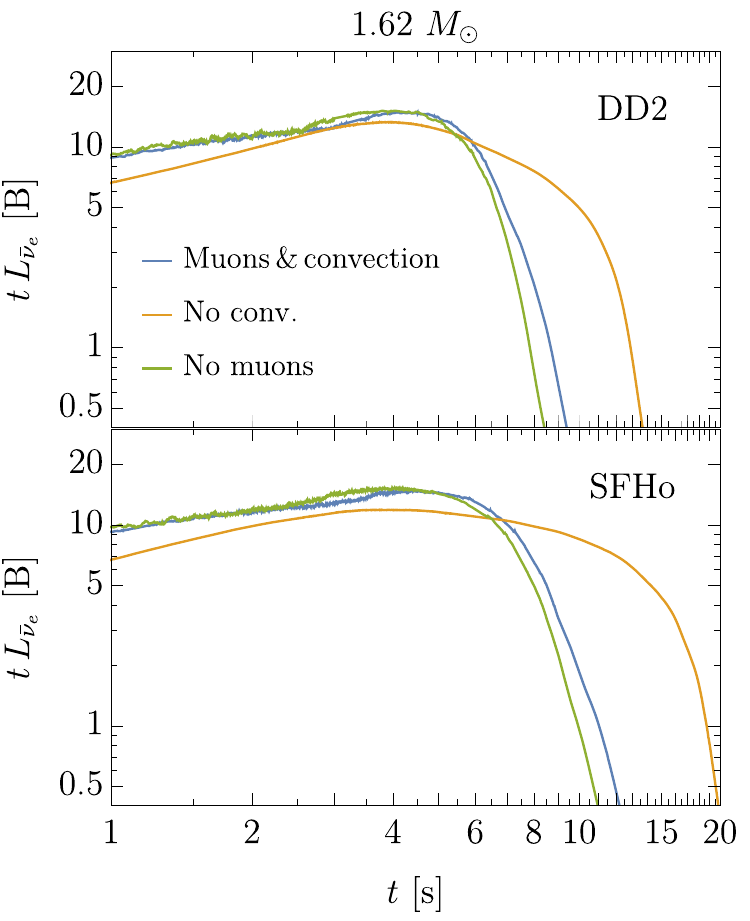}
\includegraphics[width=0.49\textwidth]{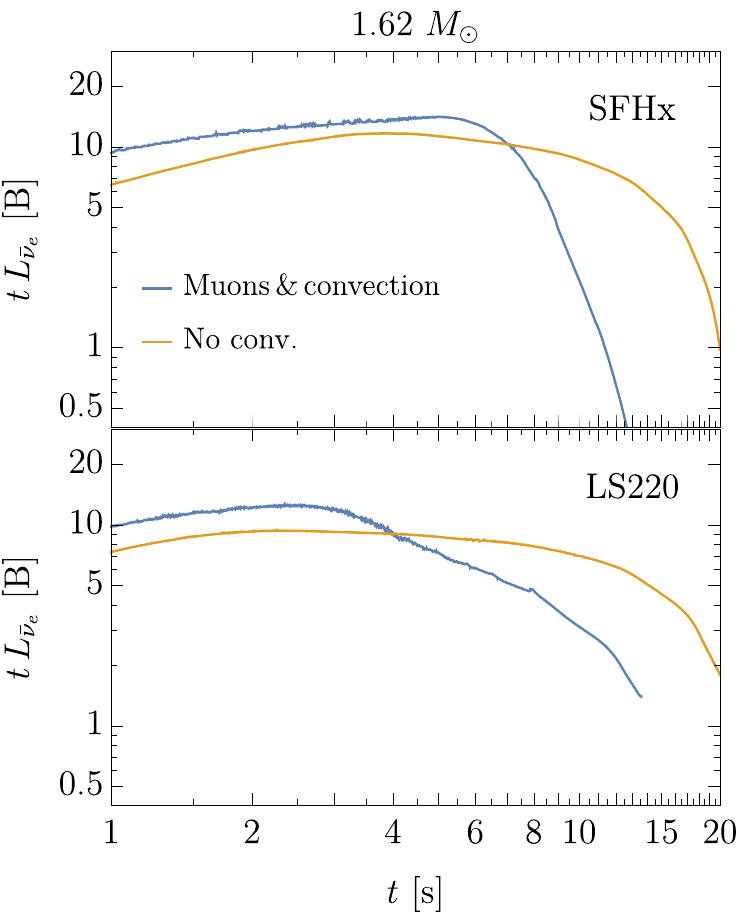}
\caption{Time evolution of the product of time and $\bar{\nu_e}$ luminosity, $t\,L_{\bar{\nu}_e}$, for $M_{\rm NS}=1.62~M_\odot$ and different EoS, namely DD2 and SFHo (left panels) and SFHx and LS220 (right panels), comparing models with both convection and muons (blue), without convection (orange), and without muons (green). Simulations using SFHx and LS220 without muons are not available.}
\label{Fig:tlnubecmEoS}
\end{figure}
\begin{figure}[t!]
\centering
\includegraphics[width=0.47\textwidth]{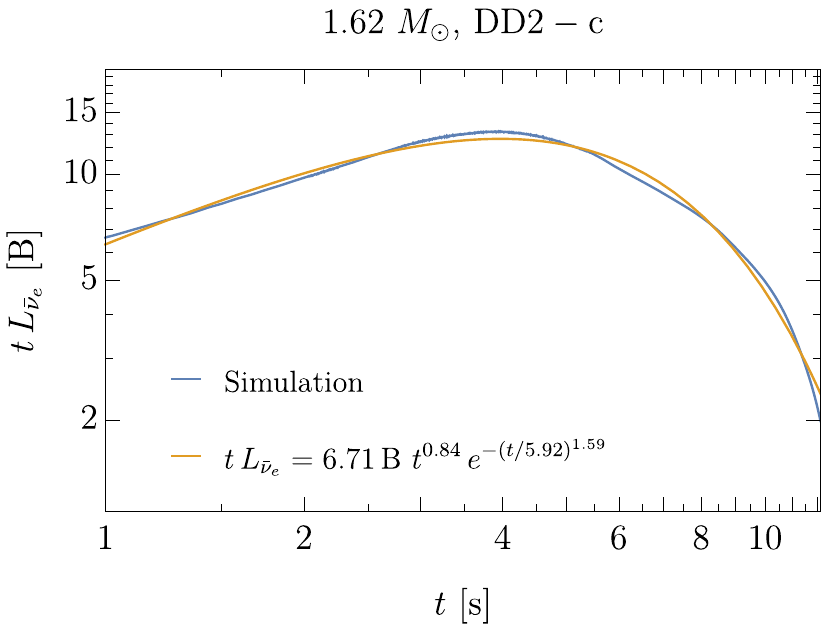}
\includegraphics[width=0.47\textwidth]{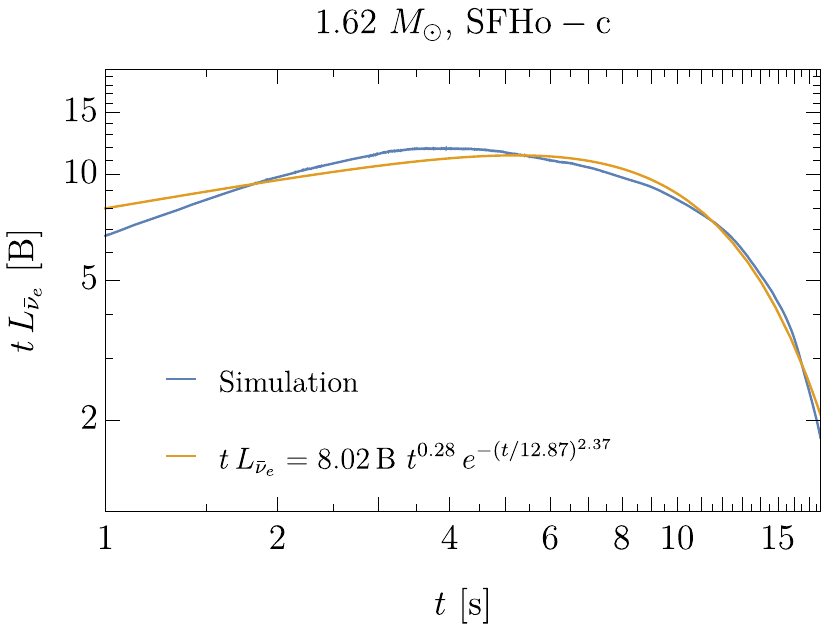}
\includegraphics[width=0.47\textwidth]{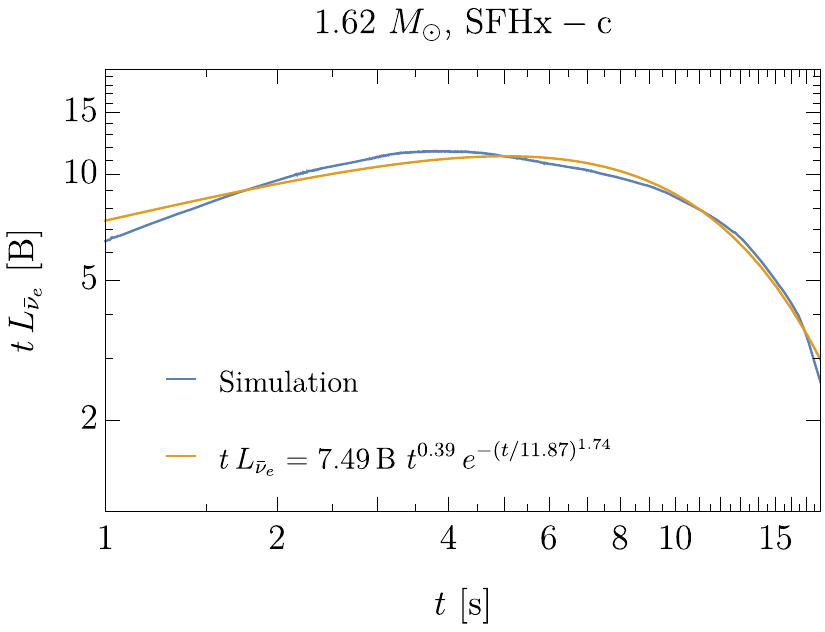}
\includegraphics[width=0.47\textwidth]{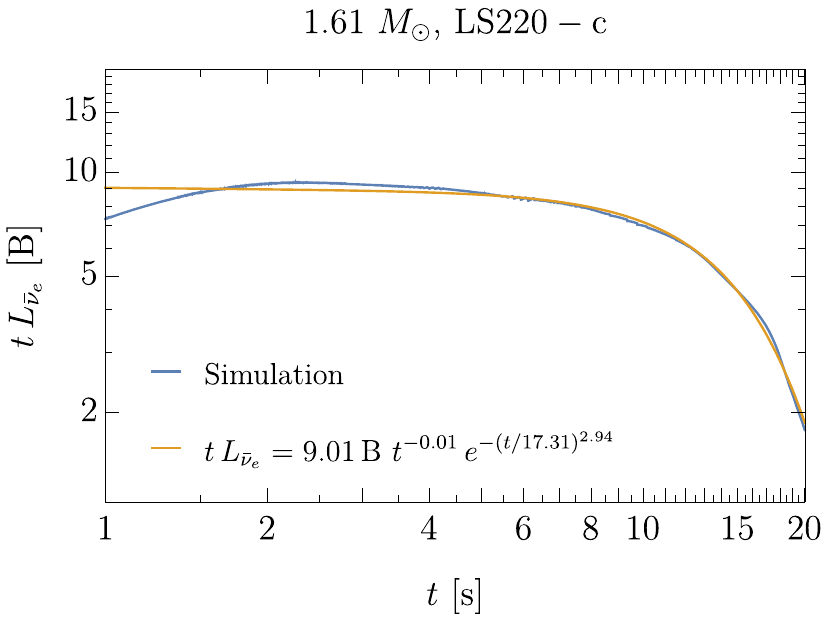}
\includegraphics[width=0.47\textwidth]{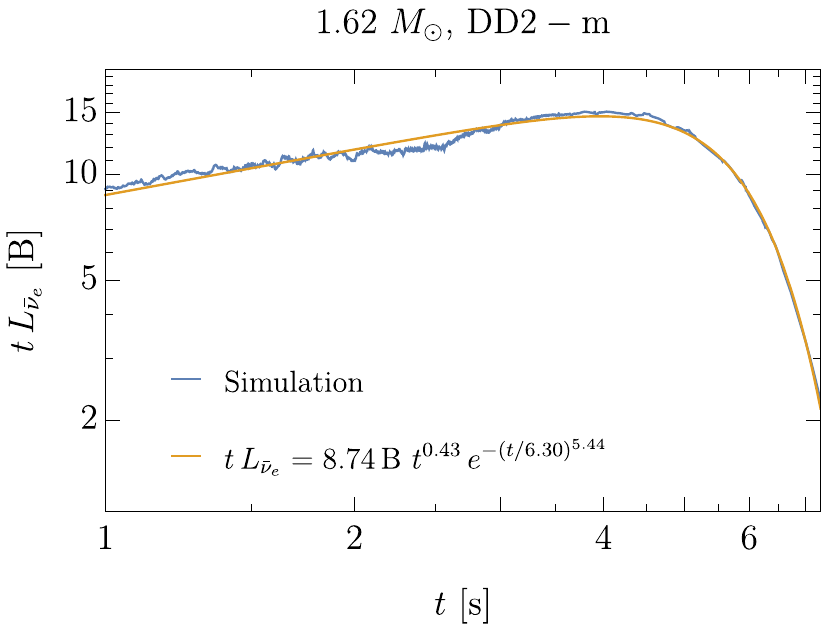}
\includegraphics[width=0.47\textwidth]{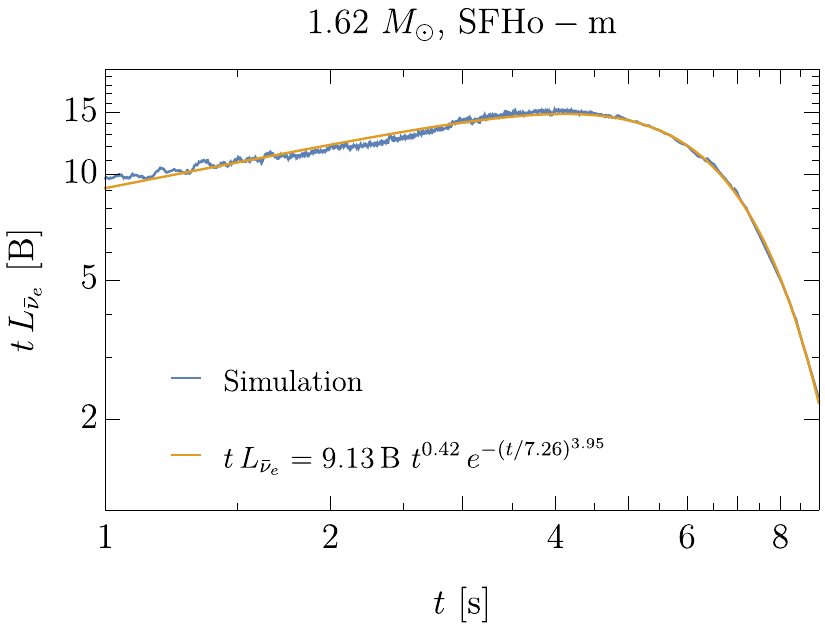}
\caption{Time evolution of the product of time and $\bar{\nu}_e$ luminosity, $t\,L_{\bar{\nu}_e}$, for simulation data (blue) and their fits (orange). The different panels show results for simulations without convection or without muons, namely of models 1.62-DD2-c (top left panel), 1.62-SFHo-c (top right), 1.62-SFHx-c (middle left), 1.61-LS220-c (middle right), 1.62-DD2-m (bottom left) and 1.62-SFHo-m (bottom right). We consider data up to $t_{\bar{\nu}_e,{\rm c}}$ for simulations with DD2 and SFHo and up to $t_{\bar{\nu}_e,{\rm Co}}$ for 1-62-SFHx-c and 1.61-LS220-c (see Appendix~\ref{App:tmax} for more details).}
\label{Fig:nuebfit162cm}
\end{figure}

\section{Impact of muons and convection}
\label{sec:muons}

In order to assess the impact of some of our physics inputs of the simulations, we investigate additional PNS cooling calculations for a PNS mass of 1.62~$M_\odot$ now, where we either omitted convection or muons. Non-convective models were computed for all considered EoSs and are denoted by a suffix ``-c'' appended to their names (e.g., 1.62-DD2-c), whereas only two models are considered without muons (suffix ``-m''), namely 1.62-DD2-m and 1.62-SFHo-m~\cite{SNarchive}. The bottom two data blocks of Table~\ref{tab:lnufinal} provide the final simulation times and the corresponding reduction factors $X_{\nu_i}^{\rm fin}$ for all neutrino species in these additional models. Bold numbers in Table~\ref{tab:lnufinal} for models 1.62-SFHx-c and 1.61-LS220-c signal that the simulations were stopped when $X_{\nu_i}^{\rm fin} > 0.15$, implying that for these models our standard cutoff time $t_{\nu_i,{\rm c}}$ is larger than $t_{\rm fin}$. In both of the simulations, the weakest suppression is obtained for $\nu_\mu$, with $X_{\nu_\mu}^{\rm fin} = 0.219$ for 1.62-SFHx-c and $X_{\nu_\mu}^{\rm fin} = 0.185$ for 1.61-LS220-c (see the values marked by a star in Table~\ref{tab:lnufinal}). Therefore, to test the impact of convection for the SFHx and LS220 EoSs we cut our luminosity data at $t_{\nu_i,{\rm Co}}$, when $X_{\nu_i}^{{\rm Co}}= t_{\nu_i,{\rm Co}}\,L_{\nu_i,{\rm Co}}/t_{\nu_i,{\rm max}}\,L_{\nu_i,{\rm max}}=0.22$ for simulations with SFHx and $X_{\nu_i}^{{\rm Co}}=0.19$ for simulations with LS220. The values of $t_{\nu_i,{\rm Co}}$ for all neutrino species of these simulations are given in Table~\ref{tab:lnumaxc} of Appendix~\ref{App:tmax}.

Figure~\ref{Fig:tlnu162cm} presents neutrino and antineutrino signals at $t > 1$\,s for the mentioned $M_{\rm NS}=1.62\,M_\odot$ simulations with modified input physics and the different EoS previously considered. The left panels display the results for our models without convection in the time interval [1,20]\,s, whereas the right panels show our cases without muons in the time interval [1,10]\,s, for $\nu_e$ (upper panels), $\bar{\nu}_e$ (central) and $\bar{\nu}_\mu$ (lower).

In the absence of convection the quantity $t\,L_\nu$ declines steeply only at $t \gtrsim 10\,\s$ for all neutrino species, with different characteristic features depending on the EoS (see the second data block from the bottom of Tables~\ref{tab:lnumax} and~\ref{tab:lnubarmax} in Appendix~\ref{App:tmax}). In particular, we witness the following:
\begin{enumerate}[(i)]
    \item DD2 (blue lines) has the shortest cooling time, with $t\,L_\nu$ peaking at $t \approx 4\,\s$ and being reduced by a factor $0.15$ of the maximum values at $t \approx 12$\,s.
    \item SFHo (orange lines) shows a peak of $t\,L_\nu$ at $t \approx 4$--6\,s and a later, steep decline, beginning roughly at $t \approx 15$\,s. 
    \item SFHx (green lines) is similar to SFHo, displaying a peak of $t\,L_\nu$ at a slightly earlier time and with a final decrease that is slightly delayed compared to SFHo.
    \item LS220 (red lines) leads to a peak in $t\,L_\nu$ at $t \approx 2$\,s and shows a more shallow decline afterward, forming a plateau-like shape in the time interval $2~\s \lesssim t \lesssim10$\,s (i.e., $L_\nu$ follows approximately $L_\nu \propto t^{-1}$) before a steeper decline sets at $t\gtrsim 15$\,s.
\end{enumerate}

In simulations without muons (see the right panels in Fig.~\ref{Fig:tlnu162cm} and the data block at the bottom of Tables~\ref{tab:lnumax}-\ref{tab:lnubarmax} in Appendix~\ref{App:tmax}), the product of time and luminosity for all $\nu$ species starts to become exponentially suppressed already at $t < 10~\s$, with DD2 (blue lines) leading to a faster cooling than SFHo (orange lines). 

To explicitly demonstrate the impact of convection and muons by means of the $\bar{\nu}_e$ luminosity, Fig.~\ref{Fig:tlnubecmEoS} displays the time evolution of $t L_{\bar{\nu}_e}$ for the $1.62~M_\odot$ models including both convection and muons (blue lines) compared to the corresponding results without convection (orange) and without muons (green, if available) for the DD2 EoS (upper left panel), SFHo EoS (lower left panel), SFHx EoS (upper right panel) and LS220 EoS (lower right panel). For all the cases
\begin{enumerate}[(i)]
    \item the ``absence of convection'' leads to a considerable stretching of the PNS Kelvin-Helmholtz neutrino cooling time, with the most moderate change for DD2,
    \item the ``omission of muons'' has a relatively mild effect on the evolution of the neutrino signals for the displayed 1.62\,$M_\odot$ models, making the suppression in the luminosity only slightly faster (because the NS becomes less compact with a lower binding energy), as visible by the green lines in the left panels.
\end{enumerate}

To quantitatively assess the impact of these variations of the input physics of our models, we also fit the neutrino and antineutrino signals of the additional simulations with the expression of Eq.~\eqref{eq:fit} and compare the best-fit parameters with those obtained in our benchmark simulations. As an example, Fig.~\ref{Fig:nuebfit162cm}  presents the simulation data (blue) and their best fits (orange) in the time interval of interest for $\bar{\nu}_e$ results from the simulations without convection (DD2 and SFHo in the top panels and SFHx and LS220 in the middle panels) and from our two simulations without muons (bottom panels). For the simulations without muons, where the neutrino signal does not experience major changes, the agreement between data and fits is similarly good as for our benchmark simulations. For the simulations without convection, the fit is still of excellent quality for DD2, although the omission of convection has altered the shape of the curve of $tL_{\bar\nu_e}$ (see Fig.~\ref{Fig:tlnubecmEoS}). In contrast, we obtain visibly larger discrepancies between fits and data for the non-convective simulations with SFHo and SFHx, for which, in particular, the shape of $tL_{\bar\nu_e}$ in the power-law dominated early phase cannot be reproduced as well as for models that include convection. Notably, the fits for $\bar\nu_e$ (and similarly for all other neutrino species) slightly overestimate $tL_\nu$ at $t \approx 1\,\s$ and tend to peak only at somewhat later times. Finally, the plateau-like region of $tL_\nu$ in the simulation results for LS220 implies a best-fit value of $\alpha \approx 1$, with the luminosity fit overestimating the data at $t\approx 1\,\s$ and following well a $t^{-1}$ power law before being exponentially suppressed at $t\gtrsim 10\,\s$. Our findings for all other neutrino species are analogous. In Tables~\ref{tab:lnucm} and~\ref{tab:lnubarcm} of Appendix~\ref{App:muons} we provide the best-fit parameter values and their $1\sigma$ errors for electron and muon neutrinos and antineutrinos, for our $1.62\,M_\odot$ models with different EoSs and varied input physics. 

As further discussed in Appendix~\ref{App:muons}, the parameters of non-convective models adopt best-fit values that are well outside the $1\sigma$ confidence bands found for the benchmark simulations. This fact underlines the strong impact of convection on the neutrino signal. As general trends, we find in simulations with Class~A EoSs in the absence of convection that
\begin{enumerate}[(i)]
\item $C$ decreases because of the lack of convective enhancement of the luminosities at early times, 
\item $\tau$ becomes larger because of the extended PNS neutrino cooling time without convective energy transport,
\item $n$ becomes smaller to account for the considerable signal stretching at late times,
\end{enumerate}
which implies that the exponential luminosity decline starts at later times and also proceeds more slowly. For $\bar\nu_e$ in model 1.62-DD2-c we notice an exception from the described general trends with respect to $\tau_{\bar\nu_e}$, which is slightly smaller than the value of the corresponding model with convection (see Table~\ref{tab:lnucm} and Fig.~\ref{Fig:nuebcm} in Appendix~\ref{App:muons}). In this case the mild decrease of $\tau_{\bar\nu_e}$ seems to be compensated by a reduction of $n_{\bar\nu_e}$ by a factor $\sim$3 compared to the non-convective model, which is by far the largest relative change for any neutrino species in all models with vs.\ without convection. Interestingly, the change in $\alpha$ depends on the EoS and neutrino species, showing, for instance, a decrease in 1.62-DD2-c compared to 1.62-DD2, an increase in 1.62-SFHo-c compared to 1.62-SFHo, and a decrease or slight increase in 1.62-SFHx-c compared to 1.62-SFHx depending on the type of neutrino. This nonuniform behavior points to differences in the influence of the EoS on PNS convection and the associated effects on the emission of different kinds of neutrinos during the early PNS cooling phase.

For simulations with the LS220 EoS, partly because of the poorer quality of the fits for the benchmark models, the omission of convection leads to radical changes in the values of the best-fit parameters. Indeed, LS220 simulations without convection show positive values of $\alpha$ (around unity), much larger values of $\tau$ and $n$ compared to the full-physics cases, and values of $C$ that are well compatible with those of simulations with the other EoSs including and excluding convection, i.e., the $C$ values are close to the luminosity values at 1\,s instead of being several 100\,B/s for our benchmark models. Finally, the weaker impact of muons on the neutrino signal is highlighted by the small changes in the best-fit parameters obtained for simulations without muons, as further detailed in Appendix~\ref{App:muons}.

\section{Counting rate in neutrino detectors}
\label{sec:count}

In order to exemplify a possible application of our luminosity fits, we discuss in this section the time evolution of the counting rate $t R_\nu(t)$ in a neutrino detector that will monitor the $t L_\nu(t)$ evolution in the case of a future Galactic SN explosion. We will demonstrate that our fitting recipe is also useful for fitting the observed neutrino signal. For this purpose, we consider as a reference case a SN at a distance of $D=10$\,kpc and evaluate the predicted signal in the water Cherenkov detector of Super-Kamiokande (SK)~\cite{Super-Kamiokande:2002weg,Abe:2013gga}, inspired by the analysis in Ref.~\cite{Tamborra:2014hga}.  

We consider the following ${\nu}$ differential flux per unit energy in $\MeV^{-1}\,\s^{-1}\,\cm^{-2}$:
\begin{equation}
    \mathcal{F}_{\nu}^0 (E_\nu)= \frac{dF_{\nu}}{dE_\nu}=\frac{L_{\nu}}{4\, \pi\, D^2 \langle E_\nu \rangle} \frac{(1+\beta_\nu)^{1+\beta_\nu}}{\Gamma(1+\beta_\nu)\, \langle E_\nu \rangle}\,\left(\frac{E_\nu}{\langle E_\nu \rangle}\right)^{\beta_\nu}\,e^{-(1+\beta_\nu)E_\nu/\langle E_\nu \rangle}\,,
    \label{eq:fbnue}
\end{equation}
where the shape parameter $\beta_\nu$ is given by
\begin{equation}
  \beta_\nu = \frac{\langle E_\nu^2 \rangle - 2\,\langle E_\nu \rangle^2}{\langle E_\nu \rangle^2 - \langle E_\nu^2 \rangle}\,,
  \label{eq:betanu}
\end{equation}
with $\langle E_\nu \rangle$ and $\langle E_\nu^2 \rangle$ being the average neutrino energy and the average squared neutrino energy, respectively.

In SK, the main detection process is inverse $\beta$ decay, $\bar{\nu}_e\,p\to n\,e^+$, where the final-state positron shows up by its Cherenkov radiation.
Because of the similarity of the electron and non-electron antineutrino luminosities and spectra in our models during PNS Kelvin-Helmholtz cooling at times $t \gtrsim 1$\,s after bounce, flavor conversions are not a major effect and can be neglected in our simplified analysis. Therefore the expected rate can be written as
\begin{equation}
    R_{\bar{\nu}_e} = N_p \int dE_e\, \int dE_\nu\, \mathcal{F}_{\bar{\nu}_e}(E_\nu) \sigma'(E_e,E_\nu)\,,
\label{eq:ratenum}
\end{equation}
where $N_p=1.51\times 10^{33}$ is the number of protons for a 22.5 kton Cherenkov detector. Here, we follow Ref.~\cite{Strumia:2003zx} for the limits of integration in $dE_\nu$ and we integrate the positron energies above the energy threshold $E_{\rm th,SK}=5~\MeV$. We mention that SK is essentially background free. Estimates for the future Hyper-Kamiokande detector with fiducial mass of 187 kton~\cite{Hyper-Kamiokande:2018ofw} can be obtained by rescaling the counting rate computed for SK by a factor $\sim 8.3$, without affecting the temporal evolution of the signal.

\begin{figure}[t!]
\centering
\includegraphics[width=0.45\textwidth]{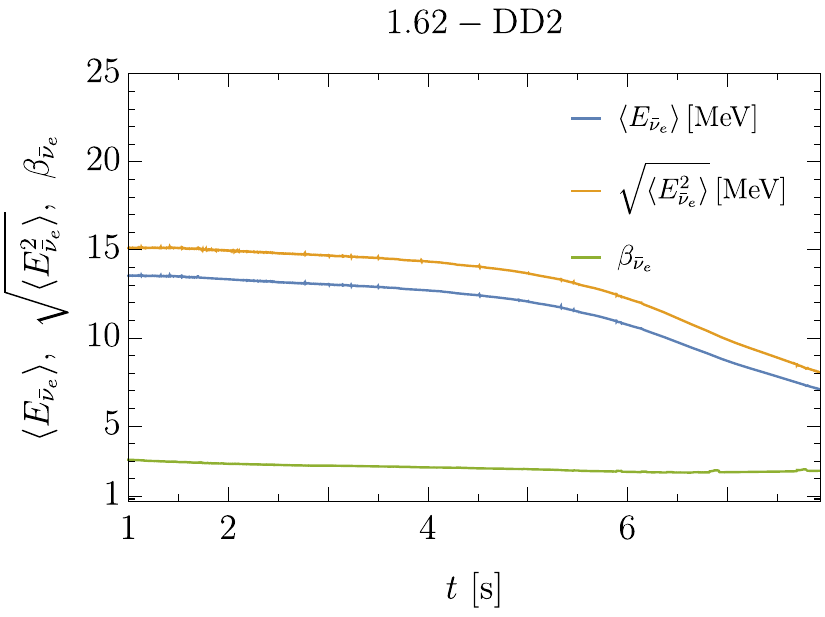}~~~~~~
\includegraphics[width=0.485\textwidth]{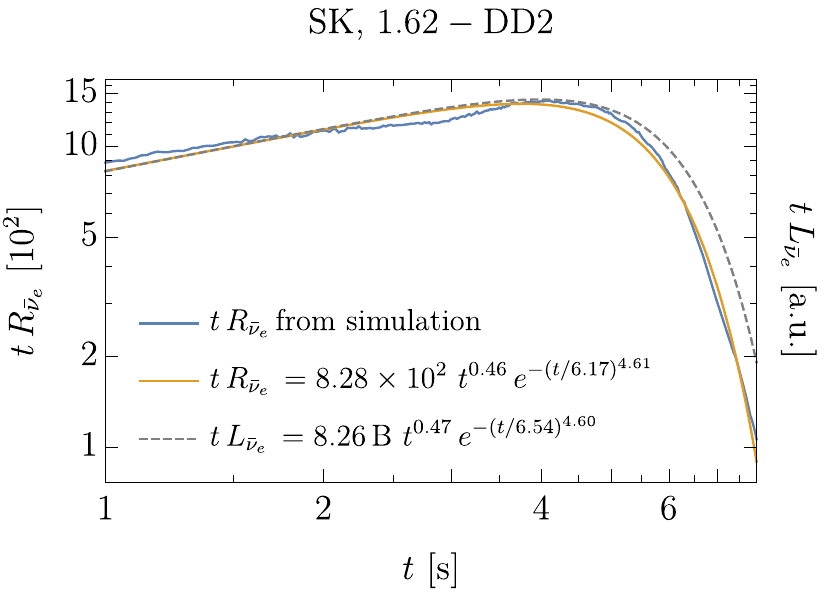}
\caption{\emph{Left panel:} Simulation results for the average energy $\langle E_{\bar{\nu}_e} \rangle$ (blue), rms energy $\sqrt{\langle E_{\bar{\nu}_e}^2 \rangle}$ (orange) and spectral shape parameter $\beta_{\bar{\nu}_e}$ (green) from model 1.62-DD2. \emph{Right panel:} $\bar{\nu}_e$-induced counting rate $t R_{{\bar\nu}_e}$ in SK computed from simulation data (blue), the fit of the counting rate (orange), and the fit of $t L_{{\bar\nu}_e}$ (dashed line).}
\label{Fig:HK}
\end{figure}

To obtain a numerical estimate, we use data from model 1.62-DD2 as an example, but the same analysis is valid for all of the models. In the left panel of Fig.~\ref{Fig:HK} we plot the average energy $\langle E_{\bar{\nu}_e} \rangle$, the root-mean-square (rms) energy $\sqrt{\langle E_{\bar{\nu}_e}^2 \rangle}$, and the shape parameter $\beta_{\bar{\nu}_e}$ as a function of the time, between 1\,s and $t_{\bar{\nu}_e,{\rm c}}=7.94\,\s$. These quantities exhibit a weak time dependence, with $\langle E_{\bar{\nu}_e} \rangle\approx (12-13.5)\,\MeV$, $\sqrt{\langle E_{\bar{\nu}_e}^2 \rangle}\approx (14-15)\,\MeV$ and $\beta_{\bar{\nu}_e}\approx 2.5-3$ at $t\lesssim 5\,\s$ and a decrease at later times. Here we focus on the 1.62-DD2 simulation as a representative case, but qualitatively similar results can be obtained with all the other models. Basic information on the time evolution of the mean neutrino energies for all considered
EoSs can be found in Ref.~\cite{Fiorillo:2023frv} (see Fig.~3 therein, where the time evolution of the average neutrino energies for the $1.44~M_\odot$ with different nuclear EoSs are shown).

The rate in SK can be simply estimated as
\begin{equation}
    R_{\bar{\nu}_e} \approx \frac{L_{\bar{\nu}_e}}{4\pi\,D^2 \langle E_{\bar{\nu}_e}\rangle}\,N_{\rm SK}\,\langle \sigma \rangle\,,
\label{eq:rate}
\end{equation}
where the average cross section is~\cite{Tamborra:2014hga}
\begin{equation}
    \langle \sigma \rangle = 7.37\times 10^{-46}\,\cm^2\,\frac{2+\beta_{\bar{\nu}_e}}{1+\beta_{\bar{\nu}_e}}\,\langle E_{\bar{\nu}_e} \rangle^{2.15}\,\times \left(\frac{76.64}{\beta_{\bar{\nu}_e}^{0.021}}-\frac{\langle E_{\bar{\nu}_e} \rangle}{\beta_{\bar{\nu}_e}^{0.24}}\right)\,\times\, \left[1-\exp\left(\frac{-0.25 + 0.55\,\langle E_{\bar{\nu}_e} \rangle}{2.2 + \beta_{\bar{\nu}_e}} - \frac{1+1.6\beta_{\bar{\nu}_e}}{1+4\beta_{\bar{\nu}_e}}\,\langle E_{\bar{\nu}_e} \rangle\right)\right],
    \label{eq:cross}
\end{equation}
dependent on $\langle E_{\bar{\nu}_e} \rangle$ 
and $\beta_{\bar{\nu}_e}$ [and also on $\sqrt{\langle E_{\bar{\nu}_e}^2 \rangle}$ via Eq.~\eqref{eq:betanu}]. Therefore, we expect that the observed rate $R_{\bar{\nu}_e}$ will follow the time dependence of the neutrino luminosity because of the weak time dependence of the other parameters, and that we can also fit it with our analytical formula in Eq.~\eqref{eq:fit}. In the right panel of Fig.~\ref{Fig:HK} we show the quantity $tR_{\bar{\nu}_e}$ obtained from Eq.~\eqref{eq:ratenum} using data from model 1.62-DD2 (blue line), its fit (orange), and the fitted $t L_{{\bar\nu}_e}$ (dashed line). As shown by the best-fit parameters for 
$t\,R_{\bar{\nu}_e}$ given in the plot, the fit leads to $C=8.28\times 10^{2}\,\s^{-1}$, 
$\alpha=0.54$, $\tau = 6.17\,\s$ and $n=4.61$, to be compared with the ones for $t L_{\bar{\nu}_e}$. The parameters $\alpha_{\bar{\nu}_e}$, $\tau_{\bar{\nu}_e}$ and $n_{\bar{\nu}_e}$ are well reconstructed, while one could get information on the normalization factor $C_{\bar{\nu}_e}$ by inverting Eq.~\eqref{eq:rate}. The slight difference between the parameters $\alpha$, $\tau$ and $n$ reconstructed from the rate and the original ones for $L_{\bar{\nu}_e}$ can be explained by the weak time dependence of $\langle E_\nu \rangle$, $\langle E_\nu^2 \rangle$, and $\beta_\nu$  entering in the computation of the rate. 
Therefore, one can fit the detected event rate with the same functional form used
for the luminosity, and from the reconstructed fitting parameters one can get information 
on the  PNS mass and on the EoS, as discussed in the previous Sections.
However, the accurate reconstruction of the parameters is beyond the scope of this analysis and will be the subject of future work dealing with their possible reconstruction using current and future neutrino detectors. There, more comprehensive information on the time evolution of $\langle E_\nu \rangle$, $\langle E_\nu^2 \rangle$ and $\beta_\nu$
will be provided.

\section{Conclusions}
\label{sec:conclusions}

In this paper we have investigated  whether the simple analytical function of Eq.~(\ref{eq:fit}) can be used as a parametric fit to the SN neutrino luminosities during the Kelvin-Helmholtz cooling phase of the PNS. For this purpose we have considered a set of several 1D simulations for different NS masses and EoSs. Our benchmark models account for PNS convection, which has a strong impact on the cooling evolution and its associated neutrino signal. In particular, we presented fits for the time-dependent neutrino luminosities
from numerical PNS cooling simulations and reported systematic dependences of the fitting-parameter values as functions of the NS mass. Future work is desirable where these fits are connected to analytic descriptions and basic PNS and EoS properties in a more formal way, e.g., similar to what was done for PNS cooling models that did not include the effects of PNS convection  (see, e.g., \cite{Suwa:2020nee,Nakazato:2019ojk}).

Our fit function employs four free parameters, namely a normalization factor $C$, a power-law exponent $\alpha$ for the time, an exponential cooling timescale $\tau$, and an exponent $n$ of $(t/\tau)$ in the exponential function. Their characteristic dependence on the PNS mass and on the EoS can be used to draw inferences on these latter properties, if the parameters are deduced from the neutrino signal of a future Galactic SN explosion.
For this purpose, we plan to investigate in future work how one can infer the parameter values of the neutrino luminosity from the SN neutrino signal measured in large underground detectors.
We have demonstrated that this possibility is facilitated by the fact that the time evolution of the detected event rate depends on the neutrino luminosity $L_\nu$, the average neutrino energy $\langle E_\nu \rangle$, and the rms energy $\sqrt{\langle E_\nu^2 \rangle}$, but the time dependence of $\langle E_\nu \rangle$ and $\sqrt{\langle E_\nu^2 \rangle}$ is weak. This allowed us to show that Eq.~(\ref{eq:fit}) provides a good functional form to also fit the time evolution of the observed neutrino signal.
Therefore, for a first estimation of the $L_\nu$-fit parameters from the event rate measured by a SN neutrino detector, one can simply apply our analytical expression for the luminosity fit and make use of the assumption that the average neutrino energy and the pinching parameter characterizing the spectral shape are constant in time.

A number of caveats of our study reported here need to be mentioned. First, the present analysis and our proposed luminosity fit are based on 1D SN and PNS cooling simulations using a fairly limited set of cases for the NS EoS. The general applicability of the fit function of Eq.~(\ref{eq:fit}) needs confirmation by testing a much larger variety of EoS models with a wide range of fundamental nuclear physics inputs that are compatible with all experimental, theoretical, and astrophysical constraints on the properties of nuclear matter and observed NSs. In particular, possible correlations of some of the fitting parameters [e.g., the relation in Eq.~(\ref{eq:tau-alpha})] require confirmation based on a wider spectrum of nuclear EoS representations. Second, our 1D SN and PNS calculations disregard 3D effects such as long-lasting accretion onto the PNS (continuing also after the onset of the explosion, when in 1D models accretion abruptly stops) and fallback of some initial explosion ejecta during the late PNS evolution \cite{Janka1996,Fiorillo:2023frv,Akaho:2023alv}. Moreover, the mixing-length treatment of PNS convection in our 1D models will have to be validated by long-time 3D simulations of PNS cooling once such calculations with good spatial resolution become available. In particular, this will also provide a test whether the fit function of Eq.~(\ref{eq:fit}) and our best-fit parameter values are compatible with 3D results for PNS cooling. If so, any deviation from the luminosity evolution described by our fit function would signal additional contributions to the neutrino emission added on top of the cooling component from the PNS. Thus, our $L_\nu$-fit could help to diagnose, disentangle, and describe such secondary neutrino emission phenomena in the neutrino measurement for a future Galactic SN. Finally, it will have to be seen how our fitting function reacts to additional, so far disregarded effects of potential importance in neutrino-cooling SN cores, for example fast flavor conversion of neutrino-antineutrino pairs, which could have a major impact on the neutrino emission properties \cite{Ehring+2023a,Ehring+2023b}, or extra cooling associated with the emission of light, weakly interacting beyond-standard-model particles (e.g., axions~\cite{Fischer:2021jfm}). Again, our neutrino luminosity fits could help to diagnose such effects beyond current standard SN modeling, once simulations including this new physics become available to be analyzed for long-time fitting.

In conclusion, we are confident that the simplicity of our fitting procedure will make it a useful tool for the neutrino community to describe the SN neutrino signal expected in a high-statistics detection, to probe a future SN neutrino measurement, and to infer valuable information on the PNS mass, nuclear EoS, and different signal components (see Ref.~\cite{Harada:2023elm} for a recent approach in this direction).

The considered model results are adopted from Ref.~\cite{Fiorillo:2023frv} and are available in the Garching Core-collapse Supernova Archive~\cite{SNarchive} upon request.

\vspace{2cm}
\acknowledgements

The work of AM  was partially supported by the research grant number 2022E2J4RK "PANTHEON: Perspectives in Astroparticle and
Neutrino THEory with Old and New messengers" under the program PRIN 2022 funded by the Italian Ministero dell’Universit\`a e della Ricerca (MUR). 
This work is (partially) supported by ICSC – Centro Nazionale di Ricerca in High Performance Computing, Big Data and Quantum Computing, funded by European Union--NextGenerationEU.
GL acknowledges support by the European Union’s Horizon 2020 Europe research and innovation programme under the Marie Skłodowska-Curie grant agreement No 860881-HIDDeN. Research at Garching received support by the German Research Foundation (DFG) through the Collaborative Research Centre ``Neutrinos and Dark Matter in Astro- and Particle Physics (NDM),'' Grant No.\ SFB-1258-283604770, and under Germany's Excellence Strategy through the Cluster of Excellence ORIGINS EXC-2094-390783311.\\

\bibliography{biblio.bib}

\appendix

\section{Tables for the final-time parameters}
\label{App:tmax}

Here we provide the values of the final simulation time $t_{\rm fin}$, the time $t_{\nu_i,{\rm max}}$ when the quantity $tL_{\nu_i}$ reaches its maximum, $t_{\nu_i,{\rm max}} \,L_{\nu_i}(t_{\nu_i,{\rm max}}) \equiv t_{\nu_i,\rm max}L_{\nu_i,\,\rm max}$, and the selected cut time $t_{\nu_i,{\rm c}}$, for neutrinos in Table~\ref{tab:lnumax} and for antineutrinos in Table~\ref{tab:lnubarmax} for all the simulations considered in our study. For each (anti)neutrino species, the selected cut time $t_{\nu_i,{\rm c}}$ is the time when the quantity $t\,L_{\nu_i}$ is reduced relative to its maximum value by a factor $X_{\nu_i}^{\rm c}=t_{\nu_i,\rm c}L_{\nu_i,\,\rm c}/t_{\nu_i,\rm max}L_{\nu_i,\,\rm max} = 0.15$, in order to take a common final time for all neutrino species. As shown by the quantities in bold print in Table~\ref{tab:lnumax} and Table~\ref{tab:lnubarmax}, $t_{\nu_i,{\rm c}}>t_{\rm fin}$ for 1.62-SFHx-c and 1.61-LS220-c, i.e. these simulations stop before reaching $t_{\nu_i,{\rm c}}$. Therefore, to check the impact of convection in these two cases we cut our simulation outputs at earlier times, $t_{\nu_i,{\rm Co}}$, when $X_{\nu_i}^{\rm Co}= t_{\nu_i,{\rm Co}}L_{\nu_i,\,{\rm Co}}/t_{\nu_i,\rm max}L_{\nu_i,\,\rm max}=0.22$ for simulations with SFHx and $X_{\nu_i}^{{\rm Co}}=0.19$ for LS220. We show $t_{\rm fin}$, $t_{\nu_i,{\rm max}}$ and $t_{\nu_i,{\rm Co}}$ used to check the impact of convection for electron and muon (anti)neutrinos in Table~\ref{tab:lnumaxc}.

\begin{table}[t!]
\centering
 \begin{tabular}{|c|c|c|c|c|c|c|c|}
\hline
Model &$t_{\rm fin}$ [s] & $t_{\nu_e,{\rm max}}$ [s]  &  $t_{\nu_e,{\rm c}}$ [s]  & $t_{\nu_\mu,{\rm max}}$ [s]  &  $t_{\nu_\mu,{\rm c}}$ [s] & $t_{\nu_\tau,{\rm max}}$ [s]  &  $t_{\nu_\tau,{\rm c}}$ [s] \\
\hline
1.36-DD2 &  8.69 & 3.14 & 6.99 & 4.03 & 6.98 & 3.81 & 7.01 \\
1.36-SFHo & 10.50 & 3.09 & 8.51 & 4.00 & 8.65 & 4.11 & 8.62  \\
1.36-SFHx & 10.06 &  3.62 & 8.69 & 4.25 & 8.91 & 4.18 & 8.86\\
1.36-LS220 & 12.36 &  1.81 & 10.72 & 1.94 & 10.52 & 1.99 & 10.44 \\
\hline
1.44-DD2 & 13.72 & 3.13 & 7.33 & 4.14 & 7.34 & 4.05 & 7.38\\
1.44-SFHo & 15.00 & 3.27 & 8.96 & 4.24 & 9.10 & 4.18 & 9.09\\
1.44-SFHx & 11.72 & 3.71 & 9.13 & 3.78 & 9.36 & 4.19 & 9.32\\
1.44-LS220 & 14.84 & 2.14 & 11.33 & 1.99 & 11.14 & 2.14 & 11.09 \\
\hline
1.62-DD2 & 10.75 & 3.94 & 8.02 & 4.26 & 8.06 & 4.52 & 8.12\\
1.62-SFHo & 14.26 & 3.46 & 9.89 & 4.71 & 10.06 & 4.73 & 10.06\\
1.62-SFHx & 13.45 &  4.28 & 10.09 & 4.27 & 10.36 & 4.98 & 10.36\\
1.62-LS220 & 13.58 & 2.35 & 12.75 & 2.43 & 12.55 & 2.44 & 12.52 \\
\hline
1.77-DD2 & 11.26 & 4.65 & 8.61 & 4.99 & 8.66 & 4.65 & 8.74\\
1.77-SFHo & 13.28 & 4.98 & 10.65 & 4.98 & 10.85 & 4.98 & 10.84\\
1.77-SFHx & 13.91 &  4.81 & 10.92 & 5.40 & 11.21 & 5.40 & 11.19\\
1.77-LS220 & 16.33 & 2.51 & 13.83 & 2.51 & 13.69 & 2.63 & 13.68 \\
\hline
1.93-DD2 & 12.81 & 5.03 & 9.23 & 5.52 & 9.32 & 5.03 & 9.39\\
1.93-SFHo & 15.52 & 5.37 & 11.55 & 5.26 & 11.79 & 5.47 & 11.82\\
1.93-SFHx & 16.38 & 5.54 & 11.86 & 5.54 & 12.19 & 6.00 & 12.26\\
1.93-LS220 & 19.95 &  2.87 & 14.79 & 3.16 & 14.69 & 3.02 & 14.90\\
\hline
1.62-DD2-c & 13.95 &  4.15 & 12.47 & 4.33 & 12.54 & 4.24 & 12.54\\
1.62-SFHo-c & 19.74 & 4.38 & 18.02 & 5.20 & 18.18 & 4.89 & 18.05\\
1.62-SFHx-c & 18.75 & 4.22 & $>\textbf{18.75}$ & 4.74 & $>\textbf{18.75}$ & 4.75 & $>\textbf{18.75}$\\
1.61-LS220-c & 20.92 & 4.06 & $>\textbf{20.92}$ & 2.51 & $>\textbf{20.92}$ & 2.53 & $>\textbf{20.92}$\\
\hline
1.62-DD2-m & 9.58 & 3.90 & 7.38 & 3.90 & 7.46 & 3.90 & 7.46\\
1.62-SFHo-m & 13.55 & 3.95 & 9.15 & 4.47 & 9.38 & 4.47 & 9.38\\
		\hline
	\end{tabular}
	\caption{Times $t_{\nu_i,{\rm max}}$ when $t\,L_{\nu_i}$ adopts its maximum $t_{\nu_i,{\rm max}}\,L_{\nu_i,{\rm max}}$ and times $t_{\nu_i,{\rm c}}$ when $X_{\nu_i}^{\rm c}=0.15$ for all neutrino species $\nu_i$. Bold print marks values corresponding to $t_{\nu_i,{\rm c}}>t_{\rm fin}$, i.e., cases when the simulation was stopped before $t_{\nu_i,{\rm c}}$ was reached.}
	\label{tab:lnumax}
\end{table}

\begin{table}[t!]
\centering
 \begin{tabular}{|c|c|c|c|c|c|c|c|}
\hline
Model & $t_{\rm fin}$ [s] & $t_{\bar{\nu}_e,{\rm max}}$ [s]  &  $t_{\bar{\nu}_e,{\rm c}}$ [s]  & $t_{\bar{\nu}_\mu,{\rm max}}$ [s]  &  $t_{\bar{\nu}_\mu,{\rm c}}$ [s] & $t_{\bar{\nu}_\tau,{\rm max}}$ [s]  &  $t_{\bar{\nu}_\tau,{\rm c}}$ [s] \\
\hline
1.36-DD2 &  8.69 & 3.56 & 6.89 & 3.31 & 7.01 & 3.81 & 7.00 \\
1.36-SFHo & 10.50 & 4.08 & 8.40 & 4.07 & 8.68 & 4.10 & 8.61  \\
1.36-SFHx & 10.06 & 4.12 & 8.65 & 3.77 & 8.94 & 4.18 & 8.84\\
1.36-LS220 & 12.36 & 1.96 & 10.45 & 1.95 & 10.31 & 1.99 & 10.39 \\
\hline
1.44-DD2 & 13.72 &  3.87 & 7.22 & 3.83 & 7.37 & 4.05 & 7.37\\
1.44-SFHo & 15.00 & 4.47 & 8.20 & 4.47 & 9.11 & 4.18 & 9.07\\
1.44-SFHx & 11.72 & 4.19 & 9.08 & 4.08 & 9.40 & 4.19 & 9.30\\
1.44-LS220 & 14.84 & 1.93 & 11.08 & 1.93 & 10.95 & 2.14 & 11.03 \\
\hline
1.62-DD2 & 10.75 & 4.12 & 7.94 & 4.06 & 8.11 & 4.52 & 8.10\\
1.62-SFHo & 14.26 & 4.39 & 9.74 & 3.73 & 10.09 & 4.72 & 10.04\\
1.62-SFHx & 13.45 & 5.00 & 10.06 & 4.36 & 10.46 & 4.88 & 10.34\\
1.62-LS220 & 13.58 & 2.36 & 12.44 & 2.35 & 12.23 & 2.36 & 12.43 \\
\hline
1.77-DD2 & 11.26 & 4.45 & 8.51 & 4.49 & 8.72 & 4.65 & 8.72\\
1.77-SFHo & 13.28 & 4.98 & 10.50 & 4.50 & 10.88 & 4.98 & 10.82\\
1.77-SFHx & 13.91 & 5.40 & 10.89 & 5.17 & 11.28 & 5.40& 11.17\\
1.77-LS220 & 16.33 &  2.67 & 13.56 & 2.67 & 13.24 & 2.63 & 13.55 \\
\hline
1.93-DD2 & 12.81 & 5.03 & 9.11 & 4.97 & 9.35 & 5.03 & 9.37\\
1.93-SFHo & 15.52 & 5.47 & 11.38 & 5.50 & 11.83 & 5.47 & 11.80\\
1.93-SFHx & 16.38 & 6.04 & 11.93 & 5.97 & 12.35 & 6.00 & 12.23\\
1.93-LS220 & 19.95 & 2.74 & 14.85 & 2.74 & 14.38 & 3.02 & 14.76\\
\hline
1.62-DD2-c & 13.95 &  3.90 & 12.14 & 4.18 & 12.49 & 4.24 & 12.52\\
1.62-SFHo-c & 19.74 & 4.32 & 17.82 & 4.49 & 18.12 & 4.49 & 18.12\\
1.62-SFHx-c & 18.75 & 3.79 & $>\textbf{18.75}$ & 4.55 & $>\textbf{18.75}$ & 4.88 & $>\textbf{18.75}$\\
1.61-LS220-c & 20.92 & 2.26 & 20.85 & 2.42 & $>\textbf{20.92}$ & 2.36 & $>\textbf{20.92}$\\
\hline
1.62-DD2-m & 9.58 & 3.90 & 7.32 & 3.90 & 7.45 & 3.90 & 7.45\\
1.62-SFHo-m & 13.55 & 4.13 & 9.00 & 4.47 & 9.37 & 4.47 & 9.37\\
		\hline
	\end{tabular}
	\caption{Times $t_{\bar{\nu}_i,{\rm max}}$ when $t\,L_{\bar{\nu}_i}$ adopts its maximum $t_{\bar{\nu}_i,{\rm max}}\,L_{\bar{\nu}_i,{\rm max}}$ and times $t_{\bar{\nu}_i,{\rm c}}$ when $X_{\bar{\nu}_i}^{\rm c}=0.15$ for all antineutrino species $\bar{\nu}_i$. Bold print marks values corresponding to $t_{\bar{\nu}_i,{\rm c}}>t_{\rm fin}$, i.e., cases when the simulation was stopped before $t_{\bar{\nu}_i,{\rm c}}$ was reached.}
	\label{tab:lnubarmax}
\end{table} 

\begin{table}[t!]
\centering
 \begin{tabular}{|c|c|c|c|c|c|c|c|c|c|}
\hline
Model & $t_{\rm fin}$ [s] & $t_{\nu_e,{\rm max}}$ [s]  &  $t_{\nu_e,{\rm Co}}$ [s]  & $t_{\bar{\nu}_e,{\rm max}}$ [s]  &  $t_{\bar{\nu}_e,{\rm Co}}$ [s]  & $t_{\nu_\mu,{\rm max}}$ [s]  &  $t_{\nu_\mu,{\rm Co}}$ [s]  & $t_{\bar{\nu}_\mu,{\rm max}}$ [s]  &  $t_{\bar{\nu}_\mu,{\rm Co}}$ [s] \\
\hline
1.62-SFHx & 13.45 & 4.26 & $9.39$ & 5.00 & $9.38$ & 4.27 & $9.74$ & 4.36 & $9.86$\\
1.62-SFHx-c & 18.75 & 4.22 & $18.09$ & 3.79 & $17.99$ & 4.74 & $18.73$ & 4.55 & $18.64$\\
\hline
\hline
1.62-LS220 & 13.58 & 2.35 & $11.92$ & 2.36 & $11.58$ & 2.43 & $11.68$ & 2.35 & $10.98$\\
1.61-LS220-c & 20.92 & 4.06 & $20.62$ & 2.26 & $20.03$ & 2.51 & $20.81$ & 2.42 & $20.49$\\
\hline
	\end{tabular}
	\caption{The time $t_{{\nu}_i,{\rm max}}$ when $t\,L_{\nu_i}$ is maximum and the time $t_{\nu_i,{\rm Co}}$ for $\nu_e$, $\bar{\nu}_e$, $\nu_\mu$ and $\bar{\nu}_\mu$ adopted to test the impact of convection in simulations where $X_{\nu_i}^{\rm fin} > 0.15$. For simulations with the SFHx EoS, $t_{\nu_i,{\rm Co}}$ is the time when $X_{\nu_i}^{{\rm Co}}=0.22$, while for LS220 is the time when $X_{\nu_i}^{{\rm Co}}=0.19$.}
	\label{tab:lnumaxc}
\end{table}

\section{Tables for best-fit parameters of luminosities for neutrinos and antineutrinos of all flavors}
\label{App:bestfit}

Here we report the best-fit values with $1\sigma$ errors of the parameters characterizing the fit for the time evolution of all neutrino and antineutrino luminosities in the time interval from $1\,\s$ to $t_{\nu_i,{\rm c}}$ (see Appendix~\ref{App:tmax} for more details). The fit function is given by Eq.~(\ref{eq:fit}),
\begin{equation}
L_{\nu_i}(t) = C\, t^{-\alpha}\, e^{-(t/\tau)^{n}}\,,
\nonumber
\end{equation}
with $C$, $\alpha$, $\tau$ and $n$ being free parameters. We show values of the fit parameters for $\nu_e$ and $\bar{\nu}_e$ in Table~\ref{tab:lnue}, for $\nu_\mu$ and $\bar{\nu}_\mu$ in Table~\ref{tab:lnum} and for $\nu_\tau$ and $\bar{\nu}_\tau$ in Table~\ref{tab:lnut}, obtained with the NonlinearModelFit function in \emph{Mathematica}.

   \begin{table}[t!]
    \centering
        \begin{tabular}{|c|c|c|c|c||c|c|c|c|}
    \hline
Model & $C_{\nu_e}$ [B/s] & $\alpha_{\nu_e}$ & $\tau_{\nu_e}$ [s] & $n_{\nu_e}$ & $C_{\bar{\nu}_e}$ [B/s] & $\alpha_{\bar{\nu}_e}$ & $\tau_{\bar{\nu}_e}$ [s] & $n_{\bar{\nu}_e}$ \\
      \hline
   1.36-DD2 & $5.485\pm0.007$ & $0.410\pm0.002$ & $5.384\pm0.004$ & $3.660\pm0.008$ & $6.322\pm0.010$ & $0.550 \pm 0.002$ & $5.659\pm0.004$ & $4.462\pm0.012$ \\
 1.36-SFHo & $6.339\pm0.004$ & $0.604\pm0.001$ & $6.573\pm0.002$ & $3.435\pm0.003$ & $6.730 \pm 0.008$ & $0.644 \pm 0.001$ & $6.756 \pm 0.004$ & $3.948 \pm 0.007$ \\
1.36-SFHx & $6.345\pm0.004$ & $0.626\pm0.001$ & $6.720\pm0.003$ & $3.431\pm0.004$ & $6.973 \pm 0.007$ & $0.686 \pm 0.001$ & $6.933\pm 0.003$ & $3.881 \pm 0.006$ \\
1.36-LS220 & $9.784\pm0.063$ & $0.727\pm0.010$ & $4.410\pm0.089$ & $1.061\pm0.011$ & $10.63 \pm 0.08$ & $0.795 \pm 0.012$ & $4.471 \pm 0.106$ & $1.052 \pm 0.013$ \\
   \hline
   1.44-DD2 & $6.050\pm0.008$ & $0.428\pm0.002$ & $5.724\pm0.004$ & $3.835\pm0.008$ & $6.996 \pm 0.011$ & $0.543 \pm 0.002$ & $5.964 \pm 0.004$ & $4.582 \pm 0.011$ \\
 1.44-SFHo & $7.054\pm0.004$ & $0.606\pm0.001$ & $6.953\pm0.002$ & $3.502\pm0.003$ & $7.515 \pm 0.007$ & $0.627 \pm 0.001$ & $7.077 \pm 0.003$ & $3.928 \pm 0.006$ \\
1.44-SFHx & $7.085\pm0.004$ & $0.637\pm0.001$ & $7.148\pm0.002$ & $3.547\pm0.003$ & $7.818 \pm 0.007$ & $0.680 \pm 0.001$ & $7.301 \pm 0.003$ & $3.916 \pm 0.006$ \\
1.44-LS220 & $120.3\pm18.1$ & $-0.504\pm0.048$ & $0.113\pm0.020$ & $0.449\pm0.010$ & $112.7 \pm 15.6$ & $-0.430 \pm 0.046$ & $0.135 \pm 0.022$ & $0.460 \pm 0.010$\\
    \hline
1.62-DD2 & $7.040\pm0.010$ & $0.430\pm0.002$ & $6.319\pm0.004$ & $3.959\pm0.008$ & $8.260 \pm 0.014$ & $0.529 \pm 0.002$ & $6.538 \pm 0.004$ & $4.599 \pm 0.012$ \\
1.62-SFHo & $8.308\pm0.005$ & $0.601\pm0.001$ & $7.743\pm0.002$ & $3.639\pm0.003$ & $8.967 \pm 0.008$ & $0.612 \pm 0.001$ & $7.823 \pm 0.003$ & $3.971 \pm 0.005$ \\
 1.62-SFHx & $8.317\pm0.006$ & $0.618\pm0.001$ & $7.923\pm0.003$ & $3.647\pm0.004$ & $9.474 \pm 0.009$ & $0.672 \pm 0.001$ & $8.091 \pm 0.003$ & $3.979 \pm 0.005$ \\
 1.62-LS220 & $266.7\pm50.4$ & $-0.672\pm0.053$ & $0.052\pm0.011$ & $0.404\pm0.009$ & $253.5 \pm 37.3$ & $-0.670 \pm 0.043$ & $0.063 \pm 0.010$ & $0.417 \pm 0.007$ \\
\hline
 1.77-DD2 & $7.948\pm0.010$ & $0.448\pm0.001$ & $6.875\pm0.004$ & $4.201\pm0.008$ & $9.428 \pm 0.014$ & $0.534 \pm 0.002$ & $7.068 \pm 0.004$ & $4.755 \pm 0.011$ \\
 1.77-SFHo & $9.306\pm0.006$ & $0.585\pm0.001$ & $8.372\pm0.002$ & $3.718\pm0.003$ & $10.21 \pm 0.01$ & $0.599 \pm 0.001$ & $8.443 \pm 0.003$ & $4.000\pm0.004$ \\
 1.77-SFHx & $9.346\pm0.005$ & $0.604\pm0.001$ & $8.594\pm0.002$ & $3.678\pm0.003$ & $10.92 \pm 0.01$ & $0.669 \pm 0.001$ & $8.795 \pm 0.003$ & $4.021 \pm 0.005$ \\
 1.77-LS220 & $404.1\pm100.0$ & $-0.745\pm0.066$ & $0.037\pm0.010$ & $0.384\pm0.010$ & $359.8 \pm 60.9$ & $-0.755 \pm 0.048$ & $0.048 \pm 0.009$ & $0.401 \pm 0.008$ \\
    \hline
 1.93-DD2 & $8.865\pm0.013$ & $0.445\pm0.002$ & $7.419\pm0.004$ & $4.349\pm0.008$ & $10.73 \pm 0.02$ & $0.534 \pm 0.001$ & $7.625 \pm 0.003$ & $4.941 \pm 0.010$ \\
 1.93-SFHo & $10.474\pm0.007$ & $0.575\pm0.001$ & $9.072\pm0.003$ & $3.749\pm0.004$ & $11.45 \pm 0.01$ & $0.577 \pm 0.001$ & $9.129 \pm 0.003$ & $4.016 \pm 0.004$ \\
 1.93-SFHx & $10.664\pm0.005$ & $0.6170\pm0.0005$ & $9.542\pm0.002$ & $3.971\pm0.003$ & $12.48 \pm 0.01$ & $0.666 \pm 0.001$ & $9.681 \pm 0.003$ & $4.163 \pm 0.004$ \\
 1.93-LS220 & $608.1\pm171.7$ & $-0.882\pm0.073$ & $0.026\pm0.008$ & $0.371\pm0.010$ & $477.1 \pm 85.8$ & $-0.872 \pm 0.050$ & $0.039 \pm 0.008$ & $0.392 \pm 0.008$ \\
    \hline
    \end{tabular}
\caption{Best-fit parameters with $1\,\sigma$ errors for $L_{\nu_i} = C \,t^{-\alpha}\,e^{-(t/\tau)^n}$ in the time interval between 1 s and $t_{\nu_i,{\rm c}}$ for $\nu_e$ (left) and $\bar{\nu}_e$ (right) for our benchmark models.}
\label{tab:lnue}
\end{table}

   \begin{table}[h!]
    \centering
    \begin{tabular}{|c|c|c|c|c||c|c|c|c|c|}
    \hline
   Model & $C_{\nu_\mu}$ [B/s] & $\alpha_{\nu_\mu}$ & $\tau_{\nu_\mu}$ [s] & $n_{\nu_\mu}$ & $C_{\bar{\nu}_\mu}$ [B/s] & $\alpha_{\bar{\nu}_\mu}$ & $\tau_{\bar{\nu}_\mu}$ [s] & $n_{\bar{\nu}_\mu}$ \\
       \hline
    1.36-DD2 & $6.333\pm0.008$ & $0.421\pm0.002$ & $5.687\pm0.003$ & $4.413\pm0.009$ & $6.927\pm0.008$ & $0.479\pm0.002$ & $5.754\pm0.003$ & $4.468\pm0.008$ \\
1.36-SFHo & $7.053\pm0.006$ & $0.579\pm0.001$ & $6.945\pm0.003$ & $3.926\pm0.005$ & $7.679\pm0.004$ & $0.634\pm0.001$ & $7.067\pm0.002$ & $4.045\pm0.003$\\
1.36-SFHx & $7.113\pm0.007$ & $0.598\pm0.001$ & $7.110\pm0.003$ & $3.873\pm0.005$ & $7.868\pm0.004$ & $0.662\pm0.001$ & $7.245\pm0.002$ & $3.964\pm0.003$ \\
1.36-LS220 & $69.88\pm8.62$ & $-0.177\pm0.045$ & $0.252\pm0.040$ & $0.499\pm0.012$ & $107.4\pm17.9$ & $-0.301\pm0.057$ & $0.164\pm0.033$ & $0.473\pm0.013$ \\
    \hline
    1.44-DD2 & $7.036\pm0.009$ & $0.426\pm0.002$ & $5.999\pm0.003$ & $4.501\pm0.009$ & $7.740\pm0.008$ & $0.489\pm0.001$ & $6.084\pm0.002$ & $4.607\pm0.008$ \\
1.44-SFHo & $7.951\pm0.006$ & $0.581\pm0.001$ & $7.317\pm0.003$ & $3.953\pm0.005$ & $8.662\pm0.004$ & $0.631\pm0.001$ & $7.429\pm0.002$ & $4.025\pm 0.003$ \\
1.44-SFHx & $8.053\pm0.007$ & $0.611\pm0.001$ & $7.535\pm0.003$ & $3.998\pm0.005$ & $8.940\pm0.004$ & $0.674\pm0.001$ & $7.671\pm0.001$ & $4.053\pm0.003$ \\
1.44-LS220 & $234.9\pm47.5$ & $-0.514\pm0.058$ & $0.068\pm0.016$ & $0.418\pm0.011$ & $297.6\pm66.0$ & $-0.573\pm0.063$ & $0.059\pm0.014$ & $0.416\pm0.011$ \\
    \hline
1.62-DD2 & $8.428\pm0.011$ & $0.430\pm0.002$ & $6.595\pm0.003$ & $4.542\pm0.009$ & $9.271\pm0.012$ & $0.486\pm0.001$ & $6.676\pm0.003$ & $4.569\pm0.009$ \\
1.62-SFHo & $9.669\pm0.008$ & $0.582\pm0.001$ & $8.104\pm0.003$ & $4.022\pm0.005$ & $10.60\pm0.01$ & $0.633\pm0.001$ & $8.239\pm0.002$ & $4.094\pm0.003$\\
1.62-SFHx & $9.802\pm0.009$ & $0.603\pm0.001$ & $8.318\pm0.003$ & $3.961\pm0.005$ & $10.98\pm0.01$ & $0.666\pm0.001$ & $8.466\pm0.002$ & $3.989\pm0.004$ \\
1.62-LS220 & $430.2\pm102.3$ & $-0.612\pm0.063$ & $0.041\pm0.011$ & $0.388\pm0.010$ & $519.6\pm128.6$ & $-0.720\pm0.067$ & $0.039\pm0.010$ & $0.394\pm0.011$ \\
\hline
    1.77-DD2 & $9.685\pm0.013$ & $0.441\pm0.001$ & $7.120\pm0.003$ & $4.668\pm0.009$ & $10.72\pm0.01$ & $0.498\pm0.001$ & $7.217\pm0.003$ & $4.709\pm0.008$ \\
1.77-SFHo & $11.12\pm0.01$ & $0.574\pm0.001$ & $8.741\pm0.002$ & $4.045\pm0.004$ & $12.32\pm0.01$ & $0.6294\pm0.0004$ & $8.897\pm0.001$ & $4.105\pm0.003$ \\
1.77-SFHx & $11.43\pm0.01$ & $0.610\pm0.001$ & $9.052\pm0.003$ & $4.049\pm0.005$ & $12.92\pm0.01$ & $0.6769\pm0.0005$ & $9.236\pm0.002$ & $4.115\pm0.003$ \\
1.77-LS220 & $531.9\pm144.0$ & $-0.634\pm0.071$ & $0.036\pm0.011$ & $0.380\pm0.011$ & $859.4\pm272.3$ & $-0.832\pm0.081$ & $0.026\pm0.008$ & $0.376\pm0.012$ \\
    \hline
1.93-DD2 & $11.00\pm0.02$ & $0.437\pm0.002$ & $7.650\pm0.004$ & $4.702\pm0.010$ & $12.47\pm0.01$ & $0.515\pm0.001$ & $7.802\pm0.003$ & $4.918\pm0.008$ \\
1.93-SFHo & $12.76\pm0.01$ & $0.563\pm0.001$ & $9.431\pm0.003$ & $3.968\pm0.004$ & $14.25\pm0.01$ & $0.6202\pm0.0005$ & $9.619\pm0.002$ & $4.049\pm0.003$\\
1.93-SFHx & $13.38\pm0.01$ & $0.623\pm0.001$ & $9.976\pm0.003$ & $4.256\pm0.004$ & $15.11\pm0.01$ & $0.6803\pm0.0004$ & $10.148\pm0.002$ & $4.213\pm0.003$\\
1.93-LS220 & $785.6\pm237.0$ & $-0.767\pm0.077$ & $0.027\pm0.009$ & $0.370\pm0.011$ & $803.2\pm240.7$ & $-0.882\pm0.081$ & $0.032\pm0.010$ & $0.385\pm0.012$ \\
    \hline
    \end{tabular}
        \begin{tabular}{|c|c|c|c|c|}
    \hline
    \end{tabular}
\caption{Best-fit parameters with $1\,\sigma$ errors for $L_{\nu_i} = C \,t^{-\alpha}\,e^{-(t/\tau)^n}$ in the time interval between 1 s and $t_{\nu_i,{\rm c}}$ for $\nu_\mu$ (left) and $\bar{\nu}_\mu$ (right) for our benchmark models.}
\label{tab:lnum}
\end{table}

   \begin{table}[h!]
    \centering
    \begin{tabular}{|c|c|c|c|c||c|c|c|c|}
    \hline
   Model & $C_{\nu_\tau}$ [B/s] & $\alpha_{\nu_\tau}$ & $\tau_{\nu_\tau}$ [s] & $n_{\nu_\tau}$ & $C_{\bar{\nu}_\tau}$ [B/s] & $\alpha_{\bar{\nu}_\tau}$ & $\tau_{\bar{\nu}_\tau}$ [s] & $n_{\bar{\nu}_\tau}$ \\
       \hline
    1.36-DD2 & $6.433\pm0.007$ & $0.440\pm0.001$ & $5.730\pm0.002$ & $4.470\pm0.007$ & $6.701\pm0.007$ & $0.446\pm0.001$ & $5.726\pm0.002$ & $4.489\pm0.007$\\
1.36-SFHo & $7.155\pm0.003$ & $0.603\pm0.001$ & $7.023\pm0.001$ & $4.116\pm0.003$ & $7.465\pm0.003$ & $0.609\pm0.001$ & $7.019\pm0.001$ & $4.129\pm0.003$ \\
1.36-SFHx & $7.272\pm0.004$ & $0.626\pm0.001$ & $7.196\pm0.002$ & $4.059\pm0.003$ & $7.584\pm0.004$ & $0.632\pm0.001$ & $7.190\pm0.002$ & $4.069\pm0.003$ \\
1.36-LS220 & $87.76\pm12.74$ & $-0.285\pm0.051$ & $0.192\pm0.034$ & $0.483\pm0.012$ & $113.7\pm19.1$ & $-0.324\pm0.056$ & $0.148\pm0.030$ & $0.465\pm0.012$ \\
    \hline
    1.44-DD2 & $7.137\pm0.008$ & $0.443\pm0.001$ & $6.043\pm0.002$ & $4.532\pm0.007$ & $7.442\pm0.008$ & $0.449\pm0.001$ & $6.040\pm0.002$ & $4.557\pm0.007$ \\
1.44-SFHo & $8.028\pm0.003$ & $0.5967\pm0.0005$ & $7.381\pm0.001$ & $4.089\pm0.003$ & $8.383\pm0.003$ & $0.6032\pm0.0005$ & $7.378\pm0.001$ & $4.104\pm0.003$ \\
1.44-SFHx & $8.210\pm0.004$ & $0.6346\pm0.0005$ & $7.612\pm0.001$ & $4.145\pm0.003$ & $8.570\pm0.004$ & $0.6403\pm0.0005$ & $7.606\pm0.001$ & $4.159\pm0.003$ \\
1.44-LS220 & $397.3\pm97.3$ & $-0.696\pm0.066$ & $0.040\pm0.010$ & $0.397\pm0.011$ & $192.1\pm36.2$ & $-0.444\pm0.057$ & $0.093\pm0.020$ & $0.435\pm0.011$ \\
    \hline
1.62-DD2 & $8.499\pm0.010$ & $0.439\pm0.001$ & $6.634\pm0.003$ & $4.513\pm0.008$ & $8.878\pm0.011$ & $0.447\pm0.001$ & $6.633\pm0.003$ & $4.543\pm0.008$ \\
1.62-SFHo & $9.714\pm0.005$ & $0.5922\pm0.0005$ & $8.169\pm0.001$ & $4.121\pm0.003$ & $10.166\pm0.005$ & $0.5993\pm0.0005$ & $8.167\pm0.001$ & $4.141\pm0.003$ \\
1.62-SFHx & $9.953\pm0.006$ & $0.620\pm0.001$ & $8.395\pm0.002$ & $4.064\pm0.004$ & $10.41\pm0.01$ & $0.626\pm0.001$ & $8.389\pm0.002$ & $4.078\pm0.004$ \\
1.62-LS220 & $366.1\pm80.8$ & $-0.623\pm0.061$ & $0.050\pm0.012$ & $0.401\pm0.010$ & $947.6\pm293.1$ & $-0.839\pm0.075$ & $0.019\pm0.006$ & $0.365\pm0.011$ \\
\hline
    1.77-DD2 & $9.731\pm0.011$ & $0.445\pm0.001$ & $7.157\pm0.003$ & $4.588\pm0.008$ & $10.18\pm0.01$ & $0.453\pm0.001$ & $7.158\pm0.003$ & $4.627\pm0.008$ \\
1.77-SFHo & $11.146\pm0.005$ & $0.5816\pm0.0004$ & $8.809\pm0.001$ & $4.115\pm0.003$ & $11.69\pm0.01$ & $0.5893\pm0.0004$ & $8.809\pm0.001$ & $4.139\pm0.003$ \\
1.77-SFHx & $11.57\pm0.01$ & $0.624\pm0.001$ & $9.138\pm0.002$ & $4.147\pm0.003$ & $12.11\pm0.01$ & $0.630\pm0.001$ & $9.131\pm0.002$ & $4.162\pm0.003$ \\
1.77-LS220 & $1283\pm457$ & $-0.914\pm0.084$ & $0.015\pm0.005$ & $0.353\pm0.011$ & $1323\pm490$ & $-0.892\pm0.088$ & $0.015\pm0.006$ & $0.353\pm0.012$ \\
    \hline
1.93-DD2 & $11.11\pm0.01$ & $0.448\pm0.001$ & $7.712\pm0.003$ & $4.680\pm0.008$ & $11.65\pm0.01$ & $0.457\pm0.001$ & $7.715\pm0.003$ & $4.729\pm0.008$ \\
1.93-SFHo & $12.73\pm0.01$ & $0.5668\pm0.0005$ & $9.512\pm0.002$ & $4.022\pm0.003$ & $13.38\pm0.01$ & $0.5751\pm0.0005$ & $9.513\pm0.002$ & $4.047\pm0.003$ \\
1.93-SFHx & $13.41\pm0.01$ & $0.6261\pm0.0005$ & $10.042\pm0.002$ & $4.262\pm0.003$ & $14.07\pm0.01$ & $0.6322\pm0.0005$ & $10.036\pm0.002$ & $4.283\pm0.003$ \\
1.93-LS220 & $1818\pm672$ & $-1.029\pm0.086$ & $0.011\pm0.004$ & $0.345\pm0.011$ & $1245\pm422$ & $-0.912\pm0.083$ & $0.018\pm0.006$ & $0.359\pm0.011$ \\
    \hline
    \end{tabular}~~~~~~~~~~~~~~~~
\caption{Best-fit parameters with $1\,\sigma$ errors for $L_{\nu_i} = C \,t^{-\alpha}\,e^{-(t/\tau)^n}$ in the time interval between 1 s and $t_{\nu_i,{\rm c}}$ for $\nu_\tau$ (left) and $\bar{\nu}_\tau$ (right) for our benchmark models.}
\label{tab:lnut}
\end{table} 

\section{Tables for the linear relations between luminosity-fitting parameters and PNS mass}
\label{App:massdep}

In Table~\ref{tab:massdep} we provide the best-fit values and their $1\,\sigma$ errors for the parameters $K_0$ and $K_1$ that describe the linear dependencies of the parameter values in the $L_\nu$-fit of Eq.~\eqref{eq:fit} on the PNS mass $M_{\rm NS}$ at fixed EoS, for all neutrino and antineutrino species:
\begin{equation}
    K = K_0 + K_1\,\frac{M_{\rm NS}}{M_\odot}\,,
    \label{eq:linMdep}
\end{equation}
where $K=C,\,\alpha,\,\tau,\,n$. The larger values of the relative uncertainties on $K_0$ and $K_1$ in Table~\ref{tab:massdep} and the widths of the confidence bands in Figs.~\ref{Fig:nuefitA}, \ref{Fig:nuebfitA}, and \ref{Fig:numbfitA} suggest that the linear fits work better for Class A EoSs than for LS220. At fixed EoS (in particular for Class A EoSs), the linear fits are excellent for $C$ and $\tau$ and slightly worse for $\alpha$ and $n$, featuring larger relative errors of the best-fit values of the parameters in Eq.~\eqref{eq:linMdep} (see, e.g., the values of $\alpha_1$ and $n_1$).

\begin{table}[h!]
 \centering
     \begin{tabular}{|c c|c|c|c|c|c|c|c|c|}
    \hline
   Neutrino & EoS & $C_0$ [B/s] & $C_1$ [B/s] & $\alpha_0$ & $\alpha_1$ & $\tau_0$ [s] & $\tau_1$ [s] & $n_0$ & $n_1$ \\
    \hline
 $\nu_e$ & DD2 & $-2.46\pm0.14$ & $5.87\pm0.09$  & $0.34\pm0.03$  & $0.06\pm0.02$ & $0.59\pm0.09$ & $3.54\pm0.06$  & $2.09\pm0.15$ & $1.17\pm0.09$  \\   
 $\nu_e$ & SFHo & $-3.29\pm0.23$  & $7.13\pm0.14$  & $0.68\pm0.02$  & $-0.05 \pm 0.01$ & $0.66\pm0.06$ & $4.36\pm0.04$  & $2.69\pm0.12$  & $0.57\pm0.07$ \\ 
 $\nu_e$ & SFHx & $-3.66\pm0.30$ & $7.39\pm0.18$  & $0.68\pm0.03$ & $-0.04\pm0.02$ & $0.16\pm0.29$ & $4.82\pm0.18$  & $2.32 \pm 0.25$  & $0.82\pm0.15$   \\
  $\nu_e$ & LS220 & $-1347.9\pm82.6$  & $1003.5\pm50.5$  & $3.12\pm1.66$  & $-2.17\pm1.01$  & $9.64\pm6.03$  & $-5.36\pm3.68$ & $1.98\pm0.85$ & $-0.89 \pm 0.52$   \\
 \hline
 $\bar{\nu}_e$ & DD2 & $-4.07\pm0.20$ & $7.64\pm0.12$ & $0.58\pm0.02$ & $-0.03\pm0.01$ & $1.02\pm0.06$  & $3.42\pm0.04$  & $3.43\pm0.20$ & $0.76\pm0.12$   \\   
 $\bar{\nu}_e$ & SFHo & $-4.40\pm0.22$ & $8.23\pm0.13$  & $0.79\pm0.01$ & $-0.11\pm0.01$ & $1.09\pm0.04$ & $4.16\pm0.02$ & $3.73\pm0.05$  & $0.15\pm0.03$   \\ 
 $\bar{\nu}_e$ & SFHx & $-6.05\pm0.13$ & $9.59\pm0.08$ & $0.73\pm0.01$ & $-0.035\pm0.004$ & $0.43\pm0.22$  & $4.76\pm0.13$ & $3.25\pm0.11$  & $0.46\pm0.06$   \\
 $\bar{\nu}_e$ & LS220 & $-1050.0\pm65.1$ & $796.0\pm39.7$  & $3.37\pm1.67$  & $-2.31\pm1.02$  & $9.77\pm6.08$  & $-5.43\pm3.72$  & $1.92\pm0.83$ & $-0.85\pm0.51$   \\
 \hline
 \hline
 $\nu_\mu$ & DD2 & $-4.72\pm0.11$ & $8.14\pm0.07$  & $0.38\pm0.01$  & $0.03\pm0.01$ & $1.05\pm0.06$  & $3.43\pm0.04$ & $3.75\pm0.11$ & $0.50\pm0.07$  \\   
 $\nu_\mu$ & SFHo & $-6.37\pm0.18$ & $9.90\pm0.11$ & $0.62\pm0.02$ & $-0.03\pm0.01$ & $1.05\pm0.05$ & $4.35\pm0.03$ & $3.80\pm0.17$  & $0.11\pm0.10$  \\ 
 $\nu_\mu$ & SFHx & $-7.61\pm0.43$ & $10.82\pm0.26$ & $0.56\pm0.02$ & $0.03\pm0.01$ & $0.39\pm0.25$ & $4.93\pm0.15$ & $3.14\pm0.26$ & $0.55\pm0.16$  \\
  $\nu_\mu$ & LS220 & $-1482.0\pm155.4$ & $1165.3\pm94.9$ & $0.82\pm0.42$  & $-0.84\pm0.26$ & $0.57\pm0.26$ & $-0.30\pm0.16$ & $0.72\pm0.11$ & $-0.19\pm0.07$  \\
 \hline
 $\bar{\nu}_\mu$ & DD2 & $-6.11\pm0.42$ & $9.56\pm0.26$ & $0.40\pm0.02$ & $0.05\pm0.01$ & $0.94\pm0.08$ & $3.55\pm0.05$ & $3.56\pm0.27$ & $0.67\pm0.17$  \\   
 $\bar{\nu}_\mu$ & SFHo & $-7.85\pm0.23$ & $11.42\pm0.14$  & $0.66\pm0.01$ & $-0.02\pm0.01$ & $0.99\pm0.02$ & $4.47\pm0.01$  & $3.96\pm0.13$ & $0.06\pm0.08$   \\ 
 $\bar{\nu}_\mu$ & SFHx & $-9.22\pm0.43$ & $12.55\pm0.26$ & $0.63\pm0.02$ & $0.03\pm0.01$ & $0.41\pm0.22$  & $5.02\pm0.14$ & $3.46\pm0.20$ & $0.37\pm0.12$  \\
 $\bar{\nu}_\mu$ & LS220 & $-1608.5\pm407.4$ & $1309.1\pm248.8$  & $0.85\pm0.35$  & $-0.93\pm0.21$  & $0.37\pm0.15$  & $-0.19\pm0.09$  & $0.64\pm0.09$ & $-0.14\pm0.05$   \\
 \hline
  \hline
 $\nu_\tau$ & DD2 & $-4.60\pm0.18$ & $8.12\pm0.11$  & $0.42\pm0.01$ & $0.012\pm0.005$ & $1.05\pm0.06$  & $3.45\pm0.03$ & $4.04\pm0.13$ & $0.32\pm0.08$  \\   
 $\nu_\tau$ & SFHo & $-5.99\pm0.15$ & $9.70\pm0.09$ & $0.68\pm0.01$ & $-0.06\pm0.01$ & $1.11\pm0.02$ & $4.36\pm0.01$  & $4.27\pm0.13$  &  $-0.11\pm0.08$ \\ 
 $\nu_\tau$ & SFHx & $-7.17\pm0.32$ & $10.62\pm0.19$ & $0.64\pm0.02$ & $-0.009\pm0.012$ & $0.50\pm0.23$ & $4.91\pm0.14$ & $3.70\pm0.22$  & $0.27\pm0.13$  \\
  $\nu_\tau$ & LS220 & $-4008.7\pm943.6$ & $2955.1\pm576.3$ & $1.09\pm0.51$ & $-1.11\pm0.31$ & $0.45\pm0.20$ & $-0.24\pm0.12$ & $0.73\pm0.11$ & $-0.21\pm0.06$  \\
 \hline
 $\bar{\nu}_\tau$ & DD2 & $-4.98\pm0.20$ & $8.59\pm0.12$ & $0.42\pm0.01$ & $0.017\pm0.006$ & $1.03\pm0.06$ & $3.46\pm0.03$ & $3.99\pm0.13$  & $0.37\pm0.08$  \\   
 $\bar{\nu}_\tau$ & SFHo & $-6.49\pm0.16$ & $10.29\pm0.10$ & $0.69\pm0.01$ & $-0.06\pm0.01$ & $1.09\pm0.02$ & $4.37\pm0.01$  & $4.25\pm0.14$ & $-0.09\pm0.08$   \\ 
 $\bar{\nu}_\tau$ & SFHx & $-7.67\pm0.35$ & $11.22\pm0.21$  & $0.64\pm0.02$ & $-0.008\pm0.012$ & $0.50\pm0.23$ & $4.91\pm0.14$ & $3.69\pm0.22$  & $0.29\pm0.13$  \\
 $\bar{\nu}_\tau$ & LS220 & $-2988.6\pm809.8$ & $2310.9\pm494.6$ & $1.09\pm0.43$  & $-1.09\pm0.26$ & $0.42\pm0.12$ & $-0.22\pm0.07$ & $0.71\pm0.09$ & $-0.20\pm0.06$  \\
 \hline
 \end{tabular}
 \caption{Coefficients $K_0$ and $K_1$ with errors for the linear dependence on the PNS mass $K = K_0 + K_1\,M_{\rm NS}/M_\odot$ at fixed EoS, for electron (upper data block), muon (central data block) and tau (lower data block) neutrinos and antineutrinos,  obtained on grounds of simulation data in the time interval between 1\,s and $t_{\nu_i,{\rm c}}$.}
 \label{tab:massdep}
 \end{table}
 
\section{Tables for parameter values of the correlations between $\tau$ and $\alpha$}
\label{App:scatternu}
We report in Table~\ref{tab:scatternu} the best-fit values and the $1\,\sigma$ errors for the parameters $A$ and $B$ of the linear functions used for describing the correlations between $\tau$ (in seconds) and $\alpha$ [see Eq.~(\ref{eq:tau-alpha})]:
\begin{equation}
\tau (s) = A + B\,\alpha\,.
\nonumber
\end{equation}
As shown in Table~\ref{tab:scatternu}, the fit works better for neutrinos than for antineutrinos. Indeed, for $\bar{\nu}_e$ the error on the parameter $A$ is larger than its best-fit value (thus, $A$ is compatible with zero) for all the NS masses, and the same is true for $\bar{\nu}_\mu$ for the largest NS mass. Additionally, the quality of the fit is similar for $\nu_\mu$ and $\nu_\tau$, while it is slightly better for $\bar{\nu}_\tau$ compared to $\bar{\nu}_\mu$, since the relative error of the best-fit parameters for $\bar{\nu}_\tau$ is smaller. This reveals a small difference between the non-electron flavors. As a common trend, the parameter $B$ increases with the NS mass for all of the neutrino species. 

\begin{table}[t!]
 \centering
     \begin{tabular}{|c|c|c|c|}
    \hline
   Neutrino & $M_\mathrm{NS}$ [$M_\odot$] & $A$ [s] & $B$ [s] \\
    \hline
 $\nu_e$ & 1.36 & $2.85\pm 0.03$ & $6.17\pm 0.05$  \\
    $\nu_e$ & 1.44 & $2.80\pm 0.05$ & $6.84\pm 0.09$  \\
$\nu_e$ & 1.62 & $2.70\pm 0.08$ & $8.42\pm 0.15$  \\
$\nu_e$ & 1.77 & $1.95\pm 0.04$ & $10.98\pm 0.08$  \\ 
$\nu_e$ & 1.93 & $1.87\pm 0.17$ & $12.47\pm 0.31$  \\ 
\hline
$\nu_\mu$ & 1.36 & $2.32\pm0.04$ & $8.00\pm 0.08$  \\
$\nu_\mu$ & 1.44 & $2.45\pm0.11$ & $8.34\pm 0.19$  \\
$\nu_\mu$ & 1.62 & $2.31\pm 0.02$ & $9.96\pm 0.04$  \\
$\nu_\mu$ & 1.77 &  $2.01 \pm 0.32$ & $11.63\pm 0.59$ \\
$\nu_\mu$ & 1.93 & $2.09\pm 0.64$ & $12.80 \pm 1.17$ \\
\hline
$\nu_\tau$ & 1.36 & $2.25\pm0.04$ & $7.91\pm 0.07$  \\
$\nu_\tau$ & 1.44 & $2.37\pm0.24$ & $8.32\pm 0.42$  \\
$\nu_\tau$ & 1.62 & $2.32\pm 0.14$ & $9.83\pm 0.26$  \\
$\nu_\tau$ & 1.77 &  $2.15 \pm 0.46$ & $11.30\pm 0.83$ \\
$\nu_\tau$ & 1.93 & $1.77\pm 0.84$ & $13.38 \pm 1.53$ \\
    \hline
 \end{tabular}~~~~~~~~~~~~~~~~
      \begin{tabular}{|c|c|c|c|}
    \hline
   Neutrino & $M_\mathrm{NS}$ [$M_\odot$] & $A$ [s] & $B$ [s] \\
\hline
$\bar{\nu}_e$ & 1.36 & $0.33\pm 1.13$ & $9.76 \pm 1.79$  \\
$\bar{\nu}_e$ & 1.44 & $0.56\pm 1.53$ & $10.08 \pm 2.47$  \\
$\bar{\nu}_e$ & 1.62 & $0.79\pm1.85$ & $11.08\pm 3.04$  \\
$\bar{\nu}_e$ & 1.77 & $0.53\pm 2.82$ & $12.61\pm 4.68$  \\
$\bar{\nu}_e$ & 1.93 & $0.40\pm 4.15$ & $14.21\pm 6.98$  \\
\hline
$\bar{\nu}_\mu$ & 1.36 & $1.81\pm 0.16$ & $8.24\pm 0.27$  \\
$\bar{\nu}_\mu$ & 1.44 & $1.78\pm 0.44$ & $8.83\pm 0.73$  \\
$\bar{\nu}_\mu$ & 1.62 & $1.75\pm 0.32$ & $10.16\pm 0.54$  \\
$\bar{\nu}_\mu$ & 1.77 & $1.46 \pm 0.72$ & $11.62 \pm 1.19$  \\
$\bar{\nu}_\mu$ & 1.93 & $0.39\pm1.34$ & $14.53\pm 2.20$  \\
\hline
$\bar{\nu}_\tau$ & 1.36 & $2.20\pm0.03$ & $7.91\pm 0.06$  \\
$\bar{\nu}_\tau$ & 1.44 & $2.30\pm0.23$ & $8.34\pm 0.40$  \\
$\bar{\nu}_\tau$ & 1.62 & $2.22\pm 0.12$ & $9.88\pm 0.22$  \\
$\bar{\nu}_\tau$ & 1.77 &  $2.01 \pm 0.43$ & $11.41\pm 0.77$ \\
$\bar{\nu}_\tau$ & 1.93 & $1.56\pm 0.83$ & $13.56 \pm 1.49$ \\ 
    \hline
 \end{tabular}
\caption{Coefficients $A$ and $B$ with errors for the relation $\tau[\s] = A + B\,\alpha$, for neutrinos (left) and antineutrinos (right), for all flavors and NS masses, obtained on grounds of simulation data in the time interval between 1\,s and $t_{\nu_i,{\rm c}}$.}
\label{tab:scatternu}
\end{table}

\section{Further details on the impact of convection and muons}
\label{App:muons}

Here we give further details on the impact of convection and muons. Since models 1.62-SFHx-c and 1.61-LS220-c stop before the product $t L_{\nu_i}$ for all neutrino species is reduced to a value of 0.15 of the maximum, to make the comparison on a solid ground in this appendix we consider data up to $t_{\nu_i,{\rm c}}$ for simulations with DD2 and SFHo and up to $t_{\nu_i,{\rm Co}}$ for simulations with SFHx and LS220 (see Appendix~\ref{App:tmax} for more details). In this way, we take into account results for neutrinos and antineutrinos of all flavors up to the time when they reach the same reduction factor. Moreover, since non-electron flavors show, in general, a similar behavior and in simulations without muons, $\tau$ and $\mu$ neutrinos behave \emph{exactly} in the same way, for all the simulations considered in this Section we report values related only to $\nu_e$, $\bar{\nu}_e$, $\nu_\mu$ and $\bar{\nu}_\mu$.

\begin{figure}[t!]
\centering
\includegraphics[width=0.49\textwidth]{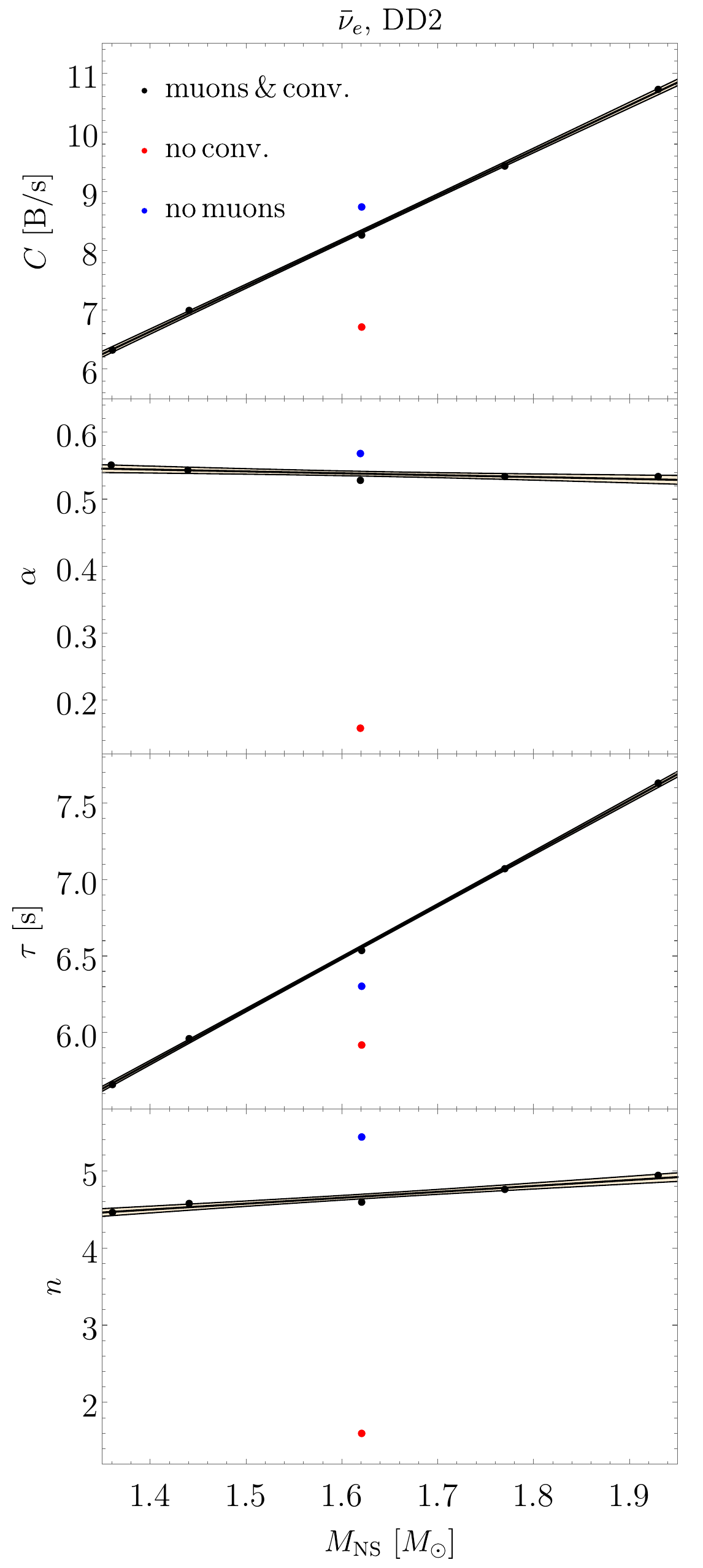}
\includegraphics[width=0.49\textwidth]{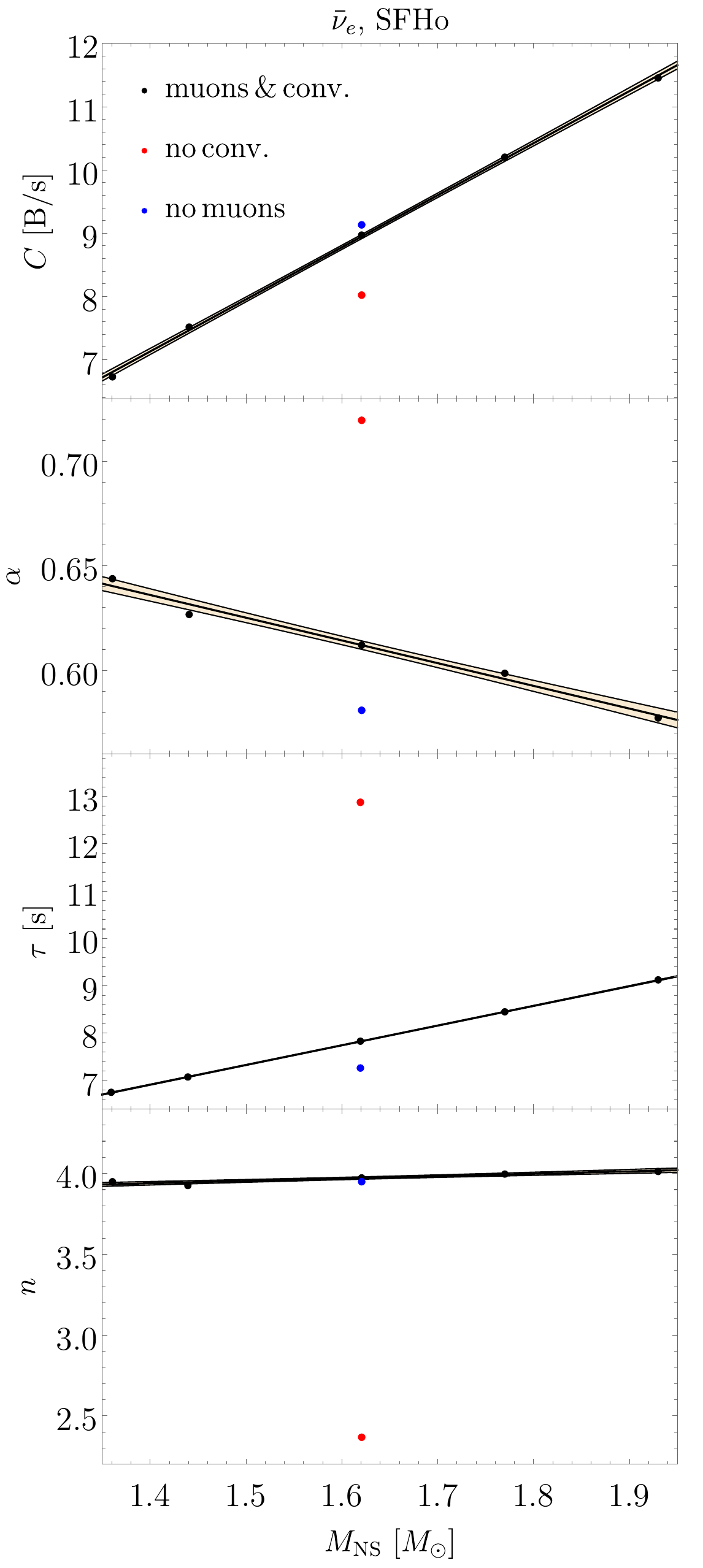}
\caption{Best-fit parameters $C$, $\alpha$, $\tau$ and $n$ as functions of the PNS mass for $\bar{\nu}_e$ and DD2 (left) respectively SFHo (right), with data up to $t_{\bar{\nu}_e,{\rm c}}$. The shaded areas represent the $1\sigma$ confidence bands. The black dots are obtained with simulations considering both convection and muons, red dots neglect convection, whereas blue dots correspond to simulations without muons.}
\label{Fig:nuebcm}
\end{figure}

We list the best-fit parameter values and their $1\,\sigma$ errors for the $1.62~M_\odot$ models with different EoS and different ingredients of the input physics for the luminosities of neutrinos in Table~\ref{tab:lnucm} and of antineutrinos in Table~\ref{tab:lnubarcm}. A better visualization of the change in the best-fit parameters can be obtained by plotting them as a function of the PNS mass for the different EoSs. As an example, we show in Fig.~\ref{Fig:nuebcm} the best-fit parameters for $\bar{\nu}_e$ as a function of $M_{\rm NS}$, for DD2 (left panels) and SFHo (right panels). Here, black dots are the values of the best-fit parameters obtained from simulations including both convection and muons, red dots are related to simulations without convection and blue dots to simulations without muons.

Tables~\ref{tab:lnucm} and \ref{tab:lnubarcm}, as well as Fig.~\ref{Fig:nuebcm}, show that, as a general trend, in simulations without convection $\tau$ becomes larger and $n$ smaller, i.e., the luminosity suppression starts at later times and it is slower. The only exception is found for $\tau_{\bar{\nu}_e}$ in model 1.62-DD2-c, which is smaller than $\tau_{\bar{\nu}_e}$ in 1.62-DD2 (see the red dot in the third panel from top on the left of Fig.~\ref{Fig:nuebcm}). This behavior is confirmed by inspecting the upper left panel in Fig.~\ref{Fig:tlnubecmEoS}, where the orange line (without convection) is peaked at earlier times compared to the blue (benchmark case) and the green (without muons) lines. As shown by the upper left panel in Fig.~\ref{Fig:nuebfit162cm}, even if $\tau_{\bar{\nu}_e}$ is smaller than in the benchmark case, the fit well reproduces the data, because the interplay between $\tau_{\bar{\nu}_e}$ and a much smaller $n_{\bar{\nu}_e}$ (compared to the complete case) well describes the slightly longer cooling time. Therefore, for $\bar{\nu}_e$ we observe a mathematical peculiarity connected to the fit function, reacting to the fact that $tL_{\bar{\nu}_e}$ is peaked at earlier times in the absence of convection (see the upper left panel in Fig.~\ref{Fig:tlnubecmEoS}), i.e.\ smaller $\tau_{\bar{\nu}_e}$, and it is characterized by a milder exponential suppression, i.e.\ smaller $n$, leading to a longer cooling time. On the other hand, in the absence of convection $\alpha$ becomes smaller in the case of DD2 (see the red dot in the second panel from top on the left of Fig.~\ref{Fig:nuebcm}) and larger in the case of SFHo (see the red dot in the second panel from top on the right of Fig.~\ref{Fig:nuebcm}), describing a change in the power-law behavior in the early cooling phase. In all the cases, the best-fit parameters in the absence of convection lie well outside the $1\,\sigma$ confidence band found for benchmark simulations, stressing the strong impact of convection on the neutrino signal. 

The weaker impact of muons on the neutrino signal is highlighted by the small changes in the best-fit parameters obtained from simulations without muons. In this case, for both DD2 and SFHo, in simulations without muons $\tau$ becomes slightly lower and $n$ is approximately equal or slightly larger for neutrinos and antineutrinos of all flavors (compare the first with the third line in the first two data blocks of Table~\ref{tab:lnucm} and Table~\ref{tab:lnubarcm}, as well as the black and blue dots in Fig.~\ref{Fig:nuebcm}). This means that in simulations without muons the suppression in the luminosity starts slightly before and it is a bit faster than in the full-physics cases. In contrast, $\alpha$ tends to increase for DD2 and to decrease for SFHo, even if the change in all cases is much smaller compared to the changes induced by the absence of convection.

Since simulations without muons are not available for SFHx and LS220 and for them $X_{\nu_i}^{\rm fin}>0.15$ in the absence of convection, we do not show the best-fit parameter values as a function of the NS mass in these two cases, but we only summarize the values of the best-fit parameters and their errors in the last two data blocks of Table~\ref{tab:lnucm} (for the electron flavor) and Table~\ref{tab:lnubarcm} (for the muon flavor), obtained by considering simulation data up to $t_{\nu_i,{\rm Co}}$. Even if the nominal values of the best-fit parameters in the benchmark simulations slightly change when switching from $t_{\nu_i,{\rm c}}$ to $t_{\nu_i,{\rm Co}}$, the impact of convection on simulations with SFHx and SFHo is similar, with an increase in $\tau$, and a decrease in $C$ and in $n$ in absence of convection. As expected, convection strongly affects also simulations with LS220. In this case, given the worse quality of the fit, neglecting convection leads to completely different values of the best-fit parameters compared to the benchmark case. Indeed, simulations without convection show positive values of $\alpha$, much larger values of $\tau$ and $n$, and drastically reduced values of $C$ compared to the complete-physics case, with all of these parameter values more closely related to the true magnitude and exponential decay time of the neutrino luminosities.

   \begin{table}[t!]
    \centering
        \begin{tabular}{|c|c|c|c|c||c|c|c|c|c|}
    \hline
   Model & $C_{\nu_e}$ [B/s] & $\alpha_{\nu_e}$ & $\tau_{\nu_e}$ [s] & $n_{\nu_e}$ & $C_{\bar{\nu}_e}$ [B/s] & $\alpha_{\bar{\nu}_e}$ & $\tau_{\bar{\nu}_e}$ [s] & $n_{\bar{\nu}_e}$ \\
    \hline
1.62-DD2 & $7.040\pm0.010$ & $0.430\pm0.002$ & $6.319\pm0.004$ & $3.959\pm0.008$ & $8.260\pm0.014$ & $0.529\pm0.002$ & $6.538\pm0.004$ & $4.599\pm0.012$ \\
1.62-DD2-c & $5.746\pm0.008$ & $0.190\pm0.003$ & $6.713\pm0.019$ & $1.763\pm0.005$ & $6.705\pm0.008$ & $0.157\pm0.004$ & $5.920\pm0.022$ & $1.594\pm0.005$ \\
1.62-DD2-m & $7.194\pm0.010$ & $0.479\pm0.002$ & $6.211\pm0.003$ & $5.208\pm0.011$ & $8.737\pm0.009$ & $0.567\pm0.001$ & $6.300\pm0.002$ & $5.443\pm0.009$ \\
    \hline
    \hline
1.62-SFHo & $8.308\pm0.005$ & $0.601\pm0.001$ & $7.743\pm0.002$ & $3.639\pm0.003$ & $8.967\pm0.008$ & $0.612\pm0.001$ & $7.823\pm0.003$ & $3.971\pm0.005$ \\
1.62-SFHo-c & $6.780\pm0.009$ & $0.625\pm0.001$ & $12.60\pm0.01$ & $2.347\pm0.005$ & $8.025\pm0.012$ & $0.720\pm0.002$ & $12.87\pm0.02$ & $2.367\pm0.006$ \\
1.62-SFHo-m & $8.713\pm0.007$ & $0.654\pm0.001$ & $7.493\pm0.002$ & $4.305\pm0.005$ & $9.128\pm0.006$ & $0.581\pm0.001$ & $7.265\pm0.002$ & $3.945\pm0.004$ \\
    \hline
    \hline
1.62-SFHx & $8.479\pm0.003$ & $0.6552\pm0.0003$ & $8.049\pm0.001$ & $4.045\pm0.002$ & $9.692\pm0.004$ & $0.7127\pm0.0005$ & $8.209\pm0.001$ & $4.544\pm0.004$ \\
1.62-SFHx-c & $6.433\pm0.007$ & $0.550\pm0.001$ & $11.97\pm0.02$ & $1.859\pm0.004$ & $7.488\pm0.009$ & $0.608\pm0.002$ & $11.87\pm0.03$ & $1.736\pm0.005$ \\
    \hline
    \hline
1.62-LS220 & $306.2\pm71.4$ & $-0.679\pm0.063$ & $0.044\pm0.011$ & $0.394\pm0.011$ & $305.5\pm56.9$ & $-0.702\pm0.052$ & $0.051\pm0.010$ & $0.406\pm0.009$ \\
1.61-LS220-c & $7.426\pm0.005$ & $0.878\pm0.001$ & $16.64\pm0.01$ & $2.779\pm0.003$  & $9.013\pm0.007$ & $1.013\pm0.001$ & $17.31\pm0.01$ & $2.942\pm0.005$ \\
    \hline
    \end{tabular}
\caption{Best-fit parameter values for $L_{\nu_i} = C \,t^{-\alpha}\,e^{-(t/\tau)^n}$ for $\nu_e$ (left) and $\bar{\nu}_e$ (right) for $M_{\rm NS}=1.62~M_\odot$ and different EoS, considering both convection and muons (upper lines), without convection (labeled with the suffix ``-c'') and without muons (labeled with the suffix ``-m''). We consider data up to $t_{\nu_i,{\rm c}}$ for simulations with DD2 and SFHo and up to $t_{\nu_i,{\rm Co}}$ for simulations with SFHx and LS220 (see Appendix~\ref{App:tmax} for more details). Simulations with SFHx and LS220 without muons are not available.}
\label{tab:lnucm}
\end{table}

   \begin{table}[t!]
    \centering
    \begin{tabular}{|c|c|c|c|c||c|c|c|c|}
    \hline
   Model & $C_{\nu_\mu}$ [B/s] & $\alpha_{\nu_\mu}$ & $\tau_{\nu_\mu}$ [s] & $n_{\nu_\mu}$ & $C_{\bar{\nu}_\mu}$ [B/s] & $\alpha_{\bar{\nu}_\mu}$ & $\tau_{\bar{\nu}_\mu}$ [s] & $n_{\bar{\nu}_\mu}$ \\
    \hline
1.62-DD2 & $8.428\pm0.011$ & $0.430\pm0.002$ & $6.595\pm0.003$ & $4.542\pm0.009$ & $9.271\pm0.012$ & $0.486\pm0.001$ & $6.676\pm0.003$ & $4.569\pm0.009$ \\
1.62-DD2-c & $6.831\pm0.010$ & $0.282\pm0.002$ & $8.025\pm0.014$ & $2.209\pm0.005$ & $7.406\pm0.010$ & $0.274\pm0.002$ & $7.635\pm0.015$ & $2.060\pm0.005$ \\
1.62-DD2-m & $8.780\pm0.007$ & $0.474\pm0.001$ & $6.407\pm0.001$ & $5.669\pm0.007$ & $9.174\pm0.007$ & $0.482\pm0.001$ & $6.406\pm0.001$ & $5.708\pm0.007$ \\
    \hline
    \hline
1.62-SFHo & $9.669\pm0.008$ & $0.582\pm0.001$ & $8.104\pm0.003$ & $4.022\pm0.005$ & $10.60\pm0.01$ & $0.633\pm0.001$ & $8.239\pm0.002$ & $4.094\pm0.003$ \\
1.62-SFHo-c & $8.394\pm0.012$ & $0.686\pm0.001$ & $14.53\pm0.01$ & $3.288\pm0.007$ & $9.037\pm0.013$ & $0.702\pm0.001$ & $14.31\pm0.01$ & $3.120\pm0.007$ \\
1.62-SFHo-m & $9.939\pm0.003$ & $0.5920\pm0.0004$ & $7.712\pm0.001$ & $4.320\pm0.002$ & $10.407\pm0.003$ & $0.6002\pm0.0004$ & $7.713\pm0.001$ & $4.344\pm0.002$ \\
    \hline
    \hline
     1.62-SFHx &  $10.00\pm0.01$ & $0.638\pm0.001$ & $8.427\pm0.002$ & $4.462\pm0.004$ & $11.125\pm0.005$ & $0.6887\pm0.0004$ & $8.540\pm0.001$ & $4.305\pm0.003$ \\
 1.62-SFHx-c &  $7.988\pm0.011$ & $0.645\pm0.001$ & $14.88\pm0.01$ & $2.736\pm0.006$ & $8.603\pm0.011$ & $0.664\pm0.001$ & $14.76\pm0.01$ & $2.647\pm0.006$ \\
    \hline
    \hline
1.62-LS220 &  $439.5\pm123.0$ & $-0.624\pm0.074$ & $0.040\pm0.012$ & $0.388\pm0.012$ & $494.5\pm139.4$ & $-0.785\pm0.077$ & $0.043\pm0.013$ & $0.404\pm0.012$ \\
 1.61-LS220-c &  $10.05\pm0.01$ & $0.950\pm0.001$ & $16.87\pm0.01$ & $2.526\pm0.004$ & $10.82\pm0.01$ & $0.976\pm0.001$ & $16.81\pm0.01$ & $2.540\pm0.004$ \\
    \hline
    \end{tabular}
\caption{Best-fit parameter values for $L_{\nu_i} = C \,t^{-\alpha}\,e^{-(t/\tau)^n}$ for $\nu_\mu$ (left) and $\bar{\nu}_\mu$ (right) for $M_{\rm NS}=1.62~M_\odot$ and different EoS, considering both convection and muons (upper lines), without convection (labeled with the suffix ``-c'') and without muons (labeled with the suffix ``-m''). We consider data up to $t_{\nu_i,{\rm c}}$ for simulations with DD2 and SFHo and up to $t_{\nu_i,{\rm Co}}$ for simulations with SFHx and LS220 (see Appendix~\ref{App:tmax} for more details). Simulations with SFHx and LS220 without muons are not available.}
\label{tab:lnubarcm}
\end{table} 

\section{Equation-of-state parameters}
\label{App:symm_en}
We report in Table~\ref{tab:symm_en} the parameters for the symmetry energies for the EoS cases used in our work. With the customary definitions of $x = (n - n_0)/(3 n_0)$ and the asymmetry parameter $\delta = 1 - 2Y_e$, the energy per nucleon can be expressed as 
\begin{equation}
    E(n) = -E_0 +\frac{1}{2}Kx^2 + \delta^2 ( J + Lx + \frac{1}{2} K_\mathrm{s} x^2 ) + ... \,,
\label{eq:enpernucl}
\end{equation}
with $E_0$ being the binding energy of symmetric matter at saturation density, $K$ the incompressibility, $J$ the symmetry energy, $L$ the slope of the symmetry energy, and $K_\mathrm{s}$ the curvature of the symmetry energy.

\begin{table}[]
    \centering
    \begin{tabular}{lcccccc}
        \hline
        EoS     & $n_0$ & $E_0$ & $K$ & $J$ & $L$ & $K_\mathrm{s}$ \\
                & [fm$^{-3}$] & [MeV] & [MeV] & [MeV] & [MeV] & [MeV] \\
        \hline
        DD2     & 0.149 & 16.0 & 243 & 31.7 & 55.0 & $-$93.2 \\
        SFHo    & 0.158 & 16.2 & 245 & 31.6 & 47.1 & $-$205 \\
        SFHx    & 0.160 & 16.2 & 239 & 28.7 & 23.2 & $-$40.0 \\
        LS220   & 0.155 & 16.0 & 220 & 28.6 & 74.0 & $-$24.0 \\
        \hline
    \end{tabular}
    \caption{Parameter values for the energy per nucleon around the nuclear saturation density $n_0$ according to Eq.~(\ref{eq:enpernucl}) for the EoSs used in the model simulations in our work. The values in this table are taken from Table~IV in \cite{Oertel:2017} and from entries for the respective EoS in the \textit{CompOSE} database \cite{Compose}.}  
    \label{tab:symm_en}
\end{table}

\end{document}